\documentclass{aa}
\usepackage[varg]{txfonts}

\usepackage{amsmath}
\usepackage{siunitx}
\usepackage{xfrac}

\usepackage{natbib}
\usepackage{color}
\usepackage{graphicx}
\usepackage{booktabs}
\usepackage{csquotes}
\makeatletter
  \renewcommand*\aa@pageof{, page \thepage{} of \pageref*{LastPage}}
\makeatother
\usepackage{hyperref}
\hypersetup{pdfpagemode = {UseNone},
            pdftitle = {Circumbinary discs: Numerical and physical behaviour},
            pdfcreator = {\LaTeX},
            pdfproducer = {pdfeTeX-0.\the\pdftexversion\pdftexrevision},
            pdfauthor = {Daniel Thun, Wilhelm Kley, Giovanni Picogna},
            pdfsubject = {},
            pdfview = {FitH},
            pdfstartview = {FitH},
            colorlinks = {true},
            linkcolor = [rgb]{0,0.35,0.7},
            citecolor = [rgb]{0,0.35,0.7},
            filecolor = [rgb]{0.61,0,0},
            urlcolor = [rgb]{0.61,0,0},
           }

\usepackage{bm}
\renewcommand{\vec}[1]{\bm{\mathrm{#1}}}

\newcommand*\dd[1]{\,\mathrm{d}#1}

\newcommand{\prt}[2]{\frac{\partial #1}{\partial #2}}
\newcommand{\Prt}[1]{\frac{\partial}{\partial #1}}

\newcommand*{\vnabla}{\vec{\nabla}}

\begin{document}

\title{Circumbinary discs: Numerical and physical behaviour}

\author{Daniel Thun \inst{1} \and
        Wilhelm Kley \inst{1} \and
        Giovanni Picogna \inst{2}}

\institute{
Institut für Astronomie und Astrophysik, Universität Tübingen,
Auf der Morgenstelle 10, D-72076 Tübingen, Germany\\
\email{\{daniel.thun@, wilhelm.kley@\}uni-tuebingen.de}\\
\and
Universitäts-Sternwarte, Ludwig-Maximilians-Universität München,
Scheinerstr. 1, D-81679 München, Germany\\
\email{picogna@usm.lmu.de}\\
}

\date{}

\abstract
{}
{
    Discs around a central binary system play an important role in star and
    planet formation and in the evolution of galactic discs. These
    circumbinary discs are strongly disturbed by the time varying potential of
    the binary system and display a complex dynamical evolution that is not well
    understood. Our goal is to investigate the impact of disc and binary
    parameters on the dynamical aspects of the disc.
}
{
    We study the evolution of circumbinary discs under the gravitational
    influence of the binary using two-dimensional hydrodynamical simulations. To
    distinguish between physical and numerical effects we apply three
    hydrodynamical codes. First we analyse in detail numerical issues concerning
    the conditions at the boundaries and grid resolution.  We then perform a
    series of simulations with different binary parameters (eccentricity, mass
    ratio) and disc parameters (viscosity, aspect ratio) starting from a
    reference model with Kepler-16 parameters.
}
{
    Concerning the numerical aspects we find that the length of the inner grid
    radius and the binary semi-major axis must be comparable, with free outflow
    conditions applied such that mass can flow onto the central binary. A closed
    inner boundary leads to unstable evolutions. 

    We find that the inner disc turns eccentric and  precesses for all
    investigated physical parameters. The precession rate is slow with periods
    ($T_\mathrm{prec}$) starting at around 500 binary orbits ($T_\mathrm{bin}$)
    for high viscosity and a high aspect ratio $H/R$ where the inner hole is
    smaller and more circular. Reducing $\alpha$ and $H/R$ increases the gap
    size and $T_\mathrm{prec}$ reaches 2500 $T_\mathrm{bin}$.  For varying
    binary mass ratios $q_\mathrm{bin}$ the gap size remains constant, whereas
    $T_\mathrm{prec}$ decreases with increasing $q_\mathrm{bin}$. 
    
    For varying binary eccentricities $e_\mathrm{bin}$ we find two separate
    branches in the gap size and eccentricity diagram. The bifurcation occurs at
    around  $e_\mathrm{crit} \approx 0.18$ where the gap is smallest with the
    shortest $T_\mathrm{prec}$. For $e_\mathrm{bin}$ lower and higher than
    $e_\mathrm{crit}$, the gap size and $T_\mathrm{prec}$ increase. Circular
    binaries create the most eccentric discs.
}
{}


\maketitle

\section{Introduction}\label{sec:intro}
Circumbinary discs are accretion discs that orbit a binary system that consists
for example of a binary star or a binary black hole system.  A very prominent
example of a circumbinary disc orbiting two stars is the system GG~Tau.  Due to
the large star separation, the whole system (stars and disc) can be directly
imaged by interferometric methods \citep{1999A&A...348..570G}.  Later data
yielded constraints on the size of the dust in the system and have indicated
that the central binary may consist of multiple stellar systems (see
\citet{2014ApJ...787..148A}, \citet{2014A&A...565L...2D} and references therein).
More recent observations have pointed to possible planet formation
\citep{2014Natur.514..600D} and streamers from the disc onto the stars
\citep{2017AJ....153....7Y}.  A summary of the properties of the GG~Tau system
is given in \citet{2016A&ARv..24....5D}.

An additional important clue for the existence and importance of circumbinary
discs is given by the observed circumbinary planets.  These are planetary
systems where the planets do not orbit one star, but instead orbit a binary star system.
These systems have been detected in recent years by the {\it Kepler Space
Mission} and have inspired an intense research activity.  The first such
circumbinary planet, Kepler-16b, orbits its host system consisting of a K-type
main-sequence star and an M-type red dwarf, in what is known as a P-type orbit with a
semi-major axis of $\SI{0.7048}{au}$ and a period of \num{228.7} days
\citep{2011Sci...333.1602D}. Presently there are ten Kepler circumbinary planets
known and a summary of the properties of the first five systems discovered is
given in \citet{2014IAUS..293..125W}
\footnote{Additional known main-sequence binaries with circumbinary planets besides
Kepler-16 are Kepler-34 and -35 \citep{2012Natur.481..475W}, Kepler-38
\citep{2012ApJ...758...87O}, Kepler-47 \citep{2012Sci...337.1511O}, Kepler-64
\citep{2013ApJ...768..127S}, Kepler-413 \citep{2014ApJ...784...14K}, Kepler-453
\citep{2015ApJ...809...26W}, and Kepler-1647 \citep{Kostov:2016}}.
The observations have shown that in these systems the stars are mutually
eclipsing each other, as well as the planets.  The simultaneous eclipses of
several objects is a clear indication of the flatness of the system because
the orbital plane of the binary star has to coincide nearly exactly with the
orbital plane of the planet. In this respect these systems are flatter than our
own solar system. Because planets form in discs, the original protoplanetary
disc in these systems had to orbit around both stars, hence it must have been a
circumbinary disc that was also coplanar with respect to the orbital plane of
the central binary.

In addition to their occurrence around younger stars, circumbinary discs are
believed to orbit supermassive black hole binaries in the centres of galaxies
\citep{1980Natur.287..307B}, where they may play an important role in the
evolution of the host galaxies \citep{2002ApJ...567L...9A}.  In the early phase
of the Universe there were many more close encounters between the young galaxies
that alread had black holes in their centres.  The gravitational tidal forces
between them often resulted in new merged objects that consisted of a central
black hole binary surrounded by a large circumbinary disc \citep[see e.g.][and
references therein]{2009MNRAS.393.1423C,2009MNRAS.398.1392L}.  The dynamical
behaviour of such a system is very similar to the one described above, where we
have two stars instead of black holes.

To understand the dynamics of discs around young binary stars such as GG~Tau or
the planet formation process in circumbinary discs or the dynamical evolution of
a central black hole binary in a galaxy, it is important to understand in detail
the behaviour of a disc orbiting a central binary. The strong gravitational
perturbation by the binary generates spiral waves in the disc; these waves transport
energy and angular momentum from the binary to the disc
\citep{1991MNRAS.248..754P} and subseqently alter the evolution of the binary
\citep{1991ApJ...370L..35A}. The most important impact of the angular momentum
transfer to the disc is the formation of an inner cavity in the disc with a very
low density whose size depends on disc parameters (such as viscosity or scale
height) and on the binary properties (mass ratio, eccentricity)
\citep{1994ApJ...421..651A}. 

The numerical simulation of these circumbinary discs is not trivial and over the
years several attempts have been made. The first simulations of circumbinary
discs used the smoothed particle hydrodynamics (SPH) method, a Lagrangian method where
the fluid is modelled as individual particles \citep{1996ApJ...467L..77A}.  On
the one hand, this method has the advantage that the whole system can be included
in the computational domain and hence the accretion onto the single binary
components can be studied. On the other hand, because of the finite mass of the
SPH-particles, the maximum density contrast that can be resolved is limited.

Later \citet{1997ASPC..121..792R} simulated circumbinary discs with the help of
a finite difference method on a polar grid. The grid allowed for a much larger
range in mass resolution than the SPH codes used to date. A polar grid is well
suited for the outer parts of a disc, but suffers from the fact that there will
always be an inner hole in the computational domain since the minimum radius
$R_\mathrm{min}$ cannot be made arbitrarily small. Decreasing the minimum radius
will also strongly reduce the required numerical time step, rendering long
simulations unfeasible. But since circumbinary disc simulations have to be
simulated for tens of thousands of binary orbits to reach a quasi-steady state,
the minimum radius is often chosen such that the motion of the central stars
is not covered by the computational domain, and the mass transfer from the disc
to the binary cannot be studied.

\citet{2002A&A...387..550G} used a special dual-grid technique to use the best
of the two worlds, the high resolution of the grid codes and the coverage of the
whole domain of the particle codes. For the outer disc they used a polar grid
and they overlaid the central hole with a Cartesian grid. In this way it was
possible to study the complex interaction between the binary and the disc in
greater detail. All these simulations indicated that despite the formation of an
inner gap, material can enter the inner regions around the stars crossing the
gap through stream like features to eventually become accreted onto the stars
and influence their evolution \citep{2002MNRAS.336..705B}.

Another interesting feature of circumbinary discs concerns their eccentricity.
As shown in hydrodynamical simulations for planetary mass companions, discs
develop a global eccentric mode even for circular binaries
\citep{2001A&A...366..263P}. The eccentricity is confined to the inner disc
region and shows a very slow precession \citep{2006A&A...447..369K}. These
results were confirmed for equal mass binaries on circular orbits
\citep{2008ApJ...672...83M}.  The disc eccentricity is excited by the 1:3 outer
eccentric Lindblad resonance \citep{1991ApJ...381..259L,2001A&A...366..263P},
which is operative for a sufficiently cleared out gap.  For circular binaries a
transition, in the disc structure from more circular to eccentric is thought
to occur for a mass ratio of secondary to primary above 1/25
\citep{2016MNRAS.459.2379D}. However, this appears to contradict the mentioned
results by \citet{2001A&A...366..263P} and \citet{2006A&A...447..369K} who show
that this transition already occurs for a secondary in the planet mass regime
with details depending on disc parameters such as pressure and viscosity.

Driven by the observation of planets orbiting eccentric binary stars and the
possibility of studying eccentric binary black holes, a large number of numerical
simulations have been performed over the last few years dealing with discs
around eccentric central binaries
\citep[e.g.][]{2013A&A...556A.134P,2014A&A...564A..72K, 2015MNRAS.448.3545D}.
The recent simulations of circumbinary discs by \citet{2013A&A...556A.134P},
\citet{2014A&A...564A..72K} and \citet{2015A&A...582A...5L} used grid-based
numerical methods on a polar grid with an inner hole. This raises questions
about the location and imposed boundary conditions at the innermost grid radius
$R_\mathrm{min}$.  In particular, the value of  $R_\mathrm{min}$ must be small
enough to capture the development of the disc eccentricity through gravitational
interaction between the binary and the disc, especially through the 1:3 Lindblad
resonance \citep{2013A&A...556A.134P}.  Therefore $R_\mathrm{min}$ has to be
chosen in a way that all important resonances lie inside the computational
domain.

In addition to the position of the inner boundary, there is no common agreement
on which numerical boundary condition better describes the disc. In simulations
concerning circumbinary planets mainly two types of boundary conditions are
used: closed inner boundaries \citep{ 2013A&A...556A.134P,2015A&A...582A...5L}
that do not allow for mass flow onto the central binary, and outflow
boundaries \citep{2014A&A...564A..72K,2015A&A...581A..20K} that do allow for
accretion onto the binary. While \citet{2015A&A...582A...5L} do not find a
significant difference between these two cases, \citet{2014A&A...564A..72K} see
a clear impact on the surface density profile, and they were also able to
construct discs with (on average) constant mass flow through the disc, which is
not possible for closed boundaries.  Driven by these discussions in the
literature we decided to perform dedicated numerical studies to test the impact
of the location of $R_\mathrm{min}$, the chosen boundary condition, and other
numerical aspects in more detail.  During our work we became aware of other
simulations that were tackling similar problems. In October 2016 alone three
publications appeared on {\tt astro-ph} that described numerical simulations of
circumbinary discs
\citep{2017MNRAS.464.3343F,2017MNRAS.466.1170M,2017MNRAS.465.4735M}. While the
first paper considered SPH-simulations, in the latter two papers grid-codes were
used and in both the boundary condition at $R_\mathrm{min}$ is discussed. 

In our paper we start out with numerical considerations and investigate the
necessary conditions at the inner boundary in detail, and we discuss
aspects of the other recent results in the presentation of our findings below.
Additionally, we present a detailed parameter study of the dynamical behaviour
of circumbinary discs as a function of binary and disc properties.

We organised this paper in the following way. In Sect.~\ref{sec:setup} we
describe the numerical and physical setup of our circumbinary simulations. In
Sect.~\ref{sec:codes} we briefly describe the numerical methods of the
different codes we used for our simulations. In Sect.~\ref{sec:num_para} we
examine the disc structure by varying the inner boundary condition and its
location. The influence of different binary parameters are examined in
Sect.~\ref{sec:bin_para}. Different disc parameters and their influence are
studied in Sect.~\ref{sec:disc_para}. Finally we summarise and discuss our
results in Sect.~\ref{sec:results}.

\section{Model setup}\label{sec:setup}
To study the evolution of circumbinary discs we perform locally isothermal
hydrodynamical simulations. As a reference system we consider a binary star having
the properties of Kepler-16, whose dynamical parameters are presented in
Table~\ref{tab:kepler_16}.  This system has a typical mass ratio and a moderate
eccentricity of the orbit, and it has been studied frequently in the literature.
In our parameter study we start from this reference system and vary the
different aspects systematically. 

\begin{table}
    \caption{Orbital parameters of the Kepler-16 system.}
    \label{tab:kepler_16}
    \centering
    \begin{tabular}{cccccc}
        \midrule\midrule
        $M_A\;[M_\sun]$ & $M_B\;[M_\sun]$ & $q_\mathrm{bin}$ &
        $a_\mathrm{bin}\;[\mathrm{au}]$ & $e_\mathrm{bin}$ &
        $T_\mathrm{bin}\;[\mathrm{d}]$ \\
        \midrule
        0.69 & 0.20 & 0.29 & 0.22 & 0.16 & 41.08 \\
        \midrule
    \end{tabular}
    \tablefoot{Values taken from \citet{2011Sci...333.1602D} (rounded to two
        decimals). The mass ratio is defined as $q_\mathrm{bin} = M_B/M_A$.}
\end{table}

\subsection{Physics and equations}\label{ssec:physics}
Inspired by the flatness of the observed circumbinary planetary systems, for
example in the Kepler-16 system the motion takes place in single plane to within
$0.5^\circ$ \citep{2011Sci...333.1602D}, we make the following two assumptions:
\begin{enumerate}
    \item The vertical thickness $H$ of the disc is small compared to the
        distance from the centre $R$;
    \item There is no vertical motion.
\end{enumerate}
It is therefore acceptable to reduce the problem to two dimensions by
vertically averaging the hydrodynamical equations. In this case it is meaningful
to work in cylindrical coordinates $(R, \varphi, z)$\footnote{With $\vec{r}$ we
denote the three-dimensional positional vector $\vec{r} = R\vec{\hat{e}_R} +
z\vec{\hat{e}_z}$ and with $\vec{R}$ the two-dimensional positional vector in
the $R-\varphi$-plane $\vec{R} = R \vec{\hat{e}_R}$} and the averaging process
is done in the $z$-direction. Furthermore, we choose the centre of mass of the
binary as the origin of our coordinate system with the vertical axis aligned
with the rotation axis of the binary. The averaged hydrodynamical equations are
then given by
\begin{align}\label{eq:hydro}
    &\Prt{t}\Sigma + \vnabla \cdot \left(\Sigma \vec{u} \right) = 0\, ,\\
    &\Prt{t}\left(\Sigma \vec{u} \right) + \vnabla \cdot \left(\Sigma \vec{u}
    \otimes \vec{u} - \vec{\Pi} \right) = -\vnabla P -\Sigma \vnabla \Phi\, ,
\end{align}
where $\Sigma = \int_{-\infty}^{\infty} \varrho \dd{z}$ is the surface density,
$P = \int_{-\infty}^{\infty} p \dd{z}$ the vertically integrated pressure, and
$\vec{u} = (u_R, u_\varphi)^\mathrm{T}$ the two-dimensional velocity vector.
We close this system of equations with a locally isothermal equation of state
\begin{equation}\label{eq:eos}
    P = c_\mathrm{s}^2(R) \Sigma \, ,
\end{equation}
with the local sound speed $c_\mathrm{s}(R)$.

The gravitational potential $\Phi$ of both stars is given by
\begin{equation}\label{eq:potential}
    \Phi = - \sum_{k=1}^2 \frac{G M_k}{\left[\left(\vec{R} - \vec{R}_k\right)^2 +
    \left(\varepsilon H \right)^2 \right]^{1/2}}\, .
\end{equation}
Here $M_k$ denotes the mass of the $k$-th star, $G$ is the gravitational
constant, and $\vec{R} - \vec{R}_k$ a vector from a point in the disc to the
primary or secondary star. From a numerical point of view, the smoothing factor
$\varepsilon H$ is not necessary if the orbit of the binary is not inside the
computational domain. We use this smoothing factor to account for the vertically
extended three-dimensional disc in our two-dimensional case
\citep{2012A&A...541A.123M}. For all simulations we use a value of $\varepsilon
= 0.6$.

In our simulations with no bulk viscosity the viscous stress tensor $\vec{\Pi}$
is given in tensor notation by
\begin{equation}\label{eq:viscous_stress}
    \vec{\Pi}_{ij} = 2\eta \left[ \frac{1}{2} \left( u_{j;i} +
    u_{i;j} \right)  - \frac{1}{3}\delta_{ij} (\vnabla \cdot \vec{u})\right]\, ,
\end{equation}
where $\eta = \Sigma \nu$ is the vertically integrated dynamical viscosity
coefficient. To model the viscosity in the disc we use the $\alpha$-disc model
by \citet{1973A&A....24..337S}, where the kinematic viscosity is given by $\nu =
\alpha c_\mathrm{s} H$. The parameter $\alpha$ is less than one, and for our
reference model we use $\alpha = 0.01$.

To calculate the aspect ratio $h = \sfrac{H}{R}$ we assume a vertical
hydrostatic equilibrium
\begin{equation}\label{eq:vert_hydro_equil}
    \frac{1}{\varrho} \prt{p}{z} = - \prt{\Phi}{z}\, ,
\end{equation}
with the density $\varrho$ and pressure $p$. To solve this equation we use the
full three-dimensional potential of the binary $\Phi = -\sum_k
\frac{GM_k}{|\vec{r} - \vec{r}_k|}$ and the isothermal equation of state $p =
c_\mathrm{s}^2(R)\varrho$ (we also assume an isothermal disc in the $z$-direction).  
Integration over $z$ yields
\begin{equation}\label{eq:dens_z}
    \varrho = \varrho_0 \exp\left\{ -\frac{1}{2} \frac{z^2}{H^2} \right\}\, ,
\end{equation}
with the disc scale height \citep{2002A&A...387..550G}
\begin{equation}\label{eq:disc_height}
    H = \left[\sum_k \frac{1}{c_\mathrm{s}^2}
        \frac{GM_k}{|\vec{R} - \vec{R}_k|^3} \right]^{-\frac{1}{2}}
\end{equation}
and the midplane density
\begin{equation}\label{eq:midplane_density}
    \varrho_0 = \frac{\Sigma}{\sqrt{2\pi} H}\, ,
\end{equation}
which can be calculated by using the definition of the surface density.

If we concentrate the mass of the binary $M_\mathrm{bin} = M_A + M_B$
in its centre of mass ($\vec{R}_k \rightarrow 0$), this reduces to
\begin{equation}\label{eq:simple_disc_height}
    H = \frac{c_\mathrm{s}}{u_\mathrm{K}} R\, ,
\end{equation}
with the Keplerian velocity $u_\mathrm{K} = \sqrt{GM_\mathrm{bin}/R}$.
Test calculations with the simpler disc height~\eqref{eq:simple_disc_height}
showed no significant difference compared to calculations with the more
sophisticated disc height~\eqref{eq:disc_height}. Therefore, we use the simpler
disc height in all our calculation to save some computation costs.

For our locally isothermal simulations we use a temperature profile of $T
\propto R^{-1}$, which corresponds to a disc with constant aspect ratio. The
local sound speed is then given by $c_\mathrm{s}(R) = h u_\mathrm{K} \propto
R^{-\sfrac{1}{2}}$. If not stated otherwise we use a disc aspect ratio of
$h = 0.05$.

\subsection{Initial disc parameters}\label{ssec:disc_para}
For the initial disc setup we follow~\citet{2015A&A...582A...5L}. The initial
surface density in all our models is given by
\begin{equation}
    \Sigma(t=0) = f_\mathrm{gap} \Sigma_\mathrm{ref} R^{-\alpha_\Sigma}\, ,
\end{equation}
with the reference surface density $\Sigma_\mathrm{ref} =
10^{-4}\,M_\sun\,\mathrm{au}^{-2}$.  The initial slope is $\alpha_\Sigma = 1.5$
and the gap function, which models the expected cavity created by the
binary-disc interaction, is given by
\begin{equation}
    f_\mathrm{gap} = \left[1 + \exp\left(- \frac{R-R_\mathrm{gap}}{
    \Delta R} \right) \right]^{-1}
\end{equation}
\citep{2002A&A...387..550G}, with the transition width $\Delta R =
0.1\,R_\mathrm{gap}$ and the estimated size of the gap $R_\mathrm{gap} =
2.5\,a_\mathrm{bin}$ \citep{1994ApJ...421..651A}.  The initial radial velocity
is set to zero $u_R(t=0) = 0$, and for the initial azimuthal velocity we choose
the local Keplerian velocity $u_\varphi(t=0) = u_\mathrm{K}$.

\subsection{Initial binary parameters}\label{ssec:bin_para}
For models run with \textsc{Pluto} or \textsc{Fargo} we start the binary at 
periastron at time $t=0$ (upper signs in equation \eqref{eq:primary} and
\eqref{eq:secondary}). Models carried out with \textsc{Rh2d} start the binary at 
apastron (lower signs in equation \eqref{eq:primary} and
\eqref{eq:secondary}).
\begin{align}
    \vec{R}_A &=
    \begin{pmatrix}K_1 a_\mathrm{bin}(1\mp e_\mathrm{bin})\\ 0 \end{pmatrix}\, , &
    \vec{v}_A &=
    \begin{pmatrix} 0 \\ K_1 \frac{2\pi}{T_\mathrm{bin}} a_\mathrm{bin}
    \sqrt{\frac{1 \pm e_\mathrm{bin}}{1 \mp e_\mathrm{bin}}}\end{pmatrix}\, ,
    \label{eq:primary} \\
    \vec{R}_B &=
    \begin{pmatrix}-K_2 a_\mathrm{bin}(1\mp e_\mathrm{bin})\\ 0 \end{pmatrix}\, ,&
    \vec{v}_B &=
    \begin{pmatrix} 0 \\ -K_2\frac{2\pi}{T_\mathrm{bin}}a_\mathrm{bin}
    \sqrt{\frac{1 \pm e_\mathrm{bin}}{1\mp e_\mathrm{bin}}}\end{pmatrix}\, ,
    \label{eq:secondary}
\end{align}
with $K_1 = {M_B}/{M_\mathrm{bin}}$, $K_2 = {M_A}/{M_\mathrm{bin}}$, and the period
of the binary $T_\mathrm{bin}=2\pi \left[a^3_\mathrm{bin}/(G
M_\mathrm{bin})\right]^{1/2}$.

\subsection{Numerics}\label{ssec:numerics}
In our simulations we use a logarithmically increasing grid in the $R$-direction and
a uniform grid in the $\varphi$-direction.  The physical and numerical parameter
for our reference system used in our extensive parameter studies in
Sect.~\ref{sec:bin_para} and Sect.~\ref{sec:disc_para} are quoted in
Table~\ref{tab:ref_setup} below. In the radial direction the computational
domain ranges from $R_\mathrm{min} = \SI{0.25}{au}$ to $R_\mathrm{max} =
\SI{15.4}{au}$ and in the azimuthal direction from $0$ to $2\pi$, with a
resolution of $762\times 582$ grid cells.  For our numerical studies we also use
also different configurations as explained below.

Because of the development of an inner cavity where the surface density drops
significantly, we use a density floor $\Sigma_\mathrm{floor} = 10^{-9}$ (in code
units) to avoid numerical difficulties with too low densities. Test 
simulations using lower floor densities did not show any differences.

At the outer radial boundary we implement a closed boundary condition where the
azimuthal velocity is set to the local Keplerian velocity. At the inner radial
boundary the standard condition is the zero-gradient outflow condition as in
\citet{2014A&A...564A..72K} or \citet{2017MNRAS.466.1170M}, but we also test
different possibilities such as closed boundaries or the viscous outflow
condition \citep{2008A&A...486..617K,2017MNRAS.465.4735M}.
 
The open boundary is implemented in such a way that gas can leave the computational
domain but cannot reenter it. This is done by using zero-gradient boundary
conditions ($\partial/\partial R = 0$) for $\Sigma$ and negative $u_R$. For
positive $u_R$ we use a reflecting boundary to prevent mass in-flow. Since there
is no well-defined Keplerian velocity at the inner boundary, due to the strong
binary-disc interaction, we also use a zero-gradient boundary condition for the
angular velocity $\Omega_\varphi = u_\varphi/R$. By using the zero-gradient
condition for the angular velocity instead of the azimuthal velocity we ensure a
zero-torque boundary.  In the $\varphi$-direction we use periodic boundary
conditions.

In all our simulations we use dimensionless units. The unit of length is
$R_0 = \SI{1}{au}$, the unit of mass is the sum of the primary and secondary
mass $M_0 = M_A + M_B$ and the unit of time is $t_0 = \sqrt{R_0^3/(G M_0)}$ so
that the gravitational constant $G$ is equal to one. The unit density is then
given by $\Sigma_0 = M_0 / R_0^2$.

\subsection{Monitored parameters}\label{ssec:monitored_para}
Since our goal is to study the evolution of the disc under the influence of the
binary we calculated the disc eccentricity
$e_\mathrm{disc}$ and the argument of the disc periastron $\varpi_\mathrm{disc}$ ten times per binary orbit. To
calculate these quantities we treat each cell as a particle with the cells mass
and velocity on an orbit around the centre of mass of the binary. Thus, the
eccentricity vector of a cell is given by
\begin{equation}\label{eq:ecc_vector}
    \vec{e}_\mathrm{cell} = \frac{\vec{u} \times \vec{j}}{G M_\mathrm{bin}} -
    \frac{\vec{R}}{|\vec{R}|}
\end{equation}
with the specific angular momentum $\vec{j} = \vec{R} \times \vec{u}$ of that
cell. We have a flat system, thus the angular momentum vector only has a
$z$-component. The eccentricity $e_\mathrm{cell}$ and longitude of periastron
$\varpi_\mathrm{cell}$ of the cell's orbit are therefore given by
\begin{align}
    &e_\mathrm{cell} = |\vec{e}_\mathrm{cell}|\, ,\\
    &\varpi_\mathrm{cell} = \mathrm{atan2}(e_y, e_x)\, .
\end{align}
The global disc values are then calculated through a mass-weighted average of
each cell's eccentricity and longitude of periastron \citep{2006A&A...447..369K}
\begin{align}
    &e_\mathrm{disc} = \left[\int_{R_1}^{R_2} \int_0^{2\pi} \!
        \Sigma e_\mathrm{cell} R \dd{\varphi}\dd{R} \right] /
        \left[\int_{R_1}^{R_2} \int_0^{2\pi} \! \Sigma R
        \dd{\varphi}\dd{R}  \right]\, ,\label{eq:e_disc} 
       \\
    &\varpi_\mathrm{disc} = \left[\int_{R_1}^{R_2} \int_0^{2\pi}
        \!  \Sigma \varpi_\mathrm{cell} R \dd{\varphi}\dd{R} \right] /
        \left[\int_{R_1}^{R_2} \int_0^{2\pi} \! \Sigma R
        \dd{\varphi}\dd{R}  \right]\, .\label{eq:peri_disc}
\end{align}
The integrals are simply evaluated by summing over all grid cells. The lower
bound is always $R_1 = R_\mathrm{min}$. For the disc eccentricity we integrate
over the whole disc ($R_2 = R_\mathrm{max}$) if not stated otherwise, whereas
for the disc's longitude of periastron it is suitable to integrate only over the
inner disc ($R_2 = \SI{1.0}{au}$) to obtain a well-defined value of
$\varpi_\mathrm{disc}$ since animations clearly show a precession of the inner
disc \citep[see e.g.][]{2015A&A...581A..20K}. The radial eccentricity
distribution of the disc is given by
\begin{equation}\label{eq:e_ring}
    \begin{split}
    e_\mathrm{ring}(R) = \left[\int_{R}^{R+\dd{R}} \int_0^{2\pi} \!
        \Sigma e_\mathrm{cell} R' \dd{\varphi}\dd{R'} \right] \\
        / \left[\int_{R}^{R+\dd{R}} \int_0^{2\pi} \! \Sigma R'
        \dd{\varphi}\dd{R'}  \right]\, .
    \end{split}
\end{equation}

\section{Hydrodynamic codes}\label{sec:codes}
Since the system under analysis is, in particular near the central binary, very
dynamical we decided to compare the results from three different hydrodynamic
codes to make sure that the observed features are physical and not numerical
artefacts. We use codes with very different numerical approaches. \textsc{Pluto}
solves the hydrodynamical equations in conservation form with a Godunov-type
shock-capturing scheme, whereas \textsc{Rh2d} and \textsc{Fargo} are second-order
upwind methods on a staggered mesh. In the following sections we describe each
code and its features briefly.

\subsection{\textsc{Pluto}}\label{ssec:pluto}
We use an in-house developed GPU version of the \textsc{Pluto} 4.2 code
\citep{2007ApJS..170..228M}. \textsc{Pluto} solves the hydrodynamic equations
using the finite-volume method which evolves volume averages in time. To evolve
the solution by one time step, three substeps are required. First, the cell
averages are interpolated to the cell interfaces, and then in the second step a
Riemann problem is solved at each interface. In the last step the averages are
evolved in time using the interface fluxes. 

For all three substeps \textsc{Pluto} offers many different numerical options.
We found that the circumbinary disc model is very sensitive to the
combination of these options. Third-order interpolation and time evolution
methods lead very quickly to a negative density from which the code cannot
recover, even though we set a density floor.  We therefore use a second-order
reconstruction of states and a second-order Runge-Kutta scheme for the time
evolution. Another important parameter is the limiter, which is used during the
reconstruction step to avoid strong oscillations. For the most diffusive
limiter, the minmod limiter, no convergence is reached for higher grid
resolutions. For the least diffusive limiter, the mc limiter, the code again produces
negative densities and aborts the calculation. Thus, we use the van Leer
limiter which, in kind of diffusion, lies between the minmod and the mc limiter.

For the binary position we solve Kepler's equation using the Newton–Raphson
method at each Runge-Kutta substep.

\subsection{\textsc{Rh2d}}\label{ssec:rh2d}
The \textsc{Rh2d} code is a two-dimensional radiation hydrodynamics code
originally designed to study boundary layers in accretion discs
\citep{1989A&A...208...98K}, but later extended to perform flat disc simulations
with embedded objects \citep{1999MNRAS.303..696K}.  It is based on the second-order
upwind method described in \citet{1984ApJS...55..211H} and
\citet{1985A&A...143...59R}. It uses a staggered grid with second-order spatial
derivatives and through operator-splitting the time integration is
semi-second order.  Viscosity can be treated explicitly or implicitly,
artificial viscosity can be applied, and the Fargo algorithm has also been
added.  The motion of the binary stars is integrated using a fourth-order
Runge-Kutta algorithm.

\subsection{\textsc{Fargo}}\label{ssec:fargo}
We adopted the \textsc{adsg} version of \textsc{Fargo} \citep{Masset:2000,
Baruteau:2008} updated by \cite{Mueller:2012}.  This code uses a staggered mesh
finite difference method to solve the hydrodynamic equations. Conceptually, the
\textsc{Fargo} code uses the same methods as \textsc{Rh2d} or the
\textsc{Zeus} code, but employs the special Fargo algorithm that avoids the time
step limitations that are due to the rotating shear flow \citep{Masset:2000}. In
our situation the application of the Fargo algorithm is not always beneficial
because of the larger deviations from pure Keplerian flow near the binary.
The position of the binary stars is calculated by a fifth-order Runge-Kutta
algorithm.

\section{Numerical considerations}\label{sec:num_para}
Before describing our results on the disc dynamics, we present 
two important numerical issues that can have a dramatic influence on
the outcome of the simulations: the inner boundary condition (open
or closed) and the location of the inner radius of the disc. We show that
\enquote{unfortunate} choices can lead to incorrect results.  In this section we use a
numerical setup different from that in Table \ref{tab:ref_setup}. Specifically,
the base model has a radial extent from $R_\mathrm{min} = \SI{0.25}{au}$ to
$R_\mathrm{max} = \SI{4.0}{au}$, which is covered with $448\times512$ gridcells
for simulations with \textsc{Rh2d} and \textsc{FARGO}, and $512\times580$ with
\textsc{Pluto} (see Appendix~\ref{sec:convergence} for an explanation of why we use
different resolutions for different codes). For simulations with varying inner
radii, the number of gridcells in the radial direction is adjusted to always give
the same resolution in the overlapping domain.

\subsection{Inner boundary condition}\label{ssec:boundary}
All simulations using a polar-coordinate grid expericence the same problem: there
is a hole in the computational domain because $R_\mathrm{min}$ cannot be
zero.  Usually this hole exceeds the binary orbit. Therefore, the area where gas
flows from the disc onto the binary and where circumstellar discs around the
binary components form is not part of the simulation. This
complex gas flow around the binary has been shown by e.g.
\citet{2002A&A...387..550G} with a special dual-grid technique that covers the 
whole inner cavity. Their code is no longer in use and 
modern efficient codes (\textsc{Fargo} and \textsc{Pluto}) that run in parallel
do not have the option of a dual-grid.

To reproduce nevertheless some results of \cite{2002A&A...387..550G}, we carried
out a simulation on a polar grid with an inner radius of $R_\mathrm{min} =
\SI{0.02}{au}$ so that both orbits of the primary and secondary lie well inside
the computational domain.
\begin{figure}
    \resizebox{\hsize}{!}{\includegraphics{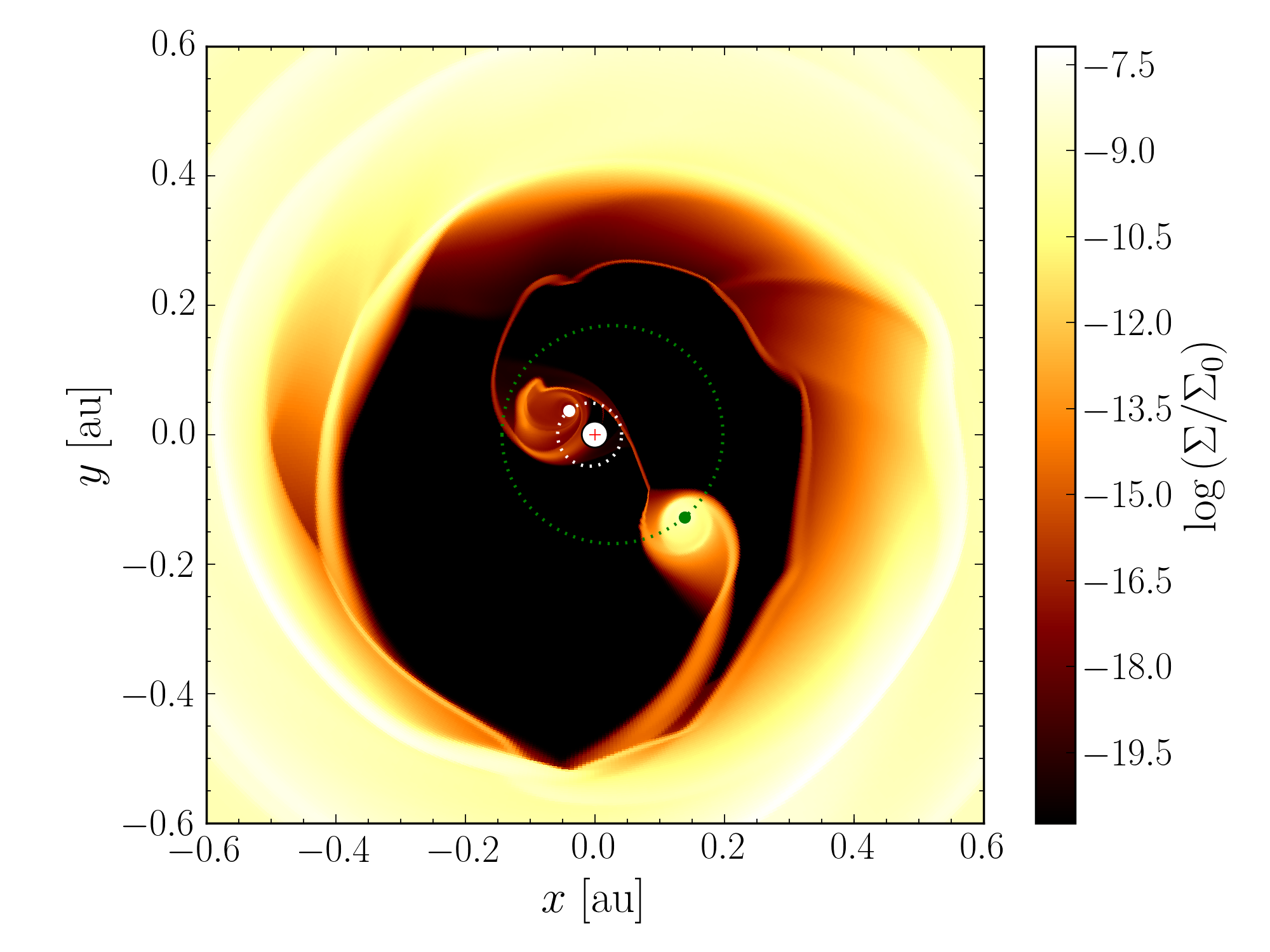}}
    \caption{Two-dimensional plot of the inner disc of one of our locally
    isothermal Kepler-16 simulations where both orbits of the binary lie inside
    the computational domain.  The logarithm of the surface density is
    colour-coded. The orbits of the primary and secondary are shown in white and
    green. The white inner region lies outside the computational domain; the red
    cross marks the centre of mass of the binary. A video of the simulation can
    be found online.}
    \label{img:2d_sigma_rmin0.02}
\end{figure}
A snapshot of this simulation is plotted in Fig.~\ref{img:2d_sigma_rmin0.02}.
The logarithm of the surface density is colour-coded, while the orbits of the
primary and secondary stars are shown in white and green. The red cross marks
the centre of mass of the binary, which lies outside the computational domain.
Figure~\ref{img:2d_sigma_rmin0.02} shows circumstellar discs
around the binary components, as well as a complex gas flow through the gap onto
the binary. Although including the binary orbit inside the computational would
be desirable, it is at the moment not feasible for long-term simulations
because of the strong time step restriction resulting from the small cell size
at the inner boundary. This means that for simulations where the binary orbit is
not included in the computational domain, the boundary condition at the inner
radius has to allow for flow into the inner cavity, at least approximately.
Two boundary conditions are usually used in circumbinary disc simulations:
closed boundaries \citep{2015A&A...582A...5L, 2013A&A...556A.134P} and outflow
boundaries \citep{2014A&A...564A..72K, 2015A&A...581A..20K,
2017MNRAS.466.1170M}. 

Given the importance of the boundary condition at $R_\mathrm{min}$, we carried
out dedicated simulations for a model adapted from \citet{2015A&A...582A...5L} and
used an inner radius of $R_\mathrm{min}= \SI{0.345}{au}$, but otherwise it was identical
to our standard model. We applied closed boundaries and open ones in
order to examine their influence on the disc structure. We also
varied different numerical parameters (resolution, radial grid spacing,
integrator) for this simulation series.
\begin{figure}
    \resizebox{\hsize}{!}{\includegraphics{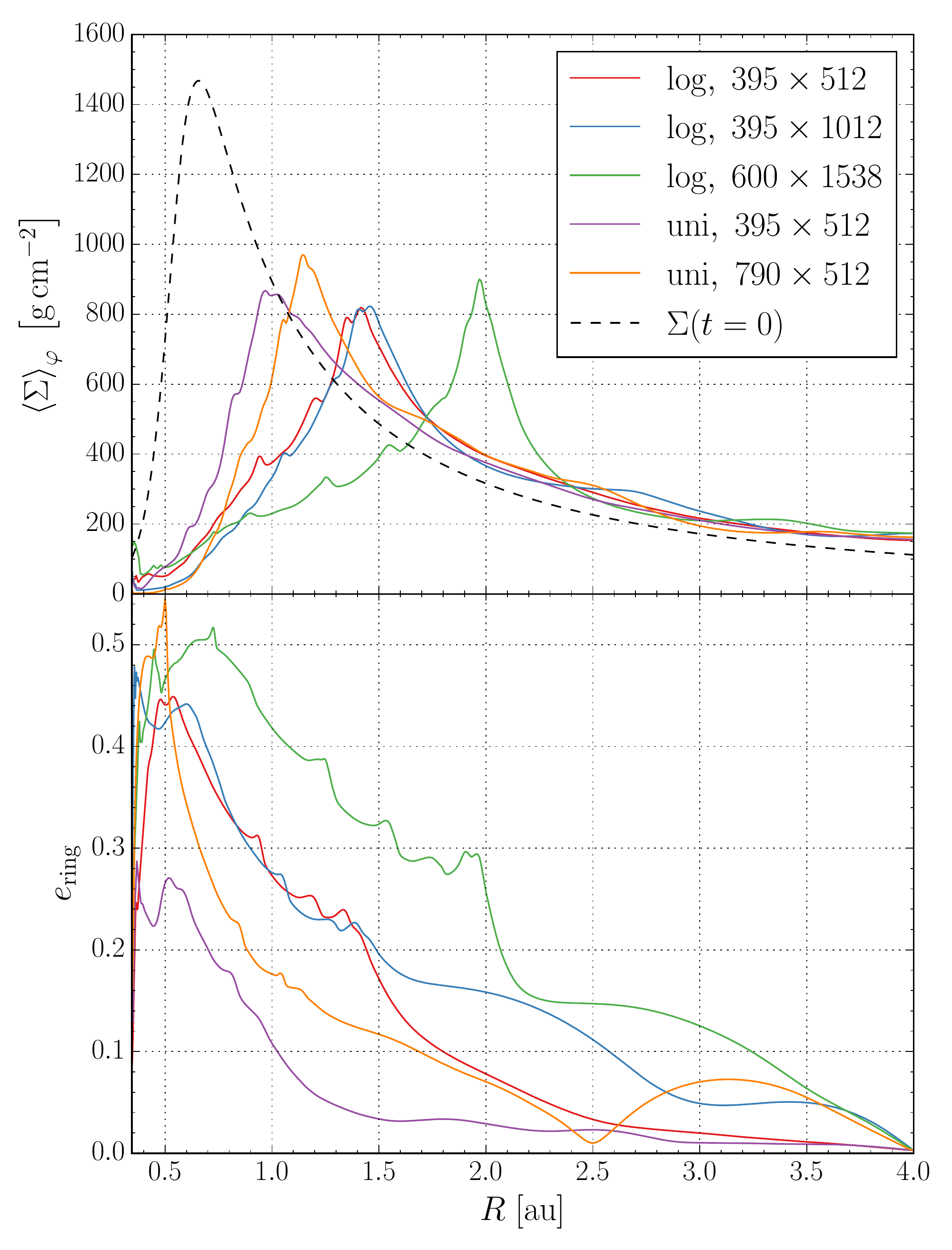}}
    \caption{Azimuthally averaged surface density profiles (top) and radial disc
    eccentricity (bottom) for our Kepler-16 reference setup (using
    $R_\mathrm{min} = \SI{0.345}{au}$) with a {\bf closed inner boundary} condition at
    $t=\num{16000}\,T_\mathrm{bin}$. Coloured solid lines show the results from
    simulations with different resolutions or different grid-spacing in the
    $R$-direction (logarithmic and uniform spacing). The initial density profile
    is also shown as a dashed black line. All simulations were performed with
    \textsc{Pluto}.}
    \label{img:inner_boundary_reflective}
\end{figure}

\begin{figure}
    \resizebox{\hsize}{!}{\includegraphics{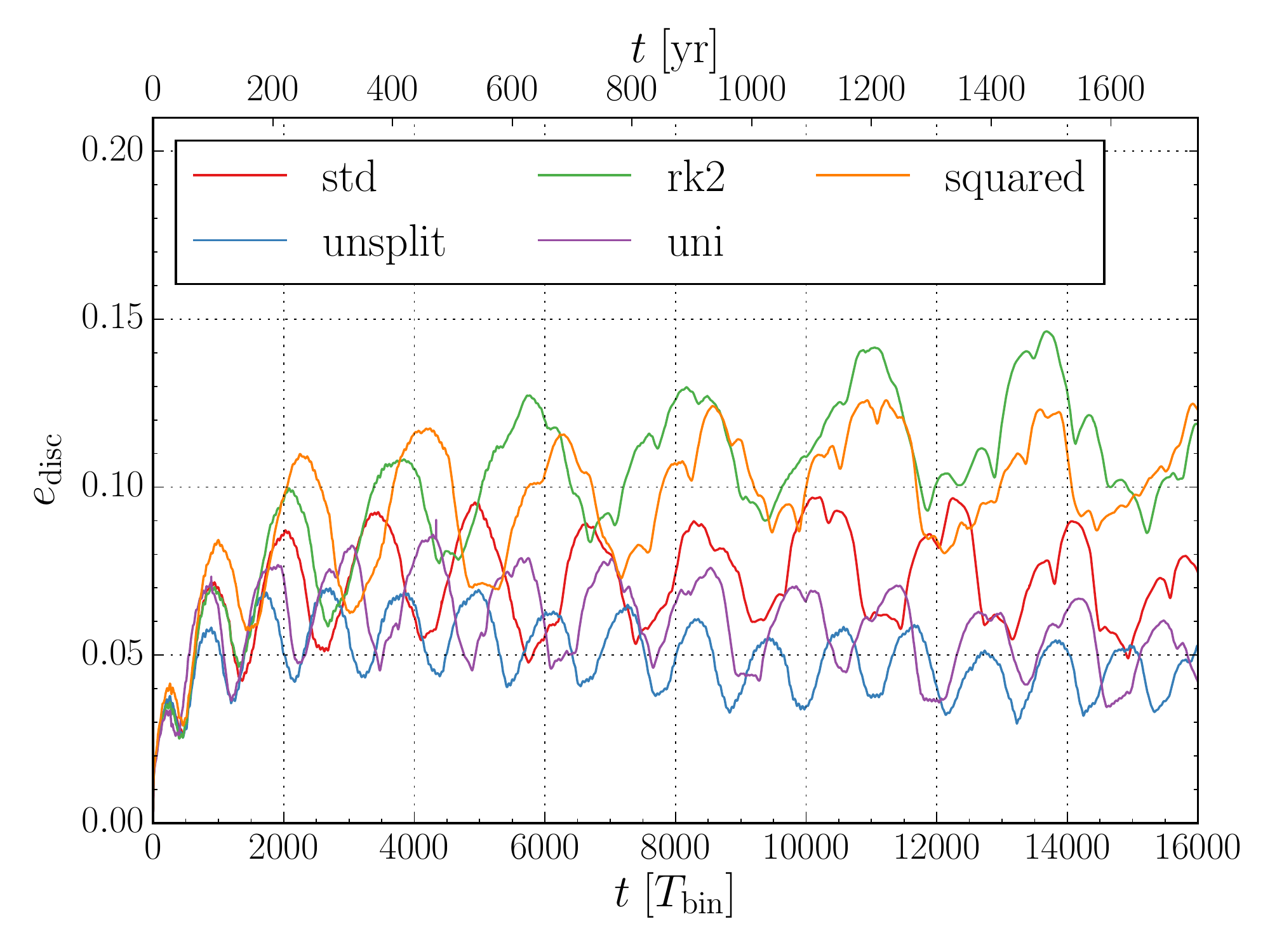}}
    \caption{Time evolution of the total disc eccentricity for simulations of
    the standard model with {\bf closed inner boundary} and $R_\mathrm{min} =
    \SI{0.345}{au}$.  Different numerical parameters such as grid spacing, time
    integrator, and operator splitting have been used. The legend refers to:
    \textsc{std} standard integrator (operator and directional splitting),
    \textsc{unsplit} no directional splitting, 
    \textsc{Rk2} second-order RK time integrator, 
    \textsc{uni} uniform grid,
    \textsc{squared} squared grid cells.
    All simulations were performed with \textsc{Rh2d}.} 
    \label{img:e_disc_rh2d_ref} 
\end{figure}

The top panel of Fig.~\ref{img:inner_boundary_reflective} shows the
azimuthally averaged density profiles for different grid resolutions and
spacings after \num{16000} binary orbits. All these simulations were performed
with \textsc{Pluto} and a closed inner boundary. The strong dependence of the
density distribution and inner gap size on the numerical setup stands out.  This
dependence on numerics is very surprising and not at all expected since the
physical setup was identical in all these simulations and therefore the density
profiles should all be similar and converge upon increasing resolution.  Not
only does the gap size show this strong dependence, but also the radial eccentricity
distribution (bottom panel of Fig.~\ref{img:inner_boundary_reflective}). To
examine this dependence further, which could in principle be the result of a bug
in our code, we reran the identical physical setup with a different code,
\textsc{Rh2d}.  The results of these \textsc{Rh2d} simulations show the same
strong numerical dependence as the \textsc{Pluto} results. In
Fig.~\ref{img:e_disc_rh2d_ref} the total disc eccentricity time evolution for
the \textsc{Rh2d} simulations is plotted. Again, one would expect that different
numerical parameters should produce approximately the same disc eccentricity, but
our simulations show a radical change in the disc eccentricity if the numerical
methods are slightly different.

We note that using a viscous outflow condition where the radial
velocity at $R_\mathrm{min}$ is fixed to $- \beta 3 \nu / (2 R)$ with $\beta=5$
(as suggested by \citet{2017MNRAS.465.4735M}) resulted in the same dynamical
behaviour as the closed boundary condition. The reason for this lies in the fact
that the viscous speed is very low in comparison to the radial velocity
induced by the perturbations of the central binary and hence there is little
difference between a closed and a viscous boundary.

\begin{figure}
    \resizebox{\hsize}{!}{\includegraphics{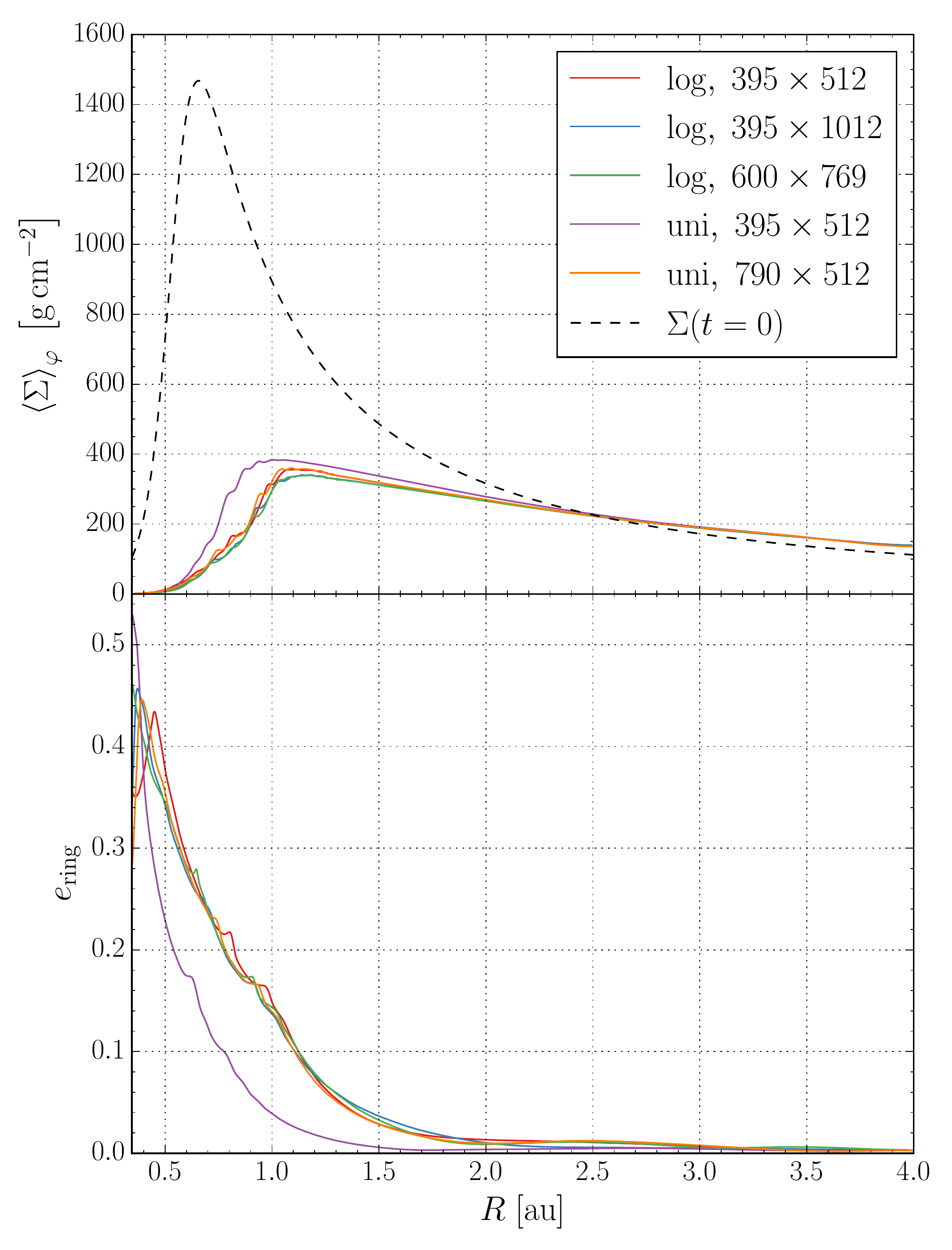}}
    \caption{Azimuthally averaged surface density profiles (top) and radial disc
    eccentricity (bottom) for our reference setup ($R_\mathrm{min} =
    \SI{0.345}{au}$)
    with a zero-gradient {\bf outflow inner boundary} condition after
    \num{16000} binary orbits. Coloured solid lines show results from
    simulations with different resolutions or different grid-spacing in the
    $R$-direction (logarithmic and uniform spacing).  The initial density
    profile is also shown as a dashed black line. All simulations were performed
    with \textsc{Pluto}.}
    \label{img:inner_boundary_outflow}
\end{figure}

In contrast to the closed inner boundary simulations, simulations with a
zero-gradient outflow inner boundary reach a quasi-steady state.  For the open
boundaries \textsc{Pluto} simulations with different numerical parameters
produce very similar surface density and radial disc eccentricity profiles
(Fig.~\ref{img:inner_boundary_outflow}) for different resolutions and grid
spacings.  Only the values for the uniform radial grid with a resolution of
$395\times512$ (purple line in Fig.~\ref{img:inner_boundary_outflow}) deviate.
Here the resolution in the inner computational domain is not high enough,
because a simulation with a higher resolution uniform grid ($790\times512$,
orange line in Fig.~\ref{img:inner_boundary_outflow}) produces approximately the
same results as the simulations with a logarithmic radial grid spacing.

We do not have a full explanation for this strong variability and
non-convergence of the flow when using the closed inner boundary, but this result
seems to imply that this problem is ill-posed.  A closed inner boundary creates
a closed cavity, which apparently implies in this case that the details of the
flow depend sensitively on numerical diffusion as introduced by different
spatial resolutions, time integrators, or codes.  As a consequence, for
circumbinary disc simulations a closed inner boundary is not recommended for two
reasons. Firstly, it produces the described numerical instabilities and
secondly, for physical reasons the inner boundary should be open because
otherwise no material can flow into the inner region and be accreted by the
stars.  This is also indicated in Fig.~\ref{img:2d_sigma_rmin0.02} which shows
mass flow through the inner gap, which cannot be modelled with a closed inner
boundary. 

\subsection{Location of the inner radius}\label{ssec:rmin}
After determining that a zero gradient outflow condition is necessary, we
use it now in all the following simulations and investigate in a second step the
optimal location of $R_\mathrm{min}$. It should be placed in such a way that it
simultaneously insures reliable results and computational efficiency.  For the
excitation of the disc eccentricity through the binary-disc interaction,
non-linear mode coupling, and the 3:1 Lindblad resonance are
important~\citep{2013A&A...556A.134P}.  Therefore, the location of the inner
boundary is an important parameter and $R_\mathrm{min}$ should be chosen such
that all major mean-motion resonances between the disc and the binary lie
inside the computational domain. To investigate the influence of the inner
boundary position, we set up simulations with an inner radius from $R_\mathrm{min}
= \SI{0.12}{au}$ to $R_\mathrm{min} = \SI{0.5}{au}$ for the standard system parameter
as noted in Table \ref{tab:ref_setup}. 
\begin{figure}
    \resizebox{\hsize}{!}{\includegraphics{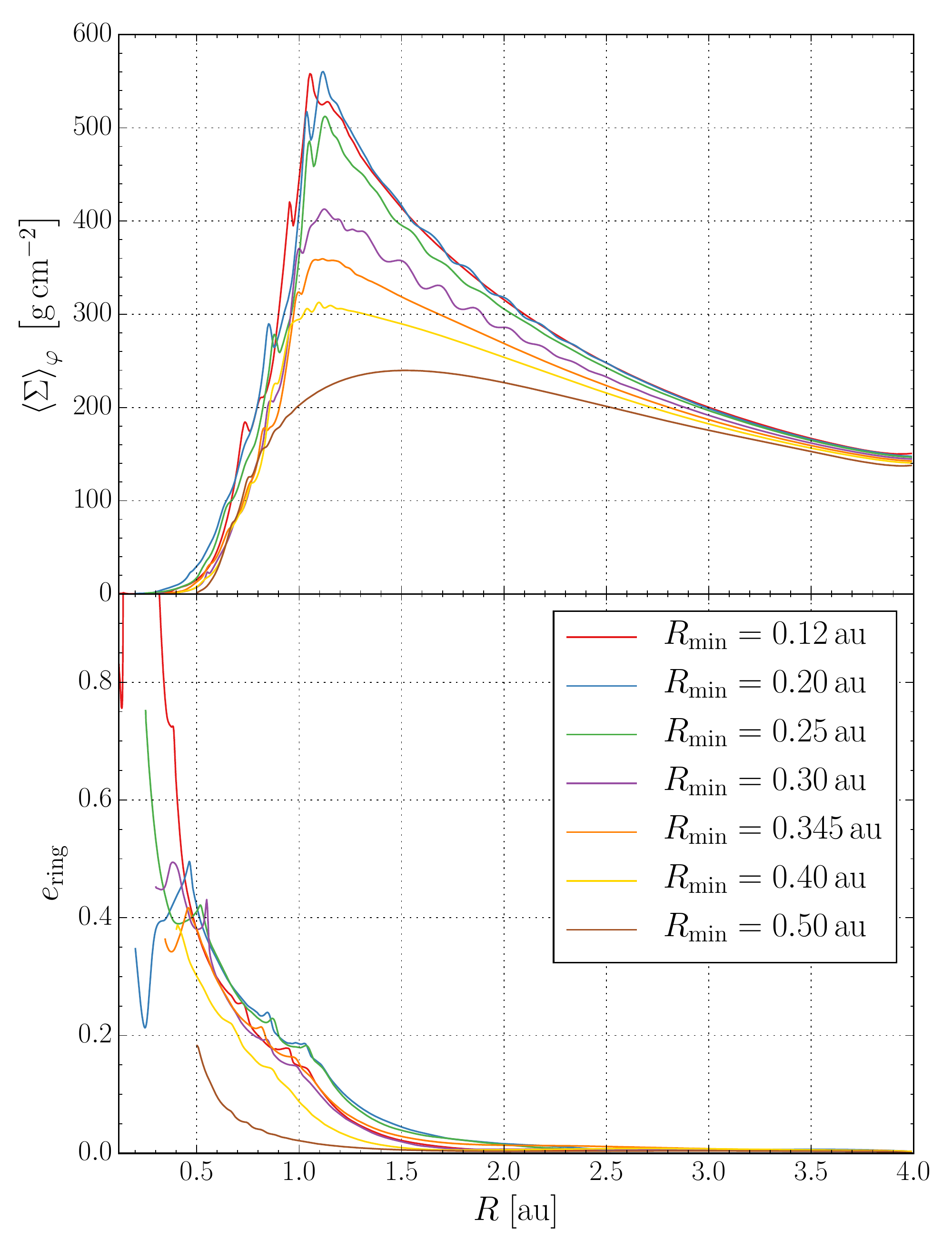}}
    \caption{Influence of different inner radii on the disc structure for
    simulations with a zero-gradient {\bf outflow inner boundary} condition.
    \emph{Top:} Azimuthally averaged surface density profiles after \num{16000}
    binary orbits. \emph{Bottom:} Radial disc eccentricity distribution at the
    same time.
    }
    \label{img:rmin}
\end{figure}
Fig.~\ref{img:rmin} shows the results of these simulations. Displayed are the
azimuthally averaged surface density and the radial disc eccentricity
distribution after \num{16000} binary orbits. In agreement
with~\citet{2014A&A...564A..72K}, we find that the radial disc eccentricity does
not depend much on the location of the inner radius and remains low in the
outer disc ($R > \SI{2}{au}$). Only for a large inner radius of $R_\mathrm{min} =
\SI{0.5}{au}$ do we observe almost no disc eccentricity growth (brown line in
Fig.~\ref{img:rmin}).  In this case the 3:1 Lindblad resonance ($R_{3:1}
=\SI{0.46}{au}$) lies outside the computational domain, confirming the
conclusion from \citet{2013A&A...556A.134P} about the importance of the 3:1
Lindblad resonance for the disc eccentricity growth. 
The variation of the radial disc eccentricity for inner radii smaller than
\SI{0.3}{au} occurs because the precessing discs are in different
phases after \num{16000} binary orbits.
Contrary to the
eccentricity distribution, the azimuthally averaged surface density shows a
stronger dependence on the inner radius. As seen in the top panel of
Fig.~\ref{img:rmin}, the maximum of the surface density increases monotonically
as the location of the inner boundary decreases because a smaller inner
radius means also that the \enquote{area} where material can leave the
computational domain decreases. For inner radii smaller than $R_\mathrm{min} \leq
\SI{0.25}{au}$ the change in the maximum surface density becomes very low upon
further reduction of $R_\mathrm{min}$.  This implies that for too large
$R_\mathrm{min}$ there is too much mass leaving the domain and it has to chosen
small enough to capture all dynamics.  Although the maximum value of the surface
density increases, the position of the maximum does not depend on the inner
radius and remains at approximately $R = \SI{1.1}{au}$ as long as the 3:1
Lindblad resonance lies inside the computational domain. Furthermore, we find
that the slope of the surface density profile at the gap's edge does not depend
on the location of the inner boundary.  These results are in contrast to
\citet{2017MNRAS.465.4735M} who find a stronger dependence of the density
profile on the location of the inner radius.  However, their results may be a
consequence of the viscous outflow condition used.

\begin{figure}
    \resizebox{\hsize}{!}{\includegraphics{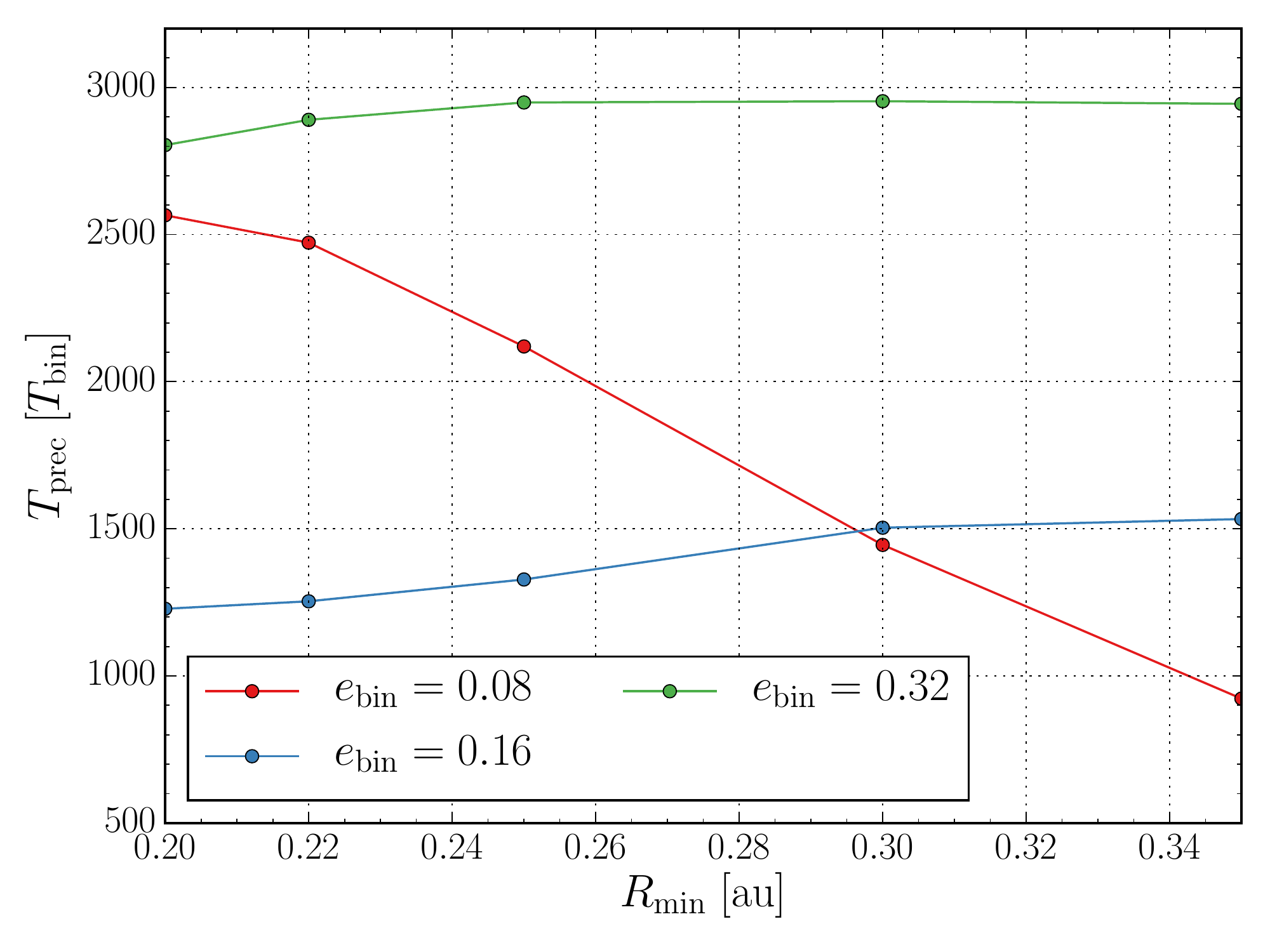}}
    \caption{Variation of precession period of the inner gap of the disc for varying
    inner radii, $R_\mathrm{min}$, and different binary eccentricities, $e_\mathrm{bin}$.}
    \label{img:rmin_convergence}
\end{figure}

\begin{figure*}
    \resizebox{\hsize}{!}{\includegraphics{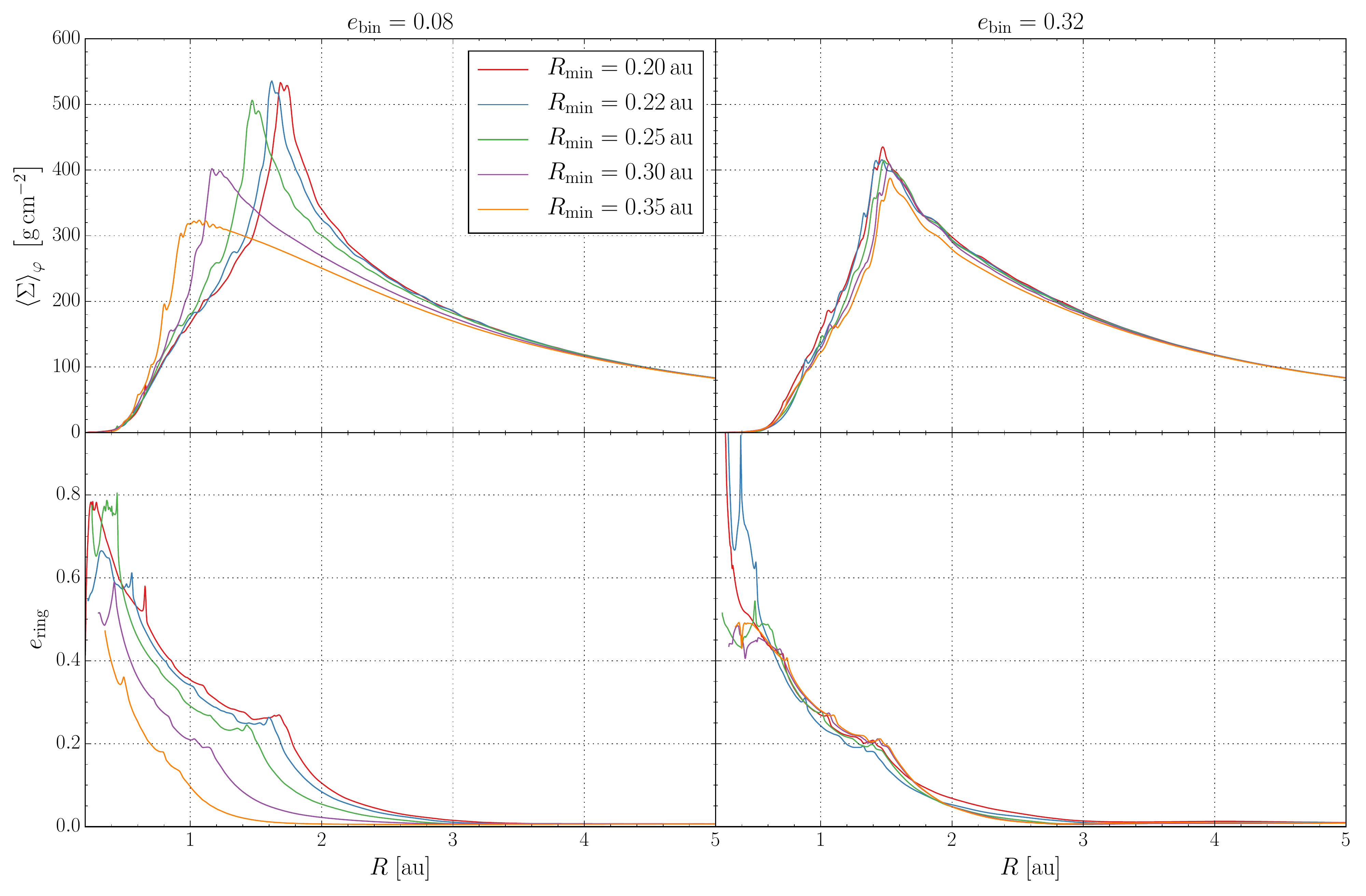}}
    \caption{Influence of different inner radii and binary
    eccentricities on the disc structure for
    simulations with a zero-gradient {\bf outflow inner boundary} condition.
    \emph{Top:} Azimuthally averaged surface density profiles after \num{16000}
    binary orbits. \emph{Bottom:} Radial disc eccentricity distribution at the
    same time. \emph{Left:} $e_\mathrm{bin} = 0.08$. \emph{Right:}
    $e_\mathrm{bin} = 0.32$}
    \label{img:sigma_ering_rmin_ebin}
\end{figure*}

As is discussed below, the disc becomes eccentric with a slow
precession that depends on the binary eccentricity, $e_\mathrm{bin}$.  To
explore how the location of the inner radius affects the disc dynamics, we ran
additional simulations with varying $R_\mathrm{min}$ for different
$e_\mathrm{bin}$. We chose $e_\mathrm{bin} = 0.08$ and $e_\mathrm{bin}
= 0.32$ in addition to the Kepler-16 value of $e_\mathrm{bin} = 0.16$, which
lies near the bifurcation point, $e_\mathrm{crit}$.
Fig.~\ref{img:rmin_convergence} shows the precession period of the inner gap for
varying inner radii and three different binary eccentricities, where the blue
curve refers to the model shown above with $e_\mathrm{bin} = 0.16$.  For a
higher binary eccentricity of $e_\mathrm{bin} = 0.32$ (green curve in
Fig.~\ref{img:rmin_convergence} and on the upper branch of the bifurcation diagram)
the disc dynamics seems to be captured well even for higher values of
$R_\mathrm{min}$. Only for values lower than $R_\mathrm{min} = \SI{0.22}{au}$
are deviations seen.  The simulation with $R_\mathrm{min} = \SI{0.20}{au}$ was more
unstable, probably because the secondary moved in and out of the computational
domain on its orbit.  On the lower branch of the bifurcation diagram the
convergence with decreasing $R_\mathrm{min}$ is slower as indicated by the case
$e_\mathrm{bin} = 0.08$ (red curve in Fig.~\ref{img:rmin_convergence}).  Here,
an inner radius of $R_\mathrm{bin} = a_\mathrm{bin}$ or even slightly lower may
be needed.  One explanation for this behaviour is that on the lower
branch (for low $e_\mathrm{bin}$) the inner gap is smaller but nevertheless
quite eccentric such that the disc is influenced more by the location of the
inner boundary.  The convergence of the results for smaller inner radii is also
visible in the azimuthally averaged surface density and radial eccentricity
profiles for $e_\mathrm{bin}= 0.08$ and $e_\mathrm{bin} = 0.32$
displayed in Fig.~\ref{img:sigma_ering_rmin_ebin}.

In summary, from the results shown in Figs.~\ref{img:rmin},
\ref{img:rmin_convergence}, and \ref{img:sigma_ering_rmin_ebin} we can conclude
that to model the circumbinary disc properly, the inner radius should correspond
to $R_\mathrm{min} \approx a_\mathrm{bin}$.  This should be used in combination
with a zero-gradient outflow (hereafter just outflow) condition.  Since a
smaller inner boundary increases the number of radial cells and imposes a
stricter condition on the time step, long-term simulations with a small inner
radius are very expensive. As a compromise to make long-term simulations
possible, we have chosen in all simulations below an inner radius of
$R_\mathrm{min} = \SI{0.25}{au}$.

\subsection{Location of the outer radius}\label{ssec:rmax}
While investigating the numerical conditions at the inner radius, we used
an outer radius of $R_\mathrm{max} = \SI{4.0}{au}$ and assumed that this value
is high enough, so that the outer boundary would not interfere with the
dynamical behaviour of the inner disc. During our study of different
binary and disc parameters, we found that in some cases this assumption is
not true.  Especially for binary eccentricities greater than $0.32$, reflections
from the outer boundary interfered with the complex inner disc structure.
Therefore, we increased the outer radius of all the following simulations to
$R_\mathrm{max} = \SI{15.4}{au} = \num{70}\,a_\mathrm{bin}$, a value used
by~\citet{2017MNRAS.466.1170M}. In order to keep the same spatial resolution as
before we also increased the resolution to $\num{762}\times\num{582}$ cells.
This ensures that in the radial direction we still have \num{512} grid cells
between \SI{0.25}{au} and \SI{4.0}{au}. To save computational time it is also
possible to use a smaller outer radius ($R_\mathrm{max} \approx 40\,a_\mathrm{bin}$) with a
damping zone where the density and radial velocity are relaxed to their initial
value. A detailed description of the implementation of such a damping zone can
be found in \citet{2006MNRAS.370..529D}.

Hence, for our subsequent simulations we use from now on
the numerical setup quoted in Table~\ref{tab:ref_setup}, unless otherwise
stated. To study the dependence on different physical parameter we start from
the standard values of the Kepler-16 system in Table~\ref{tab:ref_setup} and
vary individual parameter separately. 
\begin{table}
    \caption{Setup for our reference model.}
    \label{tab:ref_setup}
    \centering
    \begin{tabular}{lc}
        \midrule\midrule
        Numerical parameters &  \\
        \midrule
        $R_\mathrm{min}$ & $\SI{0.25}{au}$ \\
        $R_\mathrm{max}$ & $\SI{15.4}{au}$ \\
        Resolution & $762 \times 582$ \\
        Inner boundary & Outflow \\
        \\
        \midrule
        Binary parameters &  \\
        \midrule
        $a_\mathrm{bin}$ & $\SI{0.22}{au}$ \\
        $e_\mathrm{bin}$ & $0.16$ \\
        $q_\mathrm{bin}$ & $0.29$ \\
        \\
        \midrule
        Disc parameters &  \\
        \midrule
        $h$ & $0.05$ \\
        $\alpha$ & $0.01$ \\
        \midrule
    \end{tabular}
\end{table}

\begin{figure*}
    \resizebox{\hsize}{!}{\includegraphics{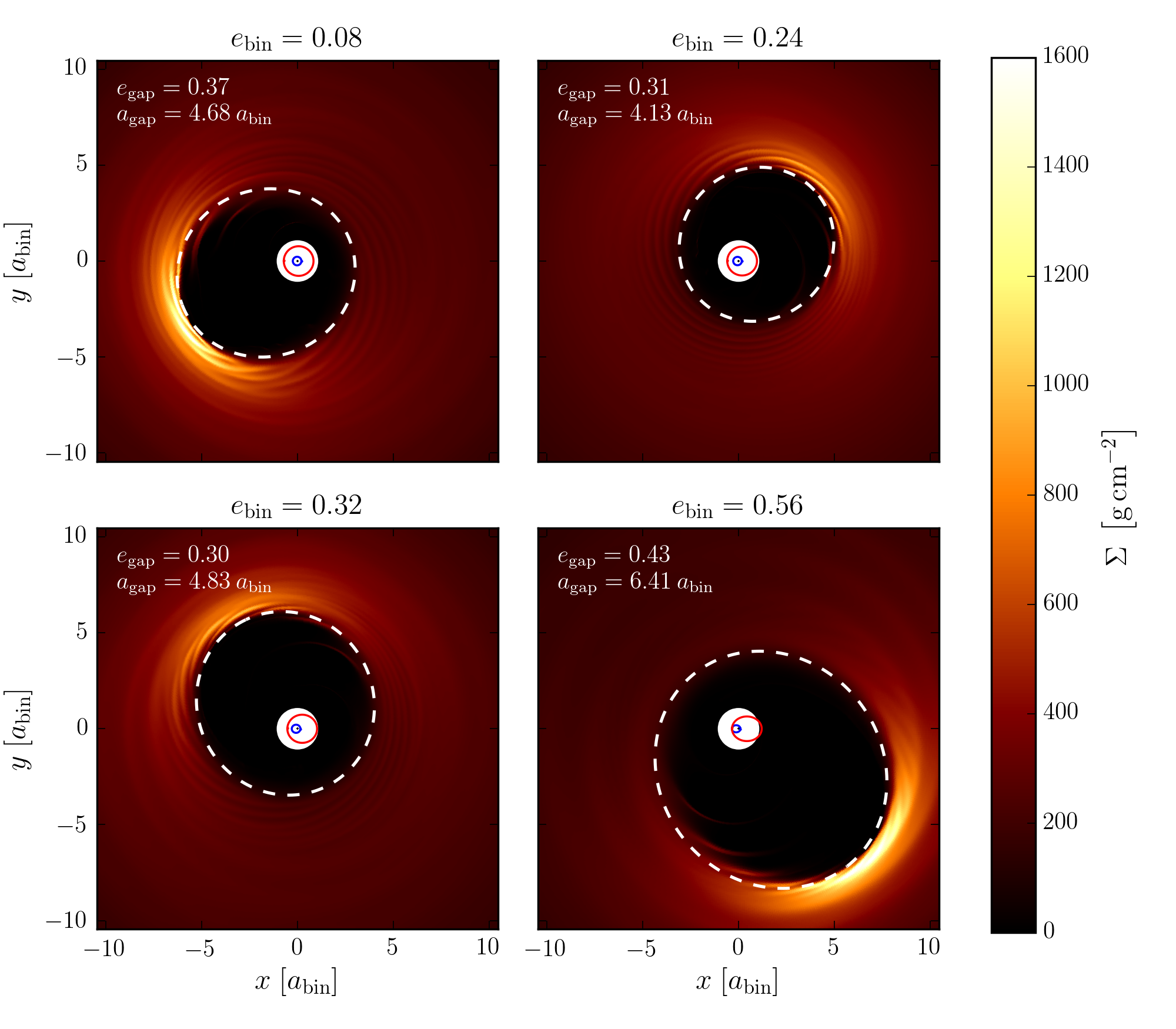}}
    \caption{Structure of the inner disc for different binary eccentricities
    after \num{16000} binary orbits. The surface density is colour-coded in
    cgs-units. The orbits of the primary and secondary around the centre of mass
    (central point) are shown in blue and red. The white dashed lines represent
    approximate ellipses fitted to match the extension of the inner gap (see
    explanation in text). Videos of these simulations can be found online.}
    \label{img:2d_sigma_ecc}
\end{figure*}
\section{Variation of binary parameters}\label{sec:bin_para}
In this section we study the influence of the central binary parameter
specifically its orbital eccentricity and mass ratio on the disc structure.
Throughout this section we use a disc aspect ratio of $h=0.05$ and a viscosity
$\alpha = 0.01$. For the inner radius and the inner boundary condition we use
the results from the previous section and apply the outflow condition at
$R_\mathrm{min}$ (see also Table~\ref{tab:ref_setup}). All the results 
in this section were obtained with \textsc{Pluto}, but comparison simulations
using \textsc{Rh2d} show very similar behaviour (see
Fig.~\ref{img:cmp_dom_ecc}).

\subsection{Dynamics of the inner cavity}\label{ssec:dynamics}
Before discussing in detail the impact of varying $e_\mathrm{bin}$ and
$q_\mathrm{bin}$, we comment first on the general dynamical behaviour of the
inner disc.  Fig.~\ref{img:2d_sigma_ecc} shows snapshots of the inner disc after
\num{16000} binary orbits for different binary eccentricities.  As seen in
several other studies (as mentioned in the introduction) we find that the inner
disc becomes eccentric and shows a coherent slow precession.  In the figure we
overplot ellipses (white dashed lines) that are approximate fits to the central
inner cavity, where the semi-major axis ($a_\textrm{gap}$) and
eccentricity ($e_\textrm{gap}$) are indicated. To calculate these parameters we
assumed first that the focus of the gap-ellipse is the centre of mass of the
binary. We then calculated from our data the coordinates
$(R_{\Sigma\mathrm{max}}, \varphi_{\Sigma\mathrm{max}})$ of the cell with the
highest surface density value, which defines the direction to the gap's
apocentre.  We defined the apastron of the gap $R_\mathrm{apa}$ as the minimum
radius along the line $(R, \varphi_{\Sigma\mathrm{max}})$ which fulfils the
condition
\begin{equation}\label{eq:cond_apastron}
    \Sigma(R, \varphi_{\Sigma\mathrm{max}}) \geq 0.1 \cdot
    \Sigma(R_{\Sigma\mathrm{max}}, \varphi_{\Sigma\mathrm{max}})\, .
\end{equation}
The periastron of the gap $R_\mathrm{peri}$ is defined analogously as the
minimum radius along the line $(R, \varphi_{\Sigma\mathrm{max}} + \pi)$ in the
opposite direction which fulfils
\begin{equation}\label{eq:cond_periastron}
    \Sigma(R, \varphi_{\Sigma\mathrm{max}}+\pi) \geq 0.1 \cdot
    \Sigma(R_{\Sigma\mathrm{max}}, \varphi_{\Sigma\mathrm{max}})\, .
\end{equation}
Using the apastron and periastron of the gap as defined above, the eccentricity and semi-major axis
of the gap are given by
\begin{align}
    a_\mathrm{gap} &= 0.5 \left(R_\mathrm{apa} + R_\mathrm{peri} \right) \,, \\
    e_\mathrm{gap} & = R_\mathrm{apa} / a_\mathrm{gap} - 1 \,.
\end{align}
As shown in Fig.~\ref{img:2d_sigma_ecc} this purely geometrical construction
matches the shape of the central cavity very well.  Even though the overall disc
behaviour shows a rather smooth slow precession, the dynamical
action of the central binary is visible as spiral waves near the gap's edges, most
clearly seen in the first and last panel. We prepared some online videos
to visualise the dynamical behaviour of the inner disc.

\begin{figure}
    \resizebox{\hsize}{!}{\includegraphics{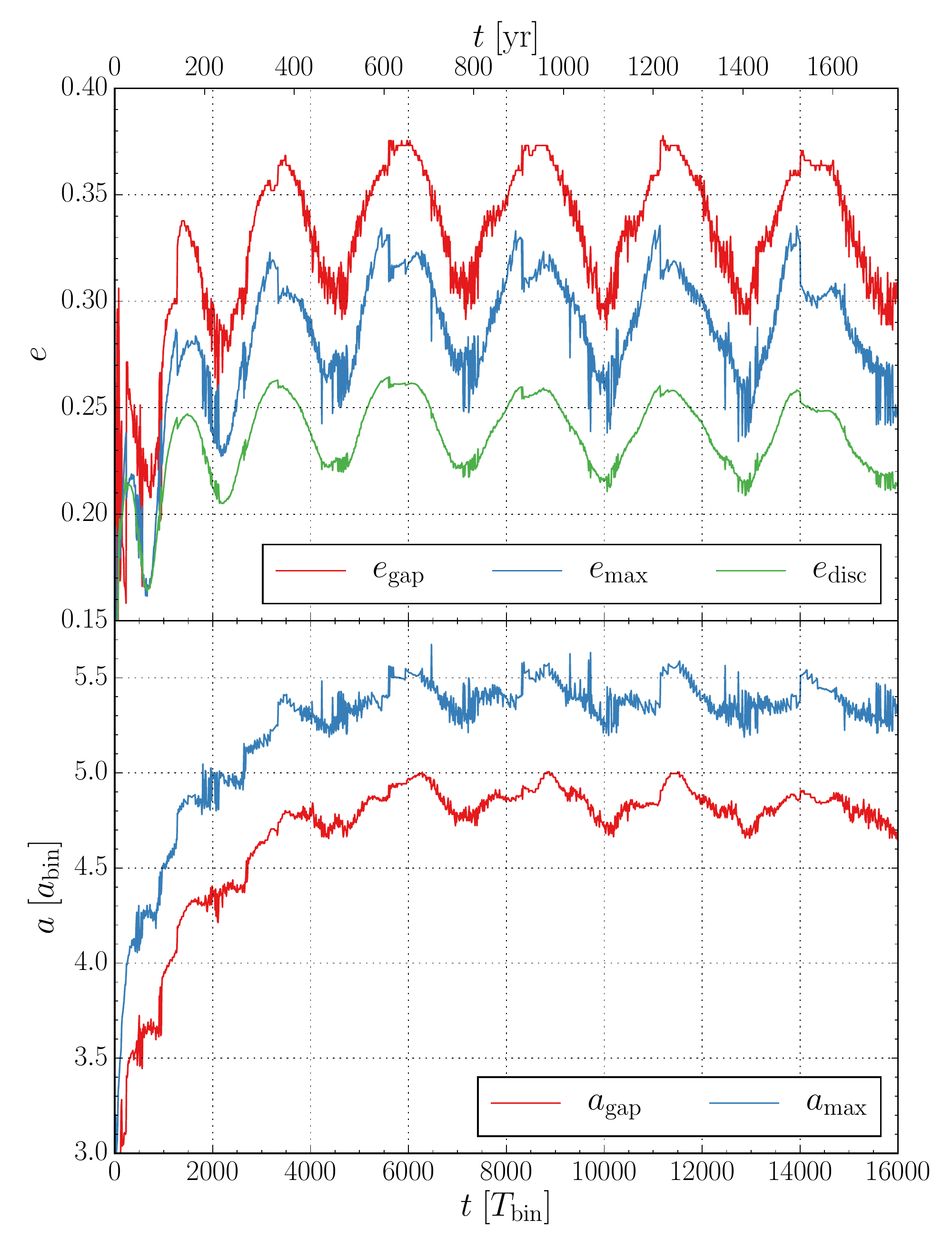}}
    \caption{Evolution of gap eccentricity (top) and semi-major axis (bottom)
    for a binary with $e_\mathrm{bin} = 0.30$. Shown are different measurements
    for the disc eccentricity. The red lines ($e_\mathrm{gap}$ and
    $a_\mathrm{gap}$) show the eccentricity and semi-major axis of the
    geometrically fitted ellipses, see Fig.~\ref{img:2d_sigma_ecc}. The blue
    lines ($e_\mathrm{max}$ and $a_\mathrm{max}$) show the eccentricity and
    semi-major axis of the cell with the maximum surface density. The green
    curve shows the averaged eccentricity of the inner disc.  Since the gap size
    varies with time we use $R_2 = R_{\Sigma\mathrm{max}}$ in
    eq.~\eqref{eq:e_disc} to calculate the inner disc eccentricity.} 
    \label{img:e_a_vs_t}
\end{figure}
An alternative way to characterise the disc gap dynamically is by using the orbital
elements ($e_\mathrm{max}, a_\mathrm{max}$) of the cell with the maximum surface
density. These orbital elements can be calculated with
equation~\eqref{eq:ecc_vector} and the vis-viva equation
\begin{equation}
    a = \left( \frac{2}{R} - \frac{\vec{u}^2}{GM_\mathrm{bin}} \right)^{-1}\,.
\end{equation}
The time evolution of these gap characteristics are plotted in
Fig.~\ref{img:e_a_vs_t}.  The gap's size and eccentricity show in phase
oscillatory behaviour with a larger amplitude in the eccentricity variations. The
period of the oscillation is identical to the precession period of the disc.
Clearly, the extension of the gap always lies inside of the location of maximum
density $a_\mathrm{gap} < a_\mathrm{max}$, but for the eccentricities the
ordering is not so clear. The radial variations of the disc eccentricity as
shown in Figs.~\ref{img:inner_boundary_outflow} and \ref{img:rmin} indicate that
the inner regions of the disc typically have a higher eccentricity. In
Fig.~\ref{img:e_a_vs_t} $e_\mathrm{disc}$ is lowest because it is weighted with
the density which is very low in the inner disc regions. The eccentricity for
the gap $e_\mathrm{gap}$ stems from a geometric fit to the very inner disc
regions and is the highest. Inside the maximum density and even slightly
beyond that radius the disc precesses coherently in the sense that the
pericentres estimated at different radii are aligned.  The data show further
that when the disc eccentricity is highest the disc is fully aligned with the
orbit of the secondary star, i.e.  the pericentre of disc and binary lie in the
same direction (see also Appendix~\ref{sec:convergence}). 

\subsection{Binary eccentricity}\label{ssec:ebin}
For simulations in this section we fixed $q_\mathrm{bin} = 0.29$ and varied
$e_\mathrm{bin}$ from $0.0$ to $0.64$. The binary eccentricity strongly
influences the size of the gap as well as the disc precession period of the
inner disc. 

\begin{figure}
    \resizebox{\hsize}{!}{\includegraphics{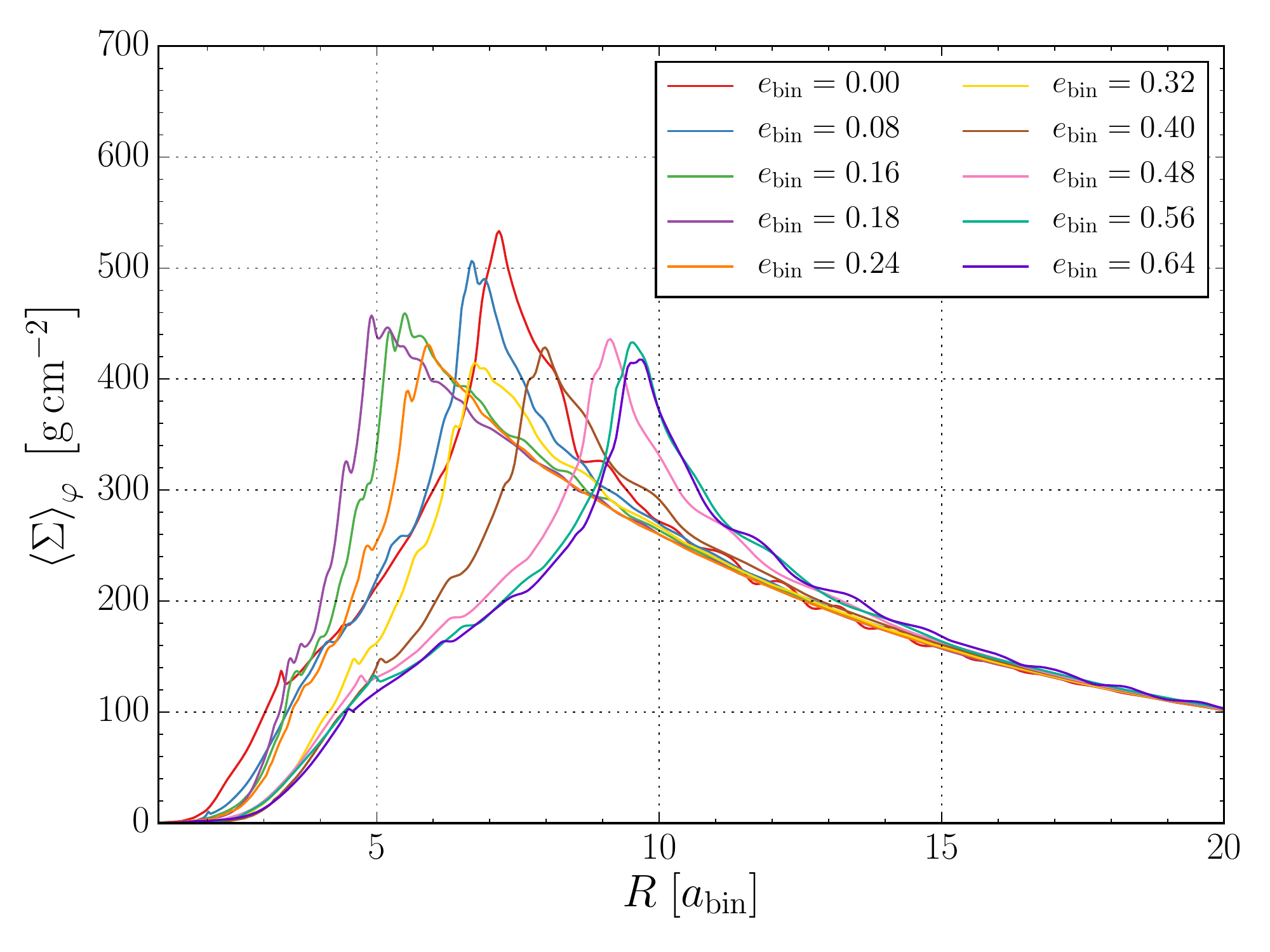}}
    \caption{Azimuthally averaged density profiles for varying binary
    eccentricities after \num{16000} binary orbits. For clarity we only show
    selected results from the simulated parameter space.}
    \label{img:sigma_ecc}
\end{figure}

In Fig.~\ref{img:sigma_ecc} the azimuthally averaged surface density profiles
are shown for the various $e_\mathrm{bin}$ at \num{16000} binary orbits.  In all
cases there is a pronounced density maximum visible which is the strongest for
the circular binary with $e_\mathrm{bin}=0$  (red curve). The position of the
peak varies systematically with binary eccentricity.  Increasing
$e_\mathrm{bin}$ from zero to higher values the peak shifts
inward. Not only does the gap size decrease, but the maximum surface density
also drops with increasing $e_\mathrm{bin}$ until a minimum at $e_\mathrm{bin} = 0.18$
is reached. Increasing $e_\mathrm{bin}$ further causes the density peak to move
outward again, whereas the maximum surface density stays roughly constant for
higher binary eccentricities.

\begin{figure} \resizebox{\hsize}{!}{\includegraphics{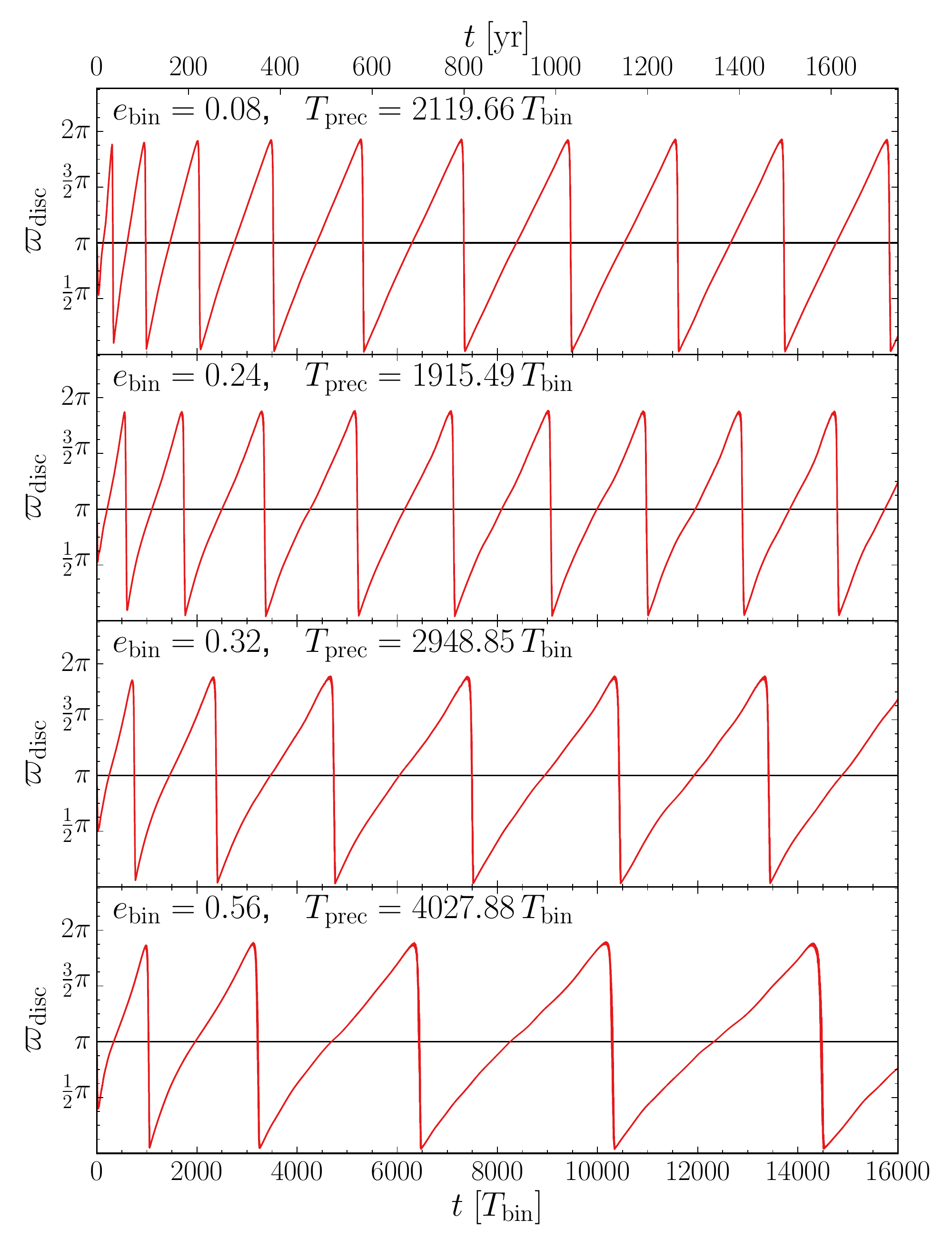}}
    \caption{Disc longitude of periastron for different binary eccentricities
    for the inner disc $R<1.0\,\mathrm{au}$.
    For all cases a clear prograde precession of the inner disc is visible
    and the precession period, $T_\mathrm{prec}$, is indicated in the legend.}
    \label{img:peri_ecc}
\end{figure}
As mentioned above, the eccentric disc precesses slowly and in Fig.~\ref{img:peri_ecc}
the longitude of the inner disc's pericentre is shown versus time over \num{16000}
binary orbits. The disc displays a slow prograde precession with inferred period
of several thousand binary orbits. 
For the measurement of the precession period we discarded the first \num{6000} binary
orbits since plots of the longitude of periastron show that during this time
span the precession period is not constant yet (see Fig.~\ref{img:peri_ecc}).
The period is then calculated by averaging over at least two full periods. To be
able to average over two periods, models with a very long $T_\mathrm{prec}$
(e.g. $e_\mathrm{bin} = 0.64$) were simulated for more than \num{16000} binary
orbits.

\begin{figure}
    \resizebox{\hsize}{!}{\includegraphics{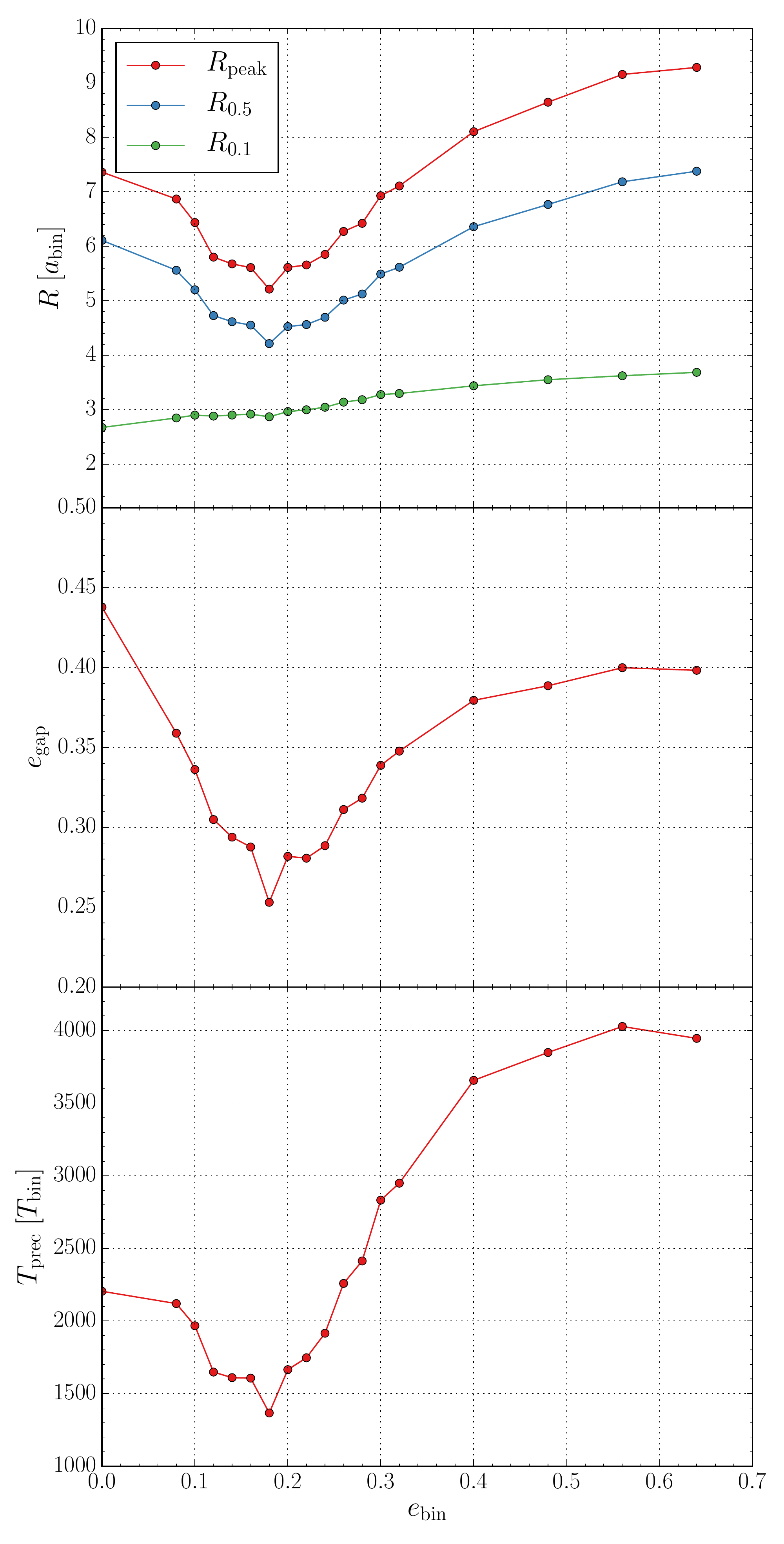}}
    \caption{Influence of different binary eccentricities on gap size,
    gap eccentricty, and precession rate. \emph{Top:} Different measures
    of the gap size are shown, averaged over several thousand binary orbits.
    The value $R_\mathrm{peak}$ refers to the location of maximum of the averaged surface
    density (see Fig.~\ref{img:sigma_ecc}), while $R_\mathrm{0.5}$ and
    $R_\mathrm{0.1}$ refer to density drops of 50 and 10 percent of the maximum
    value.  \emph{Middle}: Eccentricity of the gap. \emph{Bottom:}
    Precession period of the disc gap.}
    \label{img:gap_peri_ecc}
\end{figure}

Fig.~\ref{img:gap_peri_ecc} summarises the results from
simulations with varying binary eccentricities.
The top panel of Fig.~\ref{img:gap_peri_ecc} shows different measures for the gap
size for varying binary eccentricities. In addition to the radial position where the
azimuthally averaged surface density reaches its maximum ($R_\mathrm{peak}$), we
plot the positions where the density drops to 50 and 10 percent of its
maximum value ($R_\mathrm{0.5}$ and $R_\mathrm{0.1}$). The value $R_\mathrm{0.5}$ was used
by \citet{2014A&A...564A..72K} as a measure for the gap size, whereas
\citet{2017MNRAS.466.1170M} used $R_\mathrm{0.1}$ and \citet{2017MNRAS.465.4735M}
$R_\mathrm{peak}$.
Starting from a non-eccentric binary the curves for $R_\mathrm{peak}$ and
$R_\mathrm{0.5}$ decrease with increasing binary eccentricity. For $e_\mathrm{bin}
\approx 0.18$ the gap size reaches a minimum and then increases again for higher
binary eccentricities. In agreement with \citet{2017MNRAS.466.1170M} we see an
almost monotonic increase of $R_\mathrm{0.1}$ for increasing binary eccentricities.
The gap size, $a_\mathrm{gap}$, correlates well with $R_\mathrm{0.5}$ and is 
always about 14\% smaller. 

The middle panel of Fig.~\ref{img:gap_peri_ecc} shows the eccentricity
of the inner cavity, calculated with the method described in
Sec.~\ref{ssec:dynamics}. The disc eccentricity also changes systematically with
$e_\mathrm{bin}$. For circular binaries it reaches
$e_\mathrm{gap} \approx 0.44$, and then it drops down to about 0.25 for the
turning point $e_\mathrm{bin} = 0.18$, and increases again reaching
$e_\mathrm{gap}  \approx 0.4$ for the highest $e_\mathrm{bin} = 0.64$. Hence,
for nearly circular binaries $e_\mathrm{gap}$ can be higher than for more
eccentric binaries.  A similar variation of the disc's eccentricity and
precession rate with binary eccentricity has been noticed by
\citet{2017MNRAS.464.3343F} who attribute this to the possible excitation of
higher order resonances for higher disc eccentricities.

\begin{figure}
    \resizebox{\hsize}{!}{\includegraphics{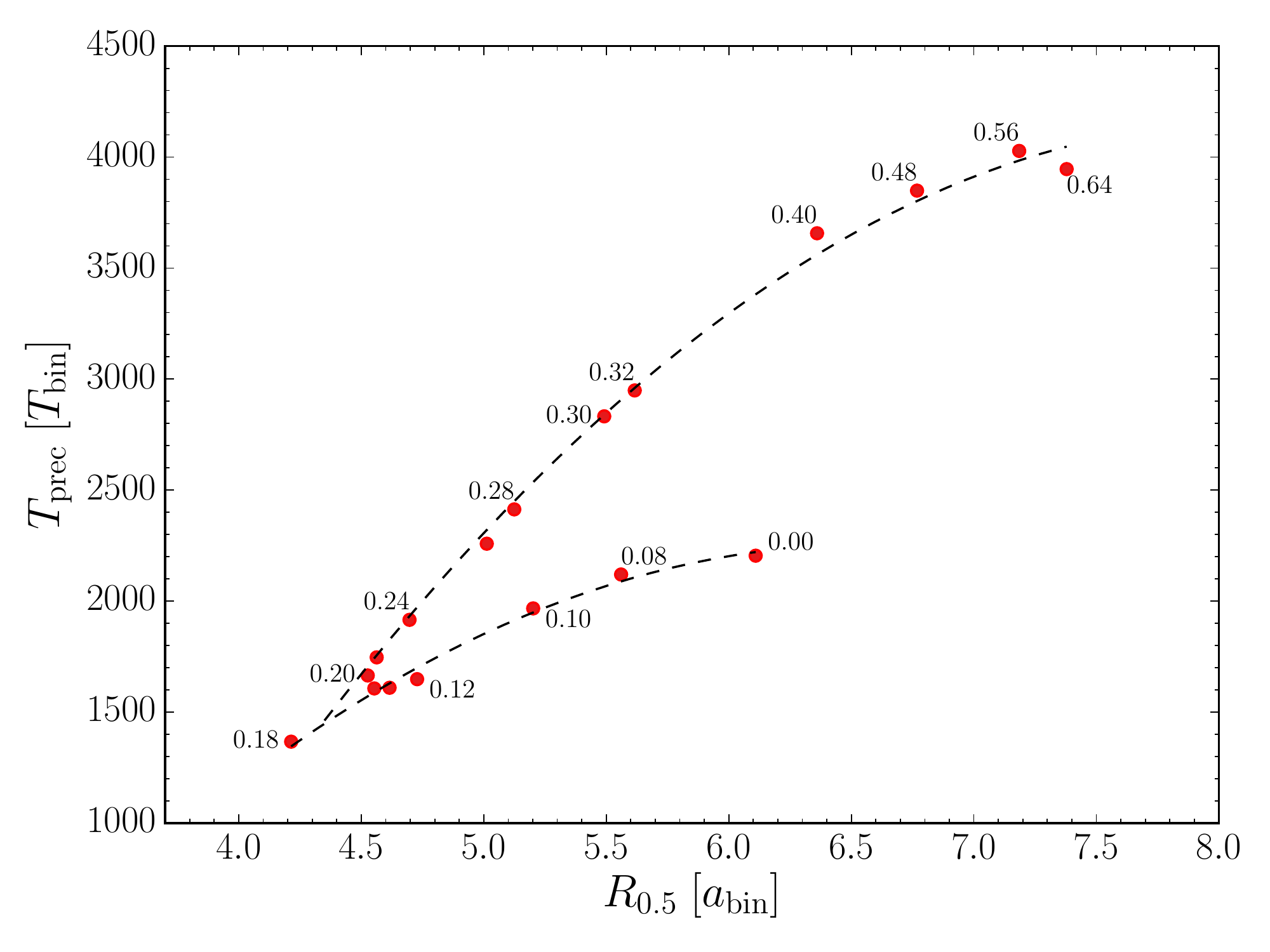}}
    \caption{Precession rate of the gap plotted against the radius where the
    azimuthally averaged surface density drops to 50 percent of its maximum
    value ($R_\mathrm{0.5}$), averaged over several thousand binary orbits.
    Different dots correspond to different binary eccentricities. The dashed
    lines are not fit curves but are just drawn to guide the eye.}
    \label{img:Tgap_vs_R}
\end{figure}

The bottom panel of Fig.~\ref{img:gap_peri_ecc} shows the precession period of
the inner disc ($R < \SI{1}{au}$) for different binary eccentricities.
The curve for the precession period shows a similar behaviour to the upper curves 
for the gap size. To investigate further the correlation of the gap size (here
represented by $R_\mathrm{0.5}$) and the
precession period, we show in Fig.~\ref{img:Tgap_vs_R} the two quantities plotted
against each other.
Two branches can be seen as indicated by the dashed curves.  One starts at
$e_\mathrm{bin} = 0.0$ and goes to $e_\mathrm{bin} = 0.18$ where the gap size
and precession period decrease with increasing binary eccentricities, and the
other branch starts at the minimum at $e_\mathrm{bin} = 0.18$ where both gap
properties increase again with increasing binary eccentricities. These two
branches may indicate two different physical processes that are responsible for
the creation of the eccentric inner gap and show the direct correlation between
precession period and gap size.

\begin{figure}
    \resizebox{\hsize}{!}{\includegraphics{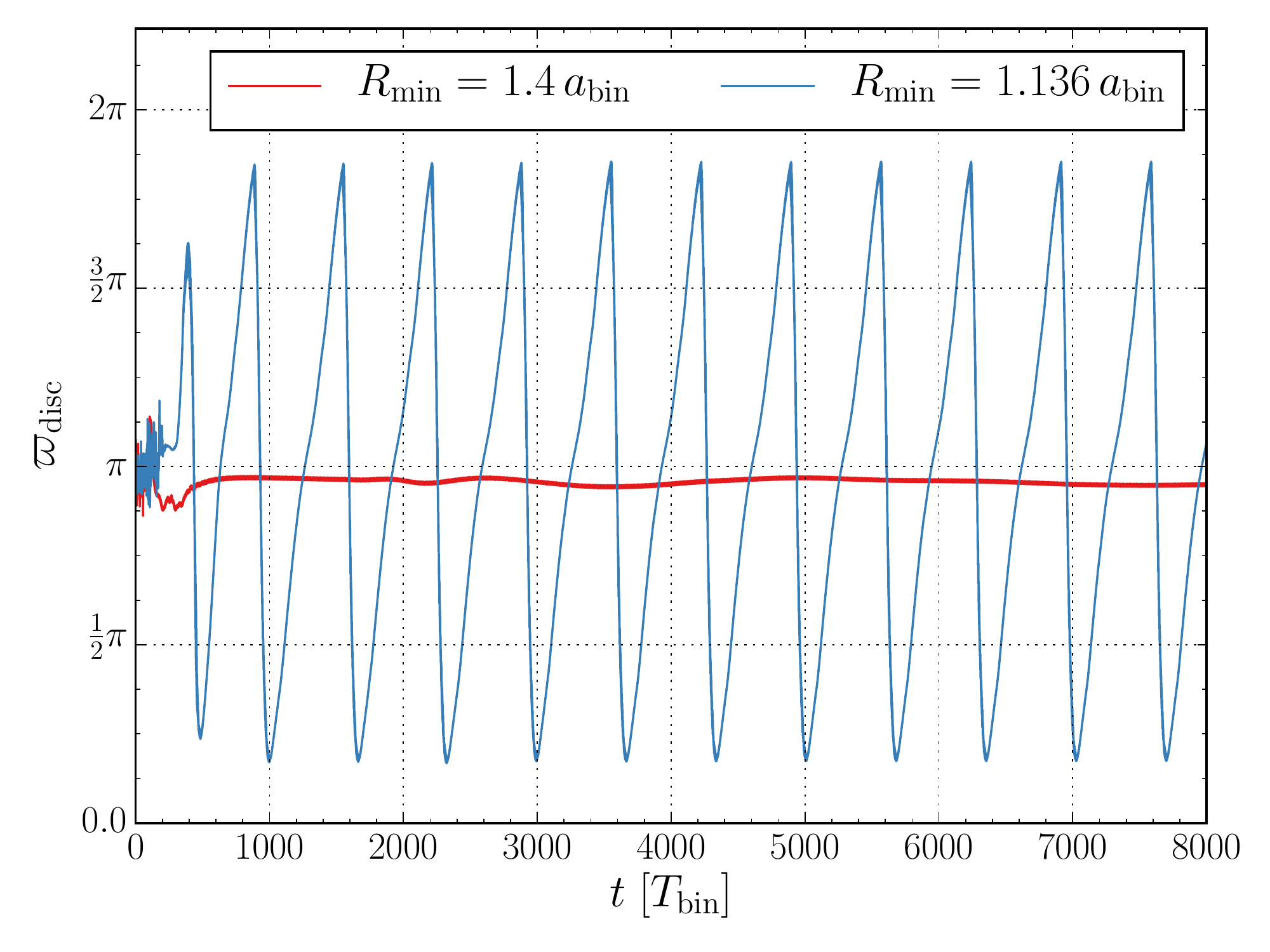}}
    \caption{Disc longitude of periastron for different inner radii of the computational domain. The red
    line shows results from a simulation with the \citet{2017MNRAS.466.1170M}
    setup where no precession of the disc is observable. The blue line shows
    results from the same physical setup but with a smaller inner disc radius.
    In this case a clear precession of the inner disc is visible.}
    \label{img:peri_stand_still}
\end{figure}

In our simulations all discs became eccentric and showed a prograde precession, 
and we did not find any indication for \enquote{stand still} discs that are 
eccentric but without any precession. 
In contrast, \citet{2017MNRAS.466.1170M} find for their disc setup ($h=0.1, \alpha=0.1$) that
the disc does not precess for binary eccentricities between $0.2$ and $0.4$. We
do not find this behaviour for our disc models, but note that we use a disc with
a lower aspect ratio and a lower $\alpha$-value.
To investigate this further we set up a simulation with the same numerical
parameters as \citet{2017MNRAS.466.1170M} and used a physical setup where they
did not observe a precession of the disc ($q_\mathrm{bin} =1.0$, $e_\mathrm{bin}
= 0.4$, $h=0.1$, and $\alpha = 0.1$). We carried out two simulations, one with an
inner radius of $R_\mathrm{min} = (1 + e_\mathrm{bin}) a_\mathrm{bin}$ and one
with an inner radius of $R_\mathrm{min} = \num{1.136}\,a_\mathrm{bin}$. The first radius
was used by \citet{2017MNRAS.466.1170M} and in this case we also observe no
precession of the disc. The second radius corresponds to $R_\mathrm{min} =
\SI{0.25}{au}$, which is the radius we established in Sect.~\ref{ssec:rmin} as
the optimum location of the inner boundary. In the case with the smaller inner
radius we observe a clear precession of the inner disc
(Fig.~\ref{img:peri_stand_still}) with a relatively short precession period
$T_\mathrm{prec}$ as expected for high viscosity and high $h$ (see below).  This
is another indication of the necessity of choosing $R_\mathrm{min}$ sufficiently
small to capture all effects properly.

\subsection{Binary mass ratio}\label{ssec:mass_ratio}
\begin{figure}
    \resizebox{\hsize}{!}{\includegraphics{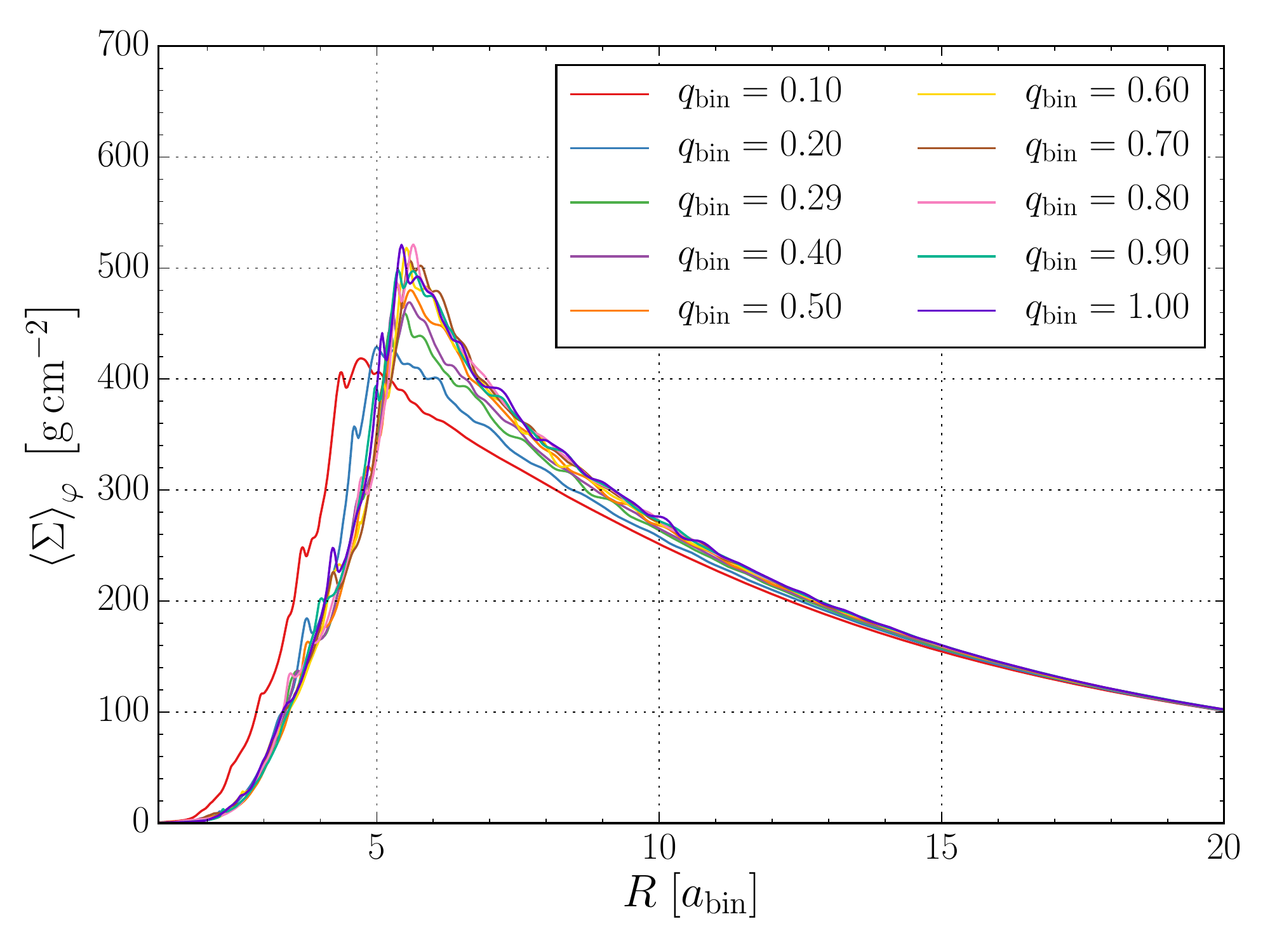}}
    \caption{Azimuthally averaged density profiles for varying binary
    mass ratio after \num{16000} binary orbits.}
    \label{img:sigma_mass_ratio}
\end{figure}

\begin{figure}
    \resizebox{\hsize}{!}{\includegraphics{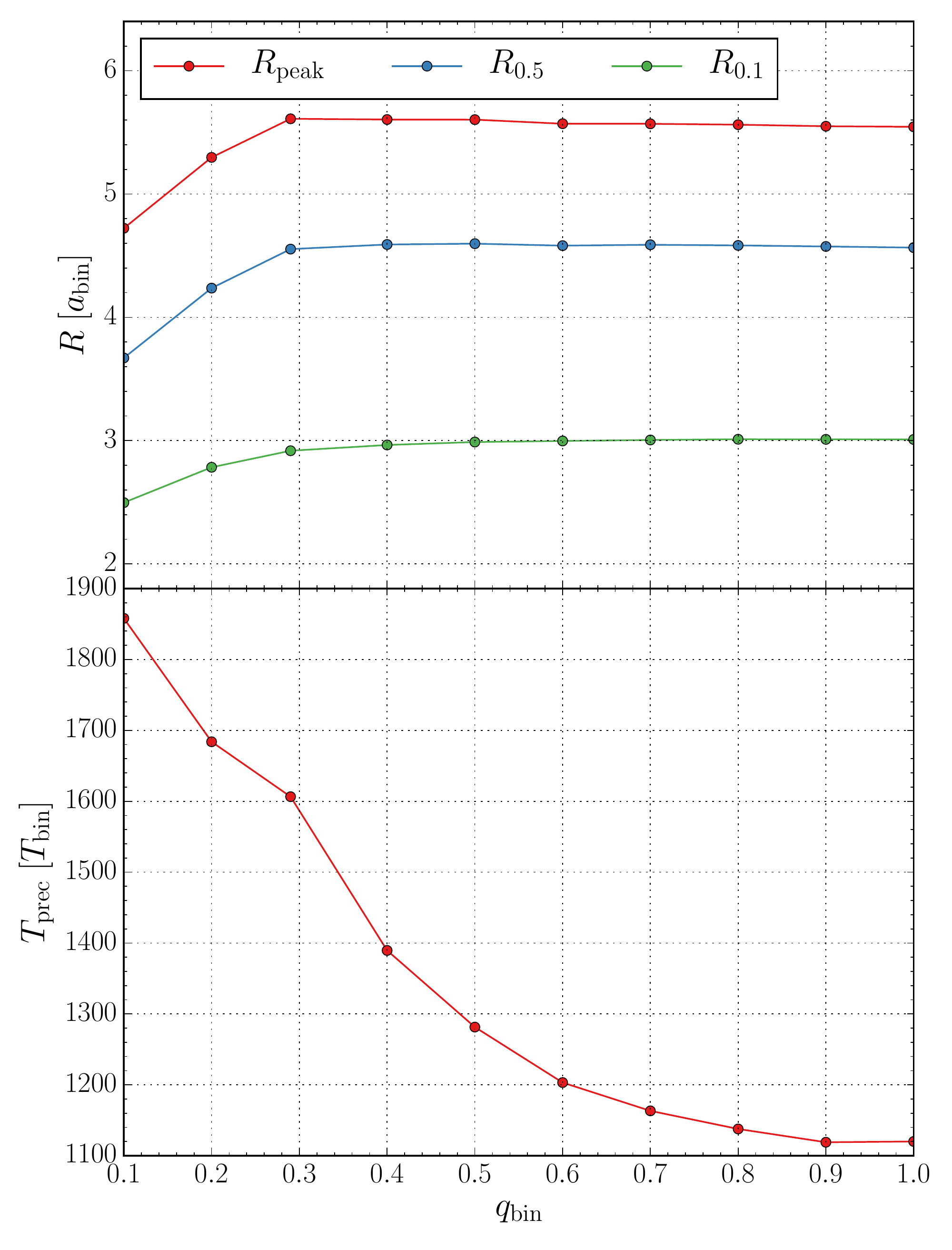}}
    \caption{Influence of different binary mass ratios on gap size and precession rate.
    \emph{Top:} Different measures of the gap size averaged over
    several thousand binary orbits. \emph{Bottom:} Precession period of the disc
    gap.}
    \label{img:gap_peri_mass_ratio}
\end{figure}

In this section we study the influence of the binary mass ratio. We carried out
simulations with a mass ratio from $q_\mathrm{bin} = 0.1$ to
$q_\mathrm{bin}=1.0$; all other parameters were set according to 
Table~\ref{tab:ref_setup}. Fig.~\ref{img:sigma_mass_ratio} shows the azimuthally
averaged density profiles for varying binary mass ratios. 
The density profiles do not show such a strong variation as in simulations
with varying binary eccentricities. Only the density profiles of the two lowest 
simulated mass ratios $q_\mathrm{bin}=0.1$ and $q_\mathrm{bin}=0.2$ differ 
significantly from the other profiles. For these models the position of the peak surface density
is closer to the binary and the maximum surface density roughly 20 percent lower. The
different gap size measures, introduced in the previous section, are shown in
the top panel of Fig.~\ref{img:gap_peri_mass_ratio}. 
In general, the variations of
all three gap-size indicators with mass ratio $q_\mathrm{bin}$ are relatively
weak. All three gap size measurements remain nearly constant for mass ratios
greater than \num{0.3}. Since the density profiles do not show a significant
variation this is not surprising. At the same time the size and eccentricity of
the gap do not vary significantly and lie in the range $a_\mathrm{gap} \approx
4\,a_\mathrm{bin}$ and $e_\mathrm{gap} \approx 0.27$.  
Overall, compared to
simulations with varying binary eccentricity the binary mass ratio does not have such
a strong influence on the inner disc structure for $q_\mathrm{bin} \gtrsim 0.3$.

In contrast, the precession period of the inner disc shows a far stronger
dependence on the binary mass ratio than the gap size (bottom panel of
Fig.~\ref{img:gap_peri_mass_ratio}).  Discs around binaries with a higher mass
ratio have a lower precession period than discs around binaries with a low mass
ratio. This dependence can be understood in terms of free particle orbits around
a binary star that also display a reduction of the precession period with
increasing $q$, as we discuss in more detail in Appendix~\ref{sec:test_particle}.

\section{Variation of disc parameters}\label{sec:disc_para}
In this section we explore the influence of discs parameters, namely the aspect
ratio and the alpha viscosity, on the inner disc structure. We use the Kepler-16
values for the binary and our standard numerical setup
(Table~\ref{tab:ref_setup}). 

While performing the simulations for this paper we observed that models with low
pressure (low $H/R$) and high viscosity (high $\alpha$) seem to be very
challenging for the Riemann-Code \textsc{Pluto}.  As a consequence we also present in
this section results using the Upwind-Code \textsc{Rh2d}.  As discussed in
Appendix~\ref{sec:convergence}, the two codes produce results in very good
agreement with each other for our standard model. For other parameters they do not necessarily
produce results with such good agreement, but the simulation results from both
codes always show the same trend.  Therefore, we decided to mix simulation
results from \textsc{Pluto} and \textsc{Rh2d} to cover a broader parameter
range.

\subsection{Disc aspect ratio}
\begin{figure}
    \resizebox{\hsize}{!}{\includegraphics{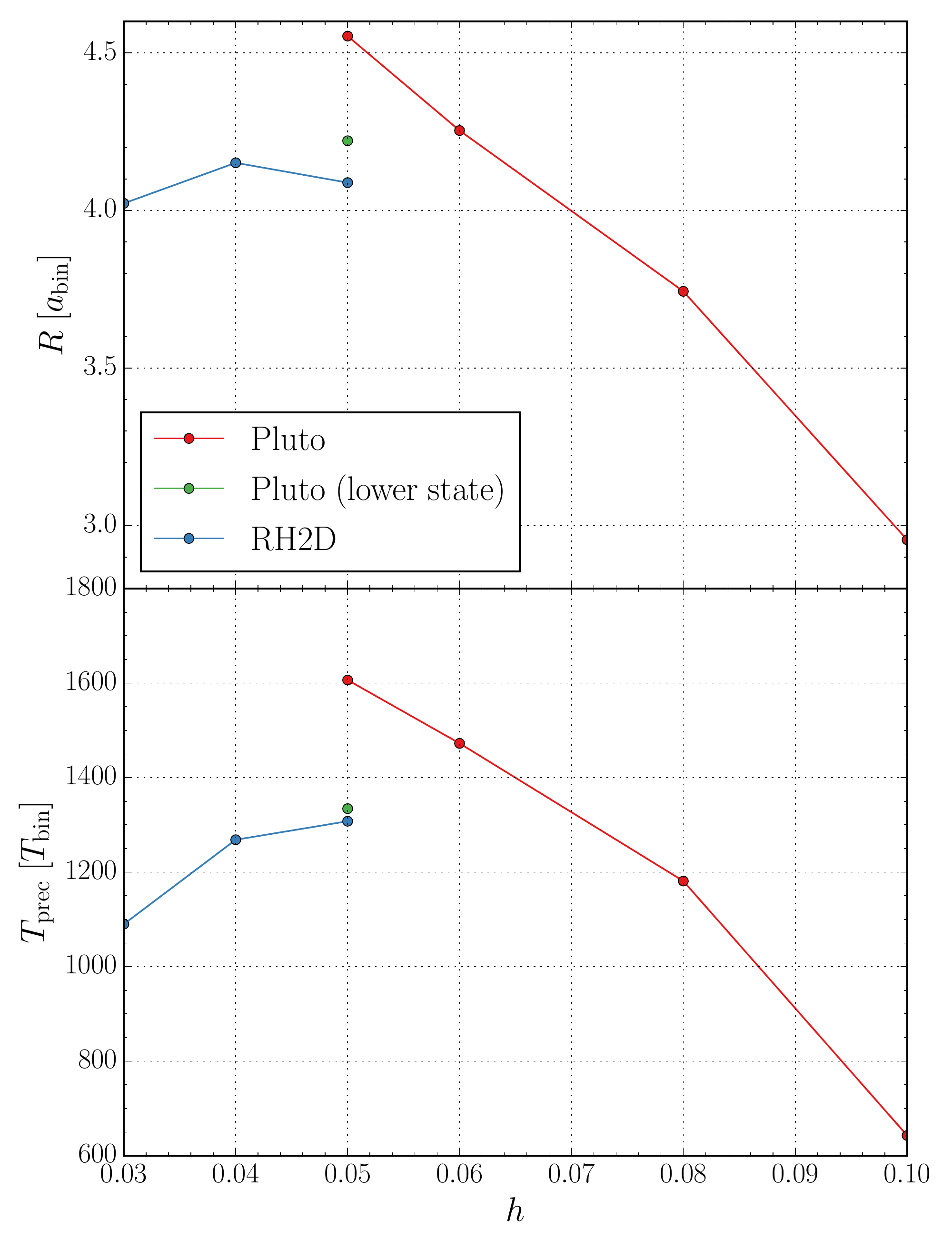}}
    \caption{Influence of different disc aspect ratios 
    on gap size and precession rate.
    \emph{Top:} Radius where the density drops to 50 percent of its
    maximum value averaged over several thousand binary orbits. Since results
    from two different codes are shown, we omit the two other measurements for
    the gap size presented in previous sections.
    \emph{Bottom:} Precession period of the disc
    gap. For $e_\mathrm{bin} \approx 0.16$ we observed that
    \textsc{Pluto} simulations can switch between two states, an upper state
    (red) and a lower state (green). For \textsc{RH2D} simulations we only
    observed the lower state (blue). This is discussed in more detail in
    Appendix~\ref{sec:convergence}.}
    \label{img:gap_peri_h}
\end{figure}

First we varied the disc aspect ratio $h$ from \num{0.03} to \num{0.1}. Our
simulation results show a decreasing gap size for higher aspect ratios
(Fig.~\ref{img:gap_peri_h}). The gaps precession period is again directly
correlated to the gap size, an observation we have already seen for different
binary eccentricities. 
In general, a drop in $T_\mathrm{prec}$ with higher $h$ is expected because an
increase in pressure will tend to reduce the gap size, which will in turn lead to
a faster precession. This trend is indeed observed in our simulations for higher
$h$.  For simulations with $h < 0.05$ we observed a decrease in gap size and
precession period. The drop in $T_\mathrm{prec}$ with lower values of $h$ may be
partly due to the lack of numerical resolution and partly to the lack of
pressure support, which may no longer allows for coherent disc precession.

\subsection{Alpha viscosity}
\begin{figure}
    \resizebox{\hsize}{!}{\includegraphics{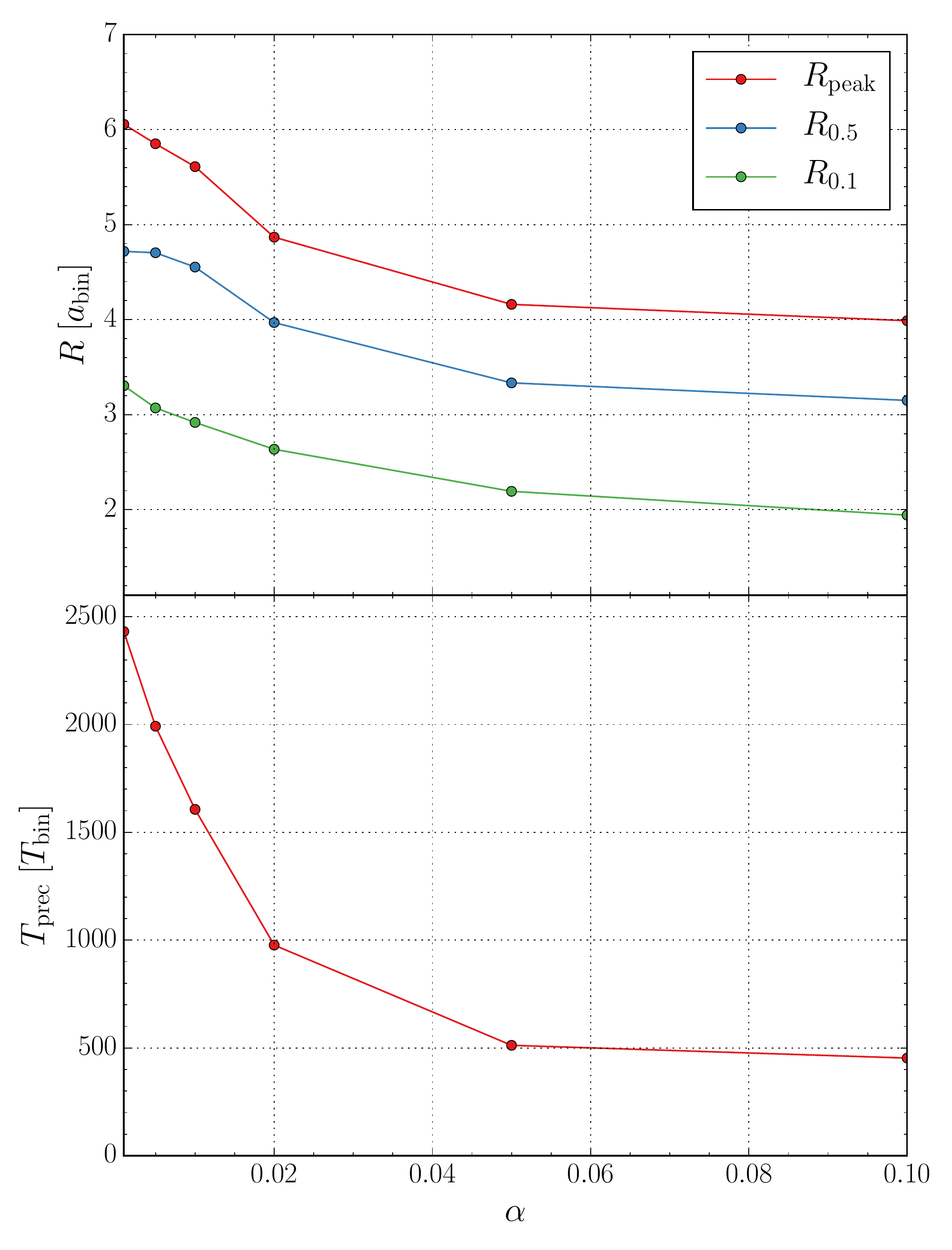}}
    \caption{Influence of different disc viscosities
   on gap size and precession rate.
    \emph{Top:} Different measures of the gap size averaged over
    several thousand binary orbits. \emph{Bottom:} Precession period of the disc
    gap.}
    \label{img:gap_peri_alpha}
\end{figure}

In this section we discuss the influence of the magnitude of viscosity on the disc
structure and therefore set up simulations with different viscosity
coefficients ranging from $\alpha = 0.001$ to $\alpha = 0.1$. We then analysed
the structure and behaviour of the inner cavity as before. The results are summarised in
Fig.~\ref{img:gap_peri_alpha}. A clear trend is visible for the gap size and for
the precession period of the gap. For higher viscosities the gap size
shrinks and the precession period decreases. Again, we see the direct correlation
of gap size and precession rate, as already observed for aspect ratio and binary
eccentricity variations. The decrease in the disc's gap size can be explained:
for higher $\alpha$ values the viscous spreading of the disc increases. This
viscous spreading counteracts the gravitational torques of the binary, which are
responsible for the gap creation.

\section{Summary and discussion}\label{sec:results}
Using two-dimensional hydrodynamical simulations, we studied the structure of
circumbinary discs for different numerical and physical parameters. Since these
simulations are, from a numerical point of view, not trivial and often it is not
easy to distinguish physical features and numerical artefacts, we checked our
results with three different numerical codes. 
In the following we summarise the most important results from our simulations
with different numerical and physical parameters.

Concerning the numerical treatment the most crucial issue with simulations
using grid codes in cylindrical coordinates is the treatment of the inner boundary condition. 
As there has been some discussion in the literature about this issue 
\citep{2013A&A...556A.134P,2014A&A...564A..72K,2015A&A...582A...5L,2017MNRAS.466.1170M,2017MNRAS.465.4735M},
we decided to perform a careful study of the location of the inner boundary and the conditions on the
hydrodynamical variables at $R_\mathrm{min}$. Our results can be summarised as follows:
\begin{itemize}
    \itemsep1.0em
    \item \textbf{Inner boundary condition}. Since there was no agreement on which
        type of inner boundary condition should be used for circumbinary disc
        simulations, especially if closed or open boundary conditions should be
        used, we investigated the influence of the inner boundary condition
        through dedicated simulations with two different codes. We observed that 
        closed boundaries can lead to numerical instabilities for all codes 
        used in our comparison and no convergence to a unique solution could be found.
        The use of a viscous outflow condition where the velocity is set to the mean
        viscous disc speed produced very similar results to the closed boundary because
        the viscous speed is very low in comparison to the velocities induced by the dynamical
        action of the binary stars. 
        Open inner boundaries, on the other hand, lead to numerically stable results and are
        also, from a physical point of view, more logical because material can
        flow onto the binary and be accreted by one of the stars.
        Hence, open boundaries with free outflow are the preferred conditions
        at $R_\mathrm{min}$.

    \item \textbf{Location of inner boundary}. The location of the inner
        boundary is also an important numerical parameter.  Since an inner hole
        in the computational domain cannot be avoided in polar coordinates, the
        inner radius has to be sufficiently small to capture all relevant physical
        effects. In particular, all important mean motion resonances, which may
        be responsible for the development of the eccentric inner cavity, should
        lie inside the computational domain. Through a parameter study we were
        able to determine that the radius of the inner boundary $R_\mathrm{min}$ has to
        be of the order of the binary separation $a_\text{bin}$ to capture all
        physical effects in a proper way. 
\end{itemize}
After having determined the best values for the numerical issues we performed, 
in the second part of the paper a careful study to determine the physical
aspects that determine the dynamics of circumbinary discs.
Starting from a reference model, where we chose the binary parameter of Kepler-16,
we varied individual parameters of the binary and the disc and studied their
impact on the disc dynamics. First of all, for all choices of our parameters 
the models produce an eccentric inner disc that shows a coherent prograde precession.
However, the size of the gap, its eccentricity, and its precession rate depend
on the physical parameters of binary and disc.
Our results can be summarised as follows:
\begin{itemize}
    \itemsep1.0em
    \item \textbf{Binary parameters} To study the influence of the binary star
        on the disc we systematically varied the eccentricity ($e_\mathrm{bin}$)
        and the mass ratio ($q_\mathrm{bin}$) of the binary.  The parameter
        study showed that $e_\mathrm{bin}$ has a strong influence on the gap
        size and on the precession period of the gap. We found that two
        regimes exist where the disc behaves differently to an increasing binary
        eccentricity (see Fig.~\ref{img:2d_sigma_ecc}).  The two branches
        bifurcate at a critical binary eccentricity, $e_\mathrm{crit} = 0.18$
        from each other. From $e_\mathrm{bin} = 0.0$ to $e_\mathrm{crit}$ the gap
        size and precession period decrease, and from $e_\mathrm{bin} = 0.18$
        onward both gap parameters become larger again, as displayed in
        Fig.~\ref{img:Tgap_vs_R}. The bifurcation of the two branches near
        $e_\mathrm{crit} = 0.18$ strongly suggests that different physical
        mechanisms, responsible for the creation of the eccentric inner cavity,
        operate in the two regimes. For the lower branch the  excitation at the
        3:1 outer Lindblad resonance may be responsible, which is supported by a
        simulation where the inner boundary was outside the 3:1 radius and no
        disc eccentricity was found. For the upper branch (high
        $e_\mathrm{bin}$) non-linear effects may be present, as suggested by
        \citet{2017MNRAS.464.3343F} who found similar behaviour. 

        As second binary parameter we varied the mass ratio of the secondary to
        the primary star, $q_\mathrm{bin}$, between $[0.1,1.0]$ and studied its
        impact on the gap size and precession period.  Overall the variation of
        $R_{0.5}$ with $q_\mathrm{bin}$ is weak.  For low $q_\mathrm{bin}$ the
        gap size increases until it becomes nearly constant for $q_\mathrm{bin}
        \gtrsim 0.3$ with $R_{0.5} \approx 4.6\,a_\mathrm{bin}$, as shown in
        Fig.~\ref{img:gap_peri_mass_ratio}.  The precession period decreases on
        average with increasing $T_\mathrm{prec}$ and for large $q_\mathrm{bin}$
        the behaviour is equivalent to a single particle at a separation of
        \SI{4.9}{au}, as we discuss in more detail in Appendix~\ref{sec:test_particle}.

    \item \textbf{Disc parameters} We also varied different disc parameters, namely the
        pressure (through the aspect ratio $h=H/R$) and the viscosity (through
        $\alpha$).  The results are displayed in Figs.~\ref{img:gap_peri_h} and
        Fig.~\ref{img:gap_peri_alpha}.  For changes in the aspect ratio no clear
        trend was visible; Both gap size and precession period first increase
        with increasing $h$ until they reach a maximum at $h=0.05$; from there
        both gap properties decrease again with increasing aspect ratio.
        For high $h$ the behaviour can be understood in term of the gap closing
        tendency of higher disc pressure. On the other hand, for very low
        pressure it may be more difficult to sustain a coherent precession of a
        large inner hole in the disc due to the reduced sound speed.
        Additionally, the damping action of the viscosity becomes more important for
        discs with lower sound speed.
 
        A clear monotonic trend was visible for the viscosity variation, where
        higher $\alpha$ leads to smaller gaps and shorter gap precession periods.
        This can be directly attributed to the gap closing tendency of
        viscosity.
\end{itemize}
\begin{figure}
    \resizebox{\hsize}{!}{\includegraphics{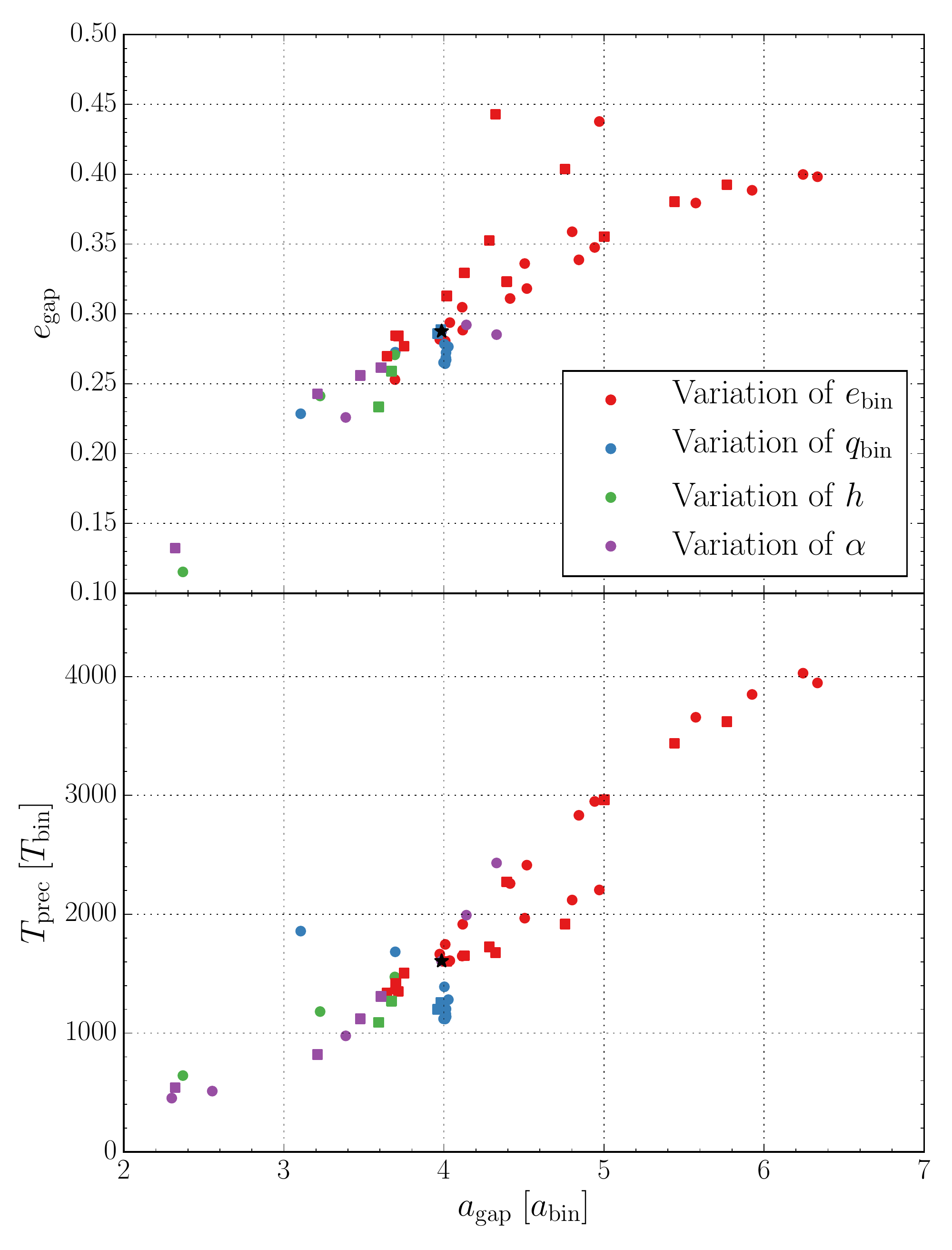}}
    \caption{Eccentricity (top) and precession period (bottom) of the inner gap
    plotted against the semi-major axis of the gap. Different simulation series
    are colour-coded. Both the measured gap eccentricity and gap semi-major axis
    are averaged over several thousand binary orbits. \textsc{Pluto} simulations
    are represented by dots and squares stand for \textsc{Rh2d} simulations. The
    reference model with Kepler-16 parameters is marked with the star.}
    \label{img:e_T_gap_vs_a_gap}
\end{figure}
To allow for an alternative view of the impact of different binary and disc
parameters on the dynamical structure of the disc we display in
Fig.~\ref{img:e_T_gap_vs_a_gap} the eccentricity (top) and the precession period
(bottom) of the gap plotted against the semi-major axis of the gap for all our
simulations where different colours stand for different model series.
\textsc{Pluto} simulations are represented by dots whereas squares stand for
\textsc{Rh2d} simulations.  The reference model is marked with a star.
Interestingly the majority of the models lie on a main-sequence branch, which has
the trend of increasing eccentricity and precession period with increasing gap
width.  The smaller branch that bifurcates near the reference model represents
the low $e_\mathrm{bin}$ sequence, which has higher $e_\mathrm{gap}$ but smaller
$T_\mathrm{prec}$.  By coincidence, our reference model with the Kepler-16
parameters lies near the bifurcation point.

How the location of the bifurcation point depends of the disc and binary
parameter needs to be explored in subsequent numerical studies. A first
simulation series with $q_\mathrm{bin} = 0.6$ suggest that this critical
eccentricity does not depend on the mass ratio of the binary. As outlined in
the Appendix, for parameters very near to the critical $e_\mathrm{bin}$ the
models can have the tendency to switch between the two states during a
single simulation. The physical cause of the bifurcation needs to be analysed in
subsequent simulations.

In addition, there are a more possibilities to further improve our model. First,
the use of a non-isothermal equation should be considered to study the effects
of viscous heating and radiative cooling. Especially for simulations, which also
evolve planets inside the disc, \citet{2014A&A...564A..72K} showed that
radiative effects should been taken into account.  Self-gravity effects have
been studied by \citet{2015A&A...582A...5L} and \citet{2017MNRAS.465.4735M} and
are important for galactic discs around a central binary black hole.  The
backreaction of the disc onto the binary could also be considered in more detail.
Angular momentum exchange with the binary occurs through gravitational torques
from the disc and direct accretion of angular momentum from the material that
enters the central cavity.  The first part always leads a decrease in the binary
semi-major axis and an increase in eccentricity \citep{2015A&A...581A..20K},
while \citet{2017MNRAS.466.1170M} pointed out that angular momentum
advection might even be dominant leading to a binary separation. 

Another interesting question is the evolution of planets inside the circumbinary
disc and their influence on the disc structure. Since an in situ formation of
the observed planets seems very unlikely, the migration and final parking
position of planets is an interesting topic that requires further
investigations.  In view of our extensive parameter studies there is the
indication that in several past studies
\citep{2013A&A...556A.134P,2014A&A...564A..72K,2015A&A...581A..20K} an inner
radius was considered that was too large and that did not allow for full disc
dynamics.

Through fully three-dimensional simulations it will be possible to study the
dynamical evolution of inclined discs.

\begin{acknowledgements}
Daniel Thun was funded by grant KL 650/26 of the German Research Foundation
(DFG), and Giovanni Picogna acknowledges DFG support grant KL 650/21 within
the collaborative research program \enquote{The first 10 Million Years of
the Solar System}.  Most of the numerical simulations were performed on the
bwForCLuster BinAC, supported by the state of Baden-Württemberg through
bwHPC, and the German Research Foundation (DFG) through grant INST 39/963-1
FUGG. All plots in this paper were made with the Python library
matplotlib \citep{Hunter:2007}.
\end{acknowledgements}

\bibliography{cb}

\begin{thebibliography}{56}
\expandafter\ifx\csname natexlab\endcsname\relax\def\natexlab#1{#1}\fi

\bibitem[{{Andrews} {et~al.}(2014){Andrews}, {Chandler}, {Isella}, {Birnstiel},
  {Rosenfeld}, {Wilner}, {P{\'e}rez}, {Ricci}, {Carpenter}, {Calvet}, {Corder},
  {Deller}, {Dullemond}, {Greaves}, {Harris}, {Henning}, {Kwon}, {Lazio},
  {Linz}, {Mundy}, {Sargent}, {Storm}, \& {Testi}}]{2014ApJ...787..148A}
{Andrews}, S.~M., {Chandler}, C.~J., {Isella}, A., {et~al.} 2014, \apj, 787,
  148

\bibitem[{{Armitage} \& {Natarajan}(2002)}]{2002ApJ...567L...9A}
{Armitage}, P.~J. \& {Natarajan}, P. 2002, \apjl, 567, L9

\bibitem[{{Artymowicz} {et~al.}(1991){Artymowicz}, {Clarke}, {Lubow}, \&
  {Pringle}}]{1991ApJ...370L..35A}
{Artymowicz}, P., {Clarke}, C.~J., {Lubow}, S.~H., \& {Pringle}, J.~E. 1991,
  \apjl, 370, L35

\bibitem[{{Artymowicz} \& {Lubow}(1994)}]{1994ApJ...421..651A}
{Artymowicz}, P. \& {Lubow}, S.~H. 1994, \apj, 421, 651

\bibitem[{{Artymowicz} \& {Lubow}(1996)}]{1996ApJ...467L..77A}
{Artymowicz}, P. \& {Lubow}, S.~H. 1996, \apjl, 467, L77

\bibitem[{{Baruteau} \& {Masset}(2008)}]{Baruteau:2008}
{Baruteau}, C. \& {Masset}, F. 2008, \apj, 672, 1054

\bibitem[{{Bate} {et~al.}(2002){Bate}, {Bonnell}, \&
  {Bromm}}]{2002MNRAS.336..705B}
{Bate}, M.~R., {Bonnell}, I.~A., \& {Bromm}, V. 2002, \mnras, 336, 705

\bibitem[{{Begelman} {et~al.}(1980){Begelman}, {Blandford}, \&
  {Rees}}]{1980Natur.287..307B}
{Begelman}, M.~C., {Blandford}, R.~D., \& {Rees}, M.~J. 1980, \nat, 287, 307

\bibitem[{{Cuadra} {et~al.}(2009){Cuadra}, {Armitage}, {Alexander}, \&
  {Begelman}}]{2009MNRAS.393.1423C}
{Cuadra}, J., {Armitage}, P.~J., {Alexander}, R.~D., \& {Begelman}, M.~C. 2009,
  \mnras, 393, 1423

\bibitem[{{de Val-Borro} {et~al.}(2006){de Val-Borro}, {Edgar}, {Artymowicz},
  {Ciecielag}, {Cresswell}, {D'Angelo}, {Delgado-Donate}, {Dirksen}, {Fromang},
  {Gawryszczak}, {Klahr}, {Kley}, {Lyra}, {Masset}, {Mellema}, {Nelson},
  {Paardekooper}, {Peplinski}, {Pierens}, {Plewa}, {Rice}, {Sch{\"a}fer}, \&
  {Speith}}]{2006MNRAS.370..529D}
{de Val-Borro}, M., {Edgar}, R.~G., {Artymowicz}, P., {et~al.} 2006, \mnras,
  370, 529

\bibitem[{{Di Folco} {et~al.}(2014){Di Folco}, {Dutrey}, {Le Bouquin},
  {Lacour}, {Berger}, {K{\"o}hler}, {Guilloteau}, {Pi{\'e}tu}, {Bary}, {Beck},
  {Beust}, \& {Pantin}}]{2014A&A...565L...2D}
{Di Folco}, E., {Dutrey}, A., {Le Bouquin}, J.-B., {et~al.} 2014, \aap, 565, L2

\bibitem[{{D'Orazio} {et~al.}(2016){D'Orazio}, {Haiman}, {Duffell},
  {MacFadyen}, \& {Farris}}]{2016MNRAS.459.2379D}
{D'Orazio}, D.~J., {Haiman}, Z., {Duffell}, P., {MacFadyen}, A., \& {Farris},
  B. 2016, \mnras, 459, 2379

\bibitem[{{Doyle} {et~al.}(2011){Doyle}, {Carter}, {Fabrycky}, {Slawson},
  {Howell}, {Winn}, {Orosz}, {Pr\v{s}a}, {Welsh}, {Quinn}, {Latham}, {Torres},
  {Buchhave}, {Marcy}, {Fortney}, {Shporer}, {Ford}, {Lissauer}, {Ragozzine},
  {Rucker}, {Batalha}, {Jenkins}, {Borucki}, {Koch}, {Middour}, {Hall},
  {McCauliff}, {Fanelli}, {Quintana}, {Holman}, {Caldwell}, {Still},
  {Stefanik}, {Brown}, {Esquerdo}, {Tang}, {Furesz}, {Geary}, {Berlind},
  {Calkins}, {Short}, {Steffen}, {Sasselov}, {Dunham}, {Cochran}, {Boss},
  {Haas}, {Buzasi}, \& {Fischer}}]{2011Sci...333.1602D}
{Doyle}, L.~R., {Carter}, J.~A., {Fabrycky}, D.~C., {et~al.} 2011, Science,
  333, 1602

\bibitem[{{Dunhill} {et~al.}(2015){Dunhill}, {Cuadra}, \&
  {Dougados}}]{2015MNRAS.448.3545D}
{Dunhill}, A.~C., {Cuadra}, J., \& {Dougados}, C. 2015, \mnras, 448, 3545

\bibitem[{{Dutrey} {et~al.}(2016){Dutrey}, {Di Folco}, {Beck}, \&
  {Guilloteau}}]{2016A&ARv..24....5D}
{Dutrey}, A., {Di Folco}, E., {Beck}, T., \& {Guilloteau}, S. 2016, \aapr, 24,
  5

\bibitem[{{Dutrey} {et~al.}(2014){Dutrey}, {di Folco}, {Guilloteau}, {Boehler},
  {Bary}, {Beck}, {Beust}, {Chapillon}, {Gueth}, {Hur{\'e}}, {Pierens},
  {Pi{\'e}tu}, {Simon}, \& {Tang}}]{2014Natur.514..600D}
{Dutrey}, A., {di Folco}, E., {Guilloteau}, S., {et~al.} 2014, \nat, 514, 600

\bibitem[{{Fleming} \& {Quinn}(2017)}]{2017MNRAS.464.3343F}
{Fleming}, D.~P. \& {Quinn}, T.~R. 2017, \mnras, 464, 3343

\bibitem[{{Georgakarakos} \& {Eggl}(2015)}]{2015ApJ...802...94G}
{Georgakarakos}, N. \& {Eggl}, S. 2015, \apj, 802, 94

\bibitem[{{Guilloteau} {et~al.}(1999){Guilloteau}, {Dutrey}, \&
  {Simon}}]{1999A&A...348..570G}
{Guilloteau}, S., {Dutrey}, A., \& {Simon}, M. 1999, \aap, 348, 570

\bibitem[{{G{\"u}nther} \& {Kley}(2002)}]{2002A&A...387..550G}
{G{\"u}nther}, R. \& {Kley}, W. 2002, \aap, 387, 550

\bibitem[{{Hawley} {et~al.}(1984){Hawley}, {Smarr}, \&
  {Wilson}}]{1984ApJS...55..211H}
{Hawley}, J.~F., {Smarr}, L.~L., \& {Wilson}, J.~R. 1984, \apjs, 55, 211

\bibitem[{Hunter(2007)}]{Hunter:2007}
Hunter, J.~D. 2007, Computing In Science \& Engineering, 9, 90

\bibitem[{{Kley}(1989)}]{1989A&A...208...98K}
{Kley}, W. 1989, \aap, 208, 98

\bibitem[{{Kley}(1999)}]{1999MNRAS.303..696K}
{Kley}, W. 1999, \mnras, 303, 696

\bibitem[{{Kley} \& {Dirksen}(2006)}]{2006A&A...447..369K}
{Kley}, W. \& {Dirksen}, G. 2006, \aap, 447, 369

\bibitem[{{Kley} \& {Haghighipour}(2014)}]{2014A&A...564A..72K}
{Kley}, W. \& {Haghighipour}, N. 2014, \aap, 564, A72

\bibitem[{{Kley} \& {Haghighipour}(2015)}]{2015A&A...581A..20K}
{Kley}, W. \& {Haghighipour}, N. 2015, \aap, 581, A20

\bibitem[{{Kley} \& {Nelson}(2008)}]{2008A&A...486..617K}
{Kley}, W. \& {Nelson}, R.~P. 2008, \aap, 486, 617

\bibitem[{{Kostov} {et~al.}(2014){Kostov}, {McCullough}, {Carter}, {Deleuil},
  {D{\'{\i}}az}, {Fabrycky}, {H{\'e}brard}, {Hinse}, {Mazeh}, {Orosz},
  {Tsvetanov}, \& {Welsh}}]{2014ApJ...784...14K}
{Kostov}, V.~B., {McCullough}, P.~R., {Carter}, J.~A., {et~al.} 2014, \apj,
  784, 14

\bibitem[{{Kostov} {et~al.}(2016){Kostov}, {Orosz}, {Welsh}, {Doyle},
  {Fabrycky}, {Haghighipour}, {Quarles}, {Short}, {Cochran}, {Endl}, {Ford},
  {Gregorio}, {Hinse}, {Isaacson}, {Jenkins}, {Jensen}, {Kane}, {Kull},
  {Latham}, {Lissauer}, {Marcy}, {Mazeh}, {M{\"u}ller}, {Pepper}, {Quinn},
  {Ragozzine}, {Shporer}, {Steffen}, {Torres}, {Windmiller}, \&
  {Borucki}}]{Kostov:2016}
{Kostov}, V.~B., {Orosz}, J.~A., {Welsh}, W.~F., {et~al.} 2016, \apj, 827, 86

\bibitem[{{Leung} \& {Lee}(2013)}]{2013ApJ...763..107L}
{Leung}, G.~C.~K. \& {Lee}, M.~H. 2013, \apj, 763, 107

\bibitem[{{Lines} {et~al.}(2015){Lines}, {Leinhardt}, {Baruteau},
  {Paardekooper}, \& {Carter}}]{2015A&A...582A...5L}
{Lines}, S., {Leinhardt}, Z.~M., {Baruteau}, C., {Paardekooper}, S.-J., \&
  {Carter}, P.~J. 2015, \aap, 582, A5

\bibitem[{{Liu} {et~al.}(2015){Liu}, {Mu{\~n}oz}, \&
  {Lai}}]{2015MNRAS.447..747L}
{Liu}, B., {Mu{\~n}oz}, D.~J., \& {Lai}, D. 2015, \mnras, 447, 747

\bibitem[{{Lodato} {et~al.}(2009){Lodato}, {Nayakshin}, {King}, \&
  {Pringle}}]{2009MNRAS.398.1392L}
{Lodato}, G., {Nayakshin}, S., {King}, A.~R., \& {Pringle}, J.~E. 2009, \mnras,
  398, 1392

\bibitem[{{Lubow}(1991)}]{1991ApJ...381..259L}
{Lubow}, S.~H. 1991, \apj, 381, 259

\bibitem[{{MacFadyen} \& {Milosavljevi{\'c}}(2008)}]{2008ApJ...672...83M}
{MacFadyen}, A.~I. \& {Milosavljevi{\'c}}, M. 2008, \apj, 672, 83

\bibitem[{{Masset}(2000)}]{Masset:2000}
{Masset}, F. 2000, \aaps, 141, 165

\bibitem[{{Mignone} {et~al.}(2007){Mignone}, {Bodo}, {Massaglia}, {Matsakos},
  {Tesileanu}, {Zanni}, \& {Ferrari}}]{2007ApJS..170..228M}
{Mignone}, A., {Bodo}, G., {Massaglia}, S., {et~al.} 2007, \apjs, 170, 228

\bibitem[{{Miranda} {et~al.}(2017){Miranda}, {Mu{\~n}oz}, \&
  {Lai}}]{2017MNRAS.466.1170M}
{Miranda}, R., {Mu{\~n}oz}, D.~J., \& {Lai}, D. 2017, \mnras, 466, 1170

\bibitem[{{Moriwaki} \& {Nakagawa}(2004)}]{2004ApJ...609.1065M}
{Moriwaki}, K. \& {Nakagawa}, Y. 2004, \apj, 609, 1065

\bibitem[{{M{\"u}ller} \& {Kley}(2012)}]{Mueller:2012}
{M{\"u}ller}, T.~W.~A. \& {Kley}, W. 2012, \aap, 539, A18

\bibitem[{{M{\"u}ller} {et~al.}(2012){M{\"u}ller}, {Kley}, \&
  {Meru}}]{2012A&A...541A.123M}
{M{\"u}ller}, T.~W.~A., {Kley}, W., \& {Meru}, F. 2012, \aap, 541, A123

\bibitem[{{Mutter} {et~al.}(2017){Mutter}, {Pierens}, \&
  {Nelson}}]{2017MNRAS.465.4735M}
{Mutter}, M.~M., {Pierens}, A., \& {Nelson}, R.~P. 2017, \mnras, 465, 4735

\bibitem[{{Orosz} {et~al.}(2012{\natexlab{a}}){Orosz}, {Welsh}, {Carter},
  {Brugamyer}, {Buchhave}, {Cochran}, {Endl}, {Ford}, {MacQueen}, {Short},
  {Torres}, {Windmiller}, {Agol}, {Barclay}, {Caldwell}, {Clarke}, {Doyle},
  {Fabrycky}, {Geary}, {Haghighipour}, {Holman}, {Ibrahim}, {Jenkins},
  {Kinemuchi}, {Li}, {Lissauer}, {Pr{\v s}a}, {Ragozzine}, {Shporer}, {Still},
  \& {Wade}}]{2012ApJ...758...87O}
{Orosz}, J.~A., {Welsh}, W.~F., {Carter}, J.~A., {et~al.} 2012{\natexlab{a}},
  \apj, 758, 87

\bibitem[{{Orosz} {et~al.}(2012{\natexlab{b}}){Orosz}, {Welsh}, {Carter},
  {Fabrycky}, {Cochran}, {Endl}, {Ford}, {Haghighipour}, {MacQueen}, {Mazeh},
  {Sanchis-Ojeda}, {Short}, {Torres}, {Agol}, {Buchhave}, {Doyle}, {Isaacson},
  {Lissauer}, {Marcy}, {Shporer}, {Windmiller}, {Barclay}, {Boss}, {Clarke},
  {Fortney}, {Geary}, {Holman}, {Huber}, {Jenkins}, {Kinemuchi}, {Kruse},
  {Ragozzine}, {Sasselov}, {Still}, {Tenenbaum}, {Uddin}, {Winn}, {Koch}, \&
  {Borucki}}]{2012Sci...337.1511O}
{Orosz}, J.~A., {Welsh}, W.~F., {Carter}, J.~A., {et~al.} 2012{\natexlab{b}},
  Science, 337, 1511

\bibitem[{{Papaloizou} {et~al.}(2001){Papaloizou}, {Nelson}, \&
  {Masset}}]{2001A&A...366..263P}
{Papaloizou}, J.~C.~B., {Nelson}, R.~P., \& {Masset}, F. 2001, \aap, 366, 263

\bibitem[{{Pierens} \& {Nelson}(2013)}]{2013A&A...556A.134P}
{Pierens}, A. \& {Nelson}, R.~P. 2013, \aap, 556, A134

\bibitem[{{Pringle}(1991)}]{1991MNRAS.248..754P}
{Pringle}, J.~E. 1991, \mnras, 248, 754

\bibitem[{{Rozyczka}(1985)}]{1985A&A...143...59R}
{Rozyczka}, M. 1985, \aap, 143, 59

\bibitem[{{Rozyczka} \& {Laughlin}(1997)}]{1997ASPC..121..792R}
{Rozyczka}, M. \& {Laughlin}, G. 1997, in Astronomical Society of the Pacific
  Conference Series, Vol. 121, IAU Colloq. 163: Accretion Phenomena and Related
  Outflows, ed. D.~T. {Wickramasinghe}, G.~V. {Bicknell}, \& L.~{Ferrario}, 792

\bibitem[{{Schwamb} {et~al.}(2013){Schwamb}, {Orosz}, {Carter}, {Welsh},
  {Fischer}, {Torres}, {Howard}, {Crepp}, {Keel}, {Lintott}, {Kaib}, {Terrell},
  {Gagliano}, {Jek}, {Parrish}, {Smith}, {Lynn}, {Simpson}, {Giguere}, \&
  {Schawinski}}]{2013ApJ...768..127S}
{Schwamb}, M.~E., {Orosz}, J.~A., {Carter}, J.~A., {et~al.} 2013, \apj, 768,
  127

\bibitem[{{Shakura} \& {Sunyaev}(1973)}]{1973A&A....24..337S}
{Shakura}, N.~I. \& {Sunyaev}, R.~A. 1973, \aap, 24, 337

\bibitem[{{Welsh} {et~al.}(2014){Welsh}, {Orosz}, {Carter}, \&
  {Fabrycky}}]{2014IAUS..293..125W}
{Welsh}, W.~F., {Orosz}, J.~A., {Carter}, J.~A., \& {Fabrycky}, D.~C. 2014, in
  IAU Symposium, Vol. 293, IAU Symposium, ed. N.~{Haghighipour}, 125--132

\bibitem[{{Welsh} {et~al.}(2012){Welsh}, {Orosz}, {Carter}, {Fabrycky}, {Ford},
  {Lissauer}, {Pr{\v s}a}, {Quinn}, {Ragozzine}, {Short}, {Torres}, {Winn},
  {Doyle}, {Barclay}, {Batalha}, {Bloemen}, {Brugamyer}, {Buchhave},
  {Caldwell}, {Caldwell}, {Christiansen}, {Ciardi}, {Cochran}, {Endl},
  {Fortney}, {Gautier}, {Gilliland}, {Haas}, {Hall}, {Holman}, {Howard},
  {Howell}, {Isaacson}, {Jenkins}, {Klaus}, {Latham}, {Li}, {Marcy}, {Mazeh},
  {Quintana}, {Robertson}, {Shporer}, {Steffen}, {Windmiller}, {Koch}, \&
  {Borucki}}]{2012Natur.481..475W}
{Welsh}, W.~F., {Orosz}, J.~A., {Carter}, J.~A., {et~al.} 2012, \nat, 481, 475

\bibitem[{{Welsh} {et~al.}(2015){Welsh}, {Orosz}, {Short}, {Cochran}, {Endl},
  {Brugamyer}, {Haghighipour}, {Buchhave}, {Doyle}, {Fabrycky}, {Hinse},
  {Kane}, {Kostov}, {Mazeh}, {Mills}, {M{\"u}ller}, {Quarles}, {Quinn},
  {Ragozzine}, {Shporer}, {Steffen}, {Tal-Or}, {Torres}, {Windmiller}, \&
  {Borucki}}]{2015ApJ...809...26W}
{Welsh}, W.~F., {Orosz}, J.~A., {Short}, D.~R., {et~al.} 2015, \apj, 809, 26

\bibitem[{{Yang} {et~al.}(2017){Yang}, {Hashimoto}, {Hayashi}, {Tamura},
  {Mayama}, {Rafikov}, {Akiyama}, {Carson}, {Janson}, {Kwon}, {de Leon}, {Oh},
  {Takami}, {Tang}, {Kudo}, {Kusakabe}, {Abe}, {Brandner}, {Brandt}, {Egner},
  {Feldt}, {Goto}, {Grady}, {Guyon}, {Hayano}, {Hayashi}, {Henning}, {Hodapp},
  {Ishii}, {Iye}, {Kandori}, {Knapp}, {Kuzuhara}, {Matsuo}, {Mcelwain},
  {Miyama}, {Morino}, {Moro-martin}, {Nishimura}, {Pyo}, {Serabyn}, {Suenaga},
  {Suto}, {Suzuki}, {Takahashi}, {Takato}, {Terada}, {Thalmann}, {Turner},
  {Watanabe}, {Wisniewski}, {Yamada}, {Takami}, \&
  {Usuda}}]{2017AJ....153....7Y}
{Yang}, Y., {Hashimoto}, J., {Hayashi}, S.~S., {et~al.} 2017, \aj, 153, 7

\end{thebibliography}
\bibliographystyle{aa}

\begin{appendix}

\section{Convergence study}\label{sec:convergence}
    The strong gravitational influence of the binary onto the disc, and the
    complex dynamics of the inner disc presents a big challenge for
    numerical schemes of grid codes. We therefore study in this appendix the
    dependence of the disc structure on numerical parameters, such as the numerical
    scheme or the grid resolution. First, we simulated our reference system,
    using an open inner boundary and an inner radius of $R_\mathrm{min} =
    \SI{0.25}{au}$, with the different codes described in Sec.~\ref{sec:codes}.

\begin{figure}
        \resizebox{\hsize}{!}{\includegraphics{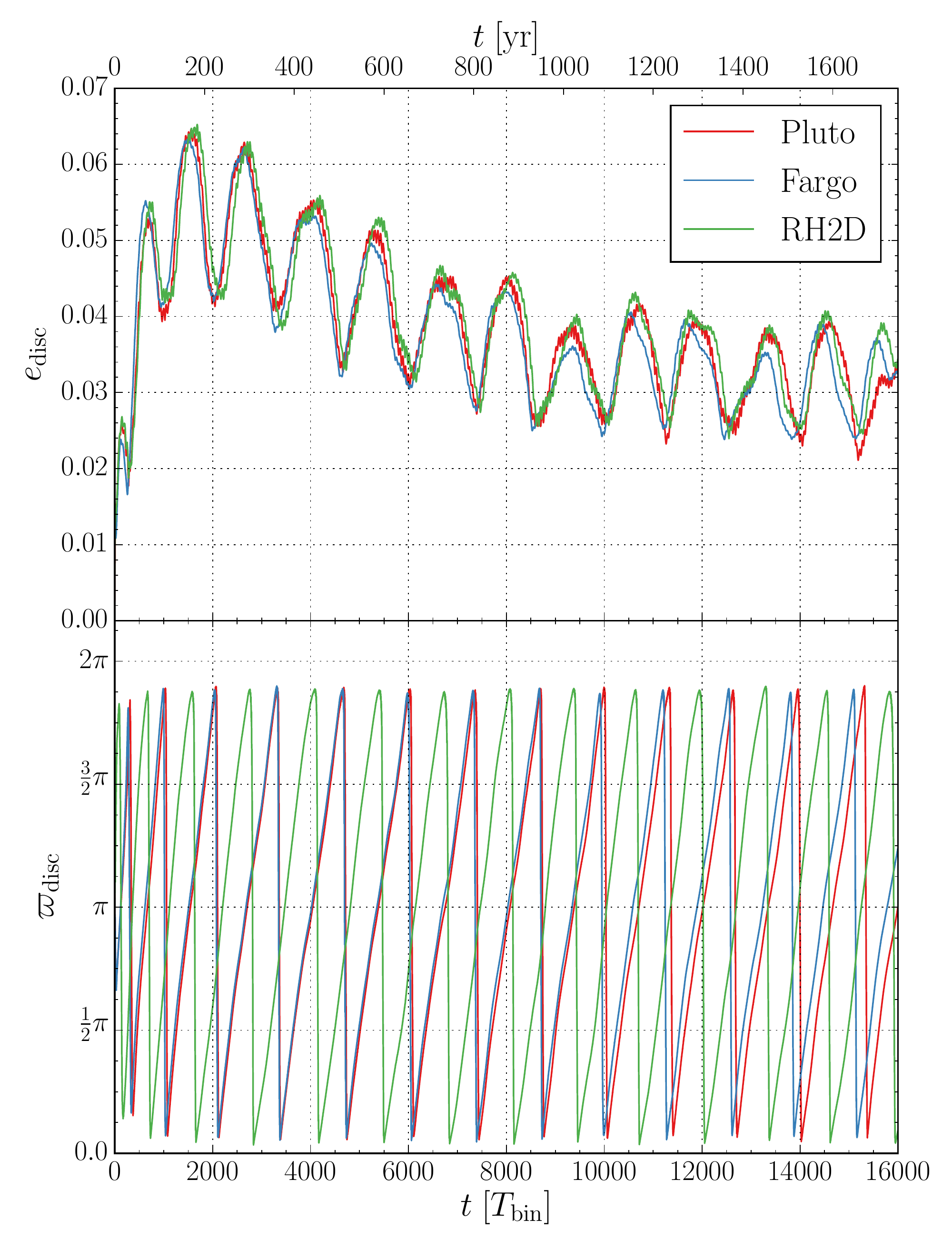}}
        \caption{Comparison of disc eccentricity (top) and disc longitude of
        pericentre (bottom) for the standard model. The disc eccentricity was
        calculated by summing over the whole disc ($R_2 = R_\mathrm{max}$ in
        equation~\eqref{eq:e_disc}),  whereas the disc longitude of pericentre
        was calculated only for the inner disc ($R_2 = \SI{1.0}{au}$ in
        equation~\eqref{eq:peri_disc}). The $\pi$-shift of the green curve in the
        bottom panel is explained in the text. We used a resolution of $512
        \times 580$ grid cells for \textsc{Pluto} simulations and for
        \textsc{Fargo} and \textsc{Rh2d} simulations a resolution of $448 \times
        512$.}
        \label{img:cmp_e_disc_peri}
    \end{figure}

\begin{figure}
        \resizebox{\hsize}{!}{\includegraphics{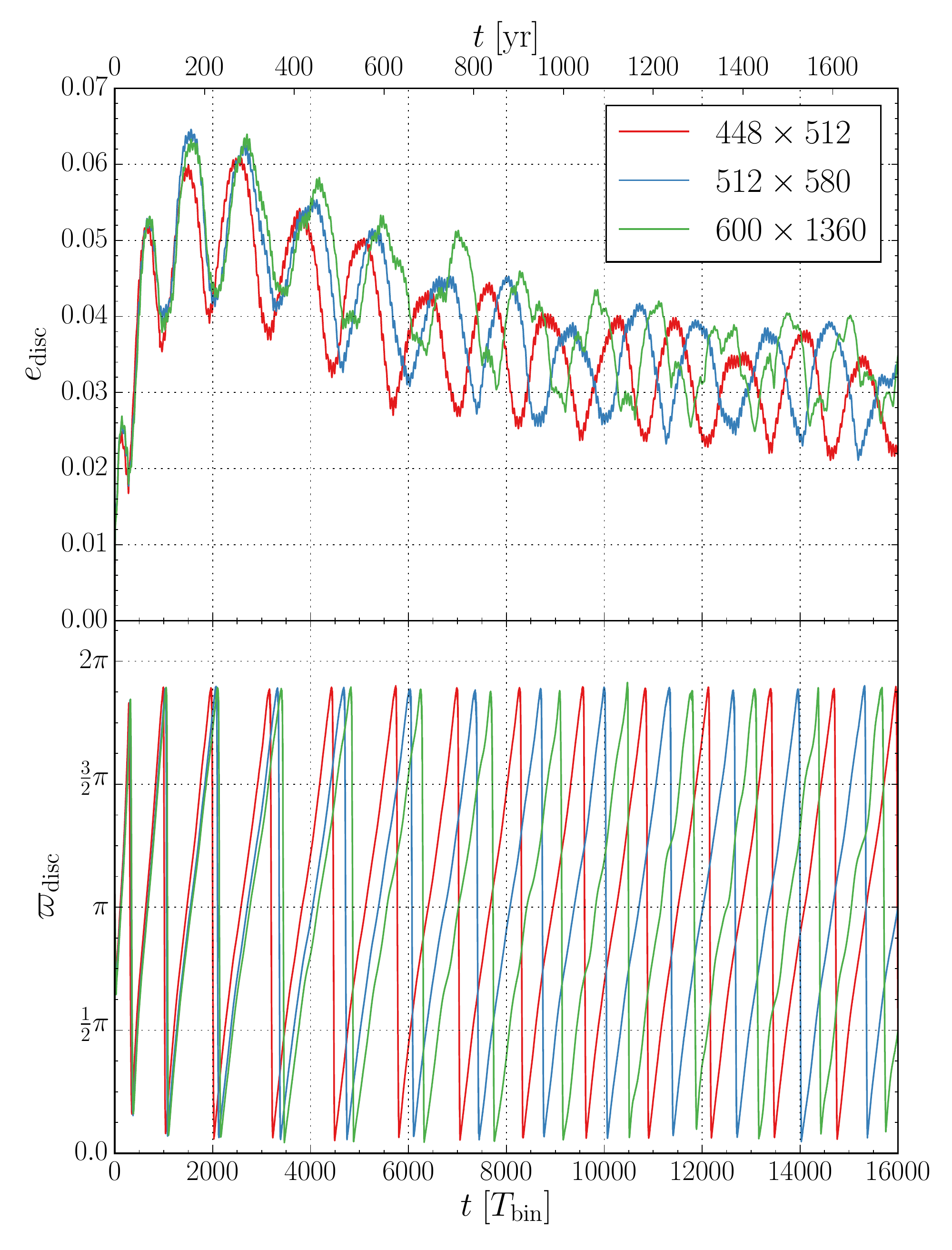}}
        \caption{Comparison of different resolutions. \emph{Top:} Disc
        eccentricity. \emph{Bottom:} Disc longitude of
        pericentre. All simulations were performed with \textsc{Pluto}.}
        \label{img:e_disc_peri_resolution}
    \end{figure}

    In Fig.~\ref{img:cmp_e_disc_peri} the long-time evolution of the disc
    eccentricity (top) and of the longitude of periastron (bottom) are shown.
    All three codes produce comparable results, although we should note that to
    get this agreement simulations performed with \textsc{Pluto} need a slightly
    higher resolution. \textsc{Pluto} simulations with a resolution of
    $448\times512$ grid cells produce a slightly smaller precession period of
    the inner disc (red line in Fig.~\ref{img:e_disc_peri_resolution}, see also
    Table~\ref{tab:res_peri}).  One possible explanation for this could be the
    size of the numerical stencil. \textsc{Fargo} and \textsc{Rh2d} have a very
    compact stencil because they use a staggered grid. In contrast,
    \textsc{Pluto} has a wider stencil because a collocated grid is used, where all
    variables are defined at the cell centres.  In particular the
    viscosity routine of \textsc{Pluto} has a very wide stencil to achieve the
    correct centring of the viscous stress tensor components. This could
    explain why for \textsc{Pluto} a slightly higher resolution is needed.

    The $\pi$-shift of the longitude of periastron calculated with \textsc{Rh2d}
    compared to the values calculated with \textsc{Pluto} and \textsc{Fargo}
    (bottom Panel of Fig.~\ref{img:cmp_e_disc_peri}) can be explained in the
    following way. Simulations performed with \textsc{Pluto} and \textsc{Fargo}
    started the binary at periastron with the pericentre of the secondary at
    $\varphi =\pi$ (see red ellipses in Fig.~\ref{img:2d_sigma_ecc}), whereas
    \textsc{Rh2d} started the binary at apastron with the pericentre of the
    secondaries orbit at $\varphi = 0$.  The data show that when the disc
    eccentricity is at its maximum the disc is aligned with the orbit of the
    secondary star.  Since the pericentre of the secondary in \textsc{Rh2d}
    simulations is shifted by $\pi$ we can also see this $\pi$-shift in the
    disc's longitude of periastron.

\begin{figure}
        \resizebox{\hsize}{!}{\includegraphics{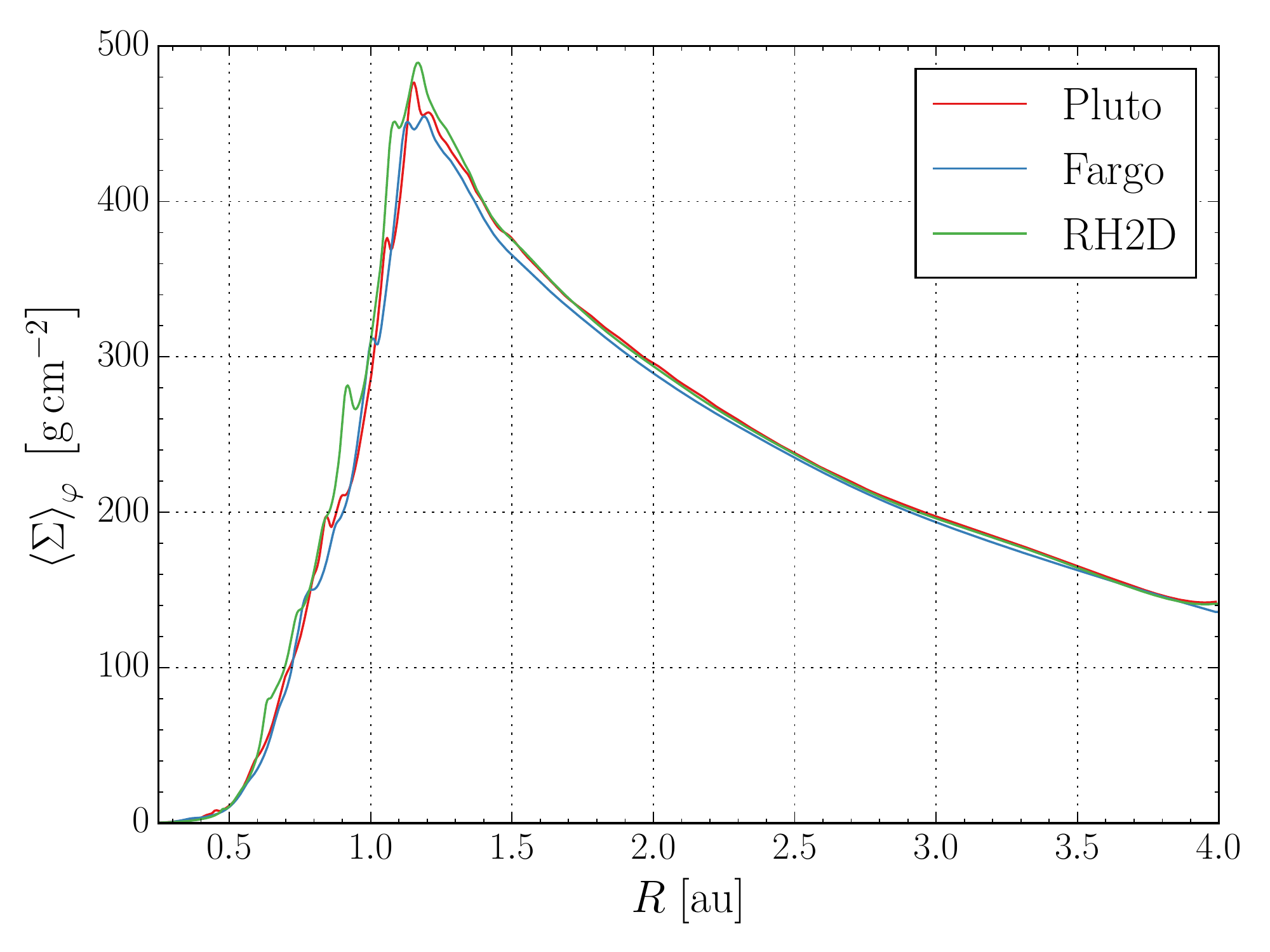}}
        \caption{Azimuthally averaged density profiles calculated with
            different codes at $t=\num{16000}\,T_\mathrm{bin}$.}
        \label{img:sigma_codes}
    \end{figure}

    Fig.~\ref{img:sigma_codes} shows the azimuthally averaged density
    profiles at $t=\num{16000}\,T_\mathrm{bin}$. Even after
    such a long time span all three codes agree very closely. All codes produce
    roughly the same density maximum and the same density slopes.

    To check the influence of the numerical resolution on the disc structure we
    run \textsc{Pluto} simulations with different resolutions, summarised in
    Table~\ref{tab:res_peri}. The results of this resolution test are shown in
    Fig.~\ref{img:e_disc_peri_resolution}.
    The disc eccentricity converges to the same value, whereas it seems that the
    disc precession rates increases with higher resolution.  A closer look at the
    data shows that after around \num{6000} orbits the precession rate for the
    higher resolution converged. Table~\ref{tab:res_peri} summarises the
    measured precession rate for different resolutions. Since there is no large
    variation between $512 \times 580$ and $600 \times 1360$ grid cells, we used a
    resolution of $512 \times 580$ grid cells for our Pluto simulations.
    \begin{table}
        \caption{Disc precession rate for different resolutions.}
        \label{tab:res_peri}
        \centering
        \begin{tabular}{lc}
            \midrule\midrule
            Resolution & $T_\mathrm{prec}\,[T_\mathrm{bin}]$ \\
            \midrule
            $448 \times 512$ & $1277.49$ \\
            $512 \times 580$ & $1321.48$ \\
            $600 \times 1360$ & $1349.50$ \\
            \midrule
        \end{tabular}
    \end{table}

    \begin{figure}
        \resizebox{\hsize}{!}{\includegraphics{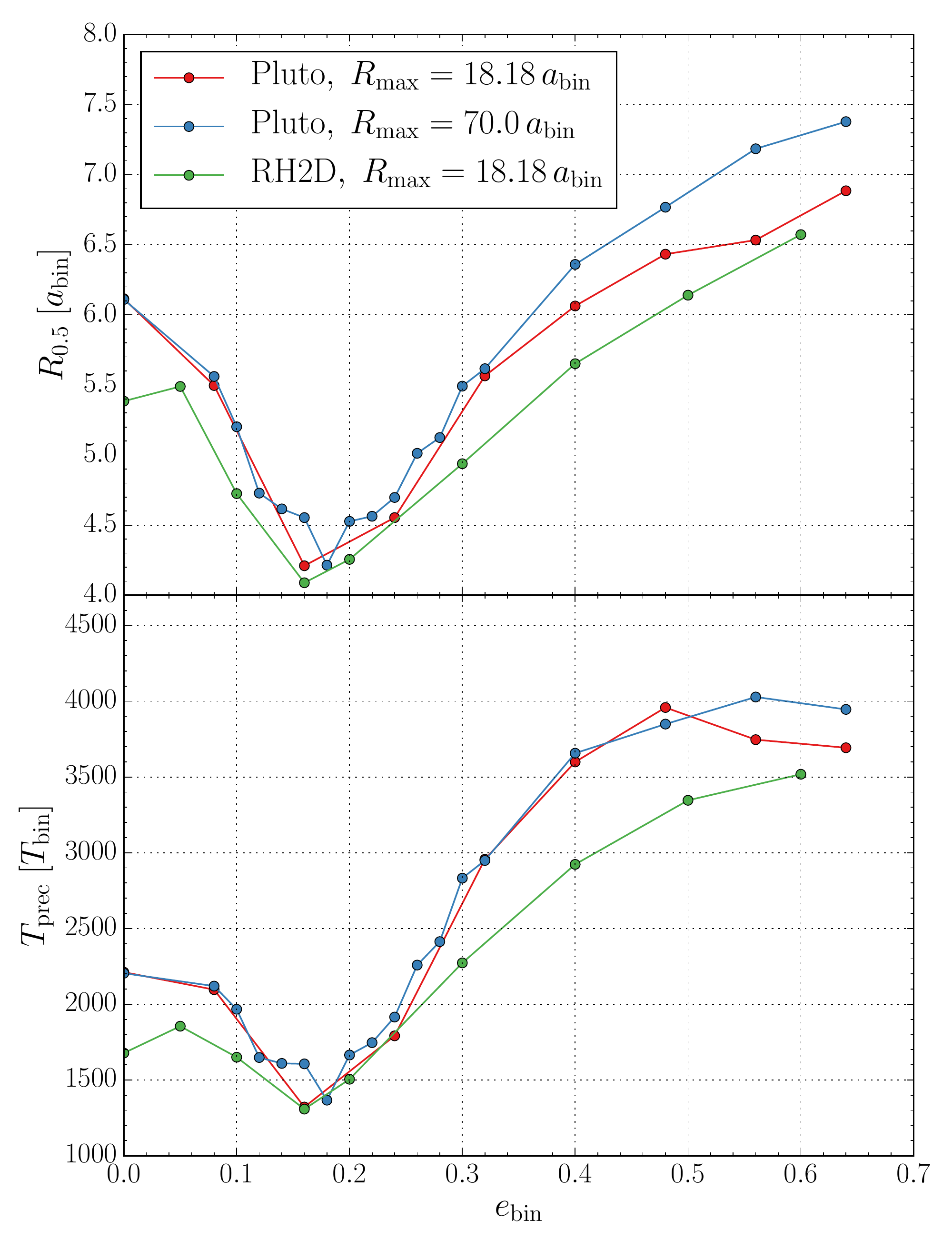}}
        \caption{Influence of different outer radii on gap size and
        precession rate while varying the binary eccentricity. 
        \emph{Top:} Radius where the azimuthally averaged surface density drops
        to 50 percent of its maximum value.
        \emph{Bottom:} Precession period of the disc gap.}
        \label{img:cmp_dom_ecc}
    \end{figure}

    As already discussed in Sect.~\ref{ssec:rmin} in some cases (for
    example high binary eccentricities) an outer radius of 
    $R_\mathrm{max} = \SI{4.0}{au}$ can lead to wave reflections which interfere with the
    inner disc. Therefore we increased the outer boundary to $R_\mathrm{max} =
    \SI{15.4}{au} = 70\,a_\mathrm{bin}$ (and to ensure the same resolution between
    $\SI{0.25}{au}$ and $\SI{4.0}{au}$ we also increased the number of grid
    points to $762 \times 582$). Figure~\ref{img:cmp_dom_ecc} compares the effect of
    different outer radii for the case of varying binary eccentricities.
    For binary eccentricities greater than \num{0.32}, a larger outer radius makes a 
    difference. Also, around the bifurcation point, at $e_\mathrm{bin} = 0.18$, a larger outer radius changes the gap size and
    precession rate of the disc. We also added results from \textsc{Rh2d}
    simulations with an outer radius of $R_\mathrm{max} =
    18.18\,a_\mathrm{bin}$. 
    These \textsc{Rh2d} results follow the same trend as the
    \textsc{Pluto} simulations, although for most binary eccentricities we could
    not reproduce the very good agreement of the $e_\mathrm{bin} = 0.16$ case.
    Around $e_\mathrm{bin} = 0.18$, which is the critical binary eccentricity
    where the two branches bifurcate, simulations tend to switch between two
    states. This jump happens after roughly \num{3000} binary
    orbits and does not depend on physical parameters. Changes in
    numerical parameters, like the Courant-Friedrichs-Lewy condition or $R_\mathrm{max}$, can trigger
    the jump. This can be seen in Fig.~\ref{img:cmp_dom_ecc} for $e_\mathrm{bin}
    = 0.16$ where simulations with $R_\mathrm{max} = 18.18\,a_\mathrm{bin}$ (red
    curve) and $R_\mathrm{max} = 70.0\,a_\mathrm{bin}$ (blue curve) produce
    different values for the gap size and the precession period. An outer
    radius of $R_\mathrm{max} = 18.18\,a_\mathrm{bin}$ is in the case of
    $e_\mathrm{bin} = 0.16$ not too small, since a simulation with
    $R_\mathrm{max} = 100\,a_\mathrm{bin}$ again produced the same values for
    the gap size and the precession period. For simulations with
    $e_\mathrm{bin}$ far away from $e_\mathrm{crit}$, we did not observe these
    jumps between states. So far we have only observed these jumps in simulations
    carried out with \textsc{Pluto}.

\section{Comparison with massless test particle}
\label{sec:test_particle}
In our circumbinary disc evolution we found that the inner disc becomes eccentric with
a constant precession rate. In this section we compare this to test particle trajectories around
the binary and investigate how a massless particle with orbital
elements similar to the disc gap (Fig.~\ref{img:e_T_gap_vs_a_gap}) would behave
under the influence of the binary potential. In particular, we would like to know at what
distance from the binary a test particle has to be positioned such that its precession period
agrees approximately with that of the disc.
Using secular perturbation theory for the coplanar motion of a massless test particle around an
eccentric binary star \citet{2004ApJ...609.1065M} derived the following formula for the
particle's precession period for low binary eccentricities
\begin{equation}\label{eq:particle_prec}
    T_\mathrm{prec} = \frac{4}{3} 
                            \frac{(q_\mathrm{bin} + 1)^2}{q_\mathrm{bin}} 
                            \left(\frac{a_\mathrm{p}}{a_\mathrm{bin}}\right)^{7/2}
                            \left(1 +\frac{3}{2} e_\mathrm{bin}^2\right)^{-1} 
                            T_\mathrm{bin} \,,
\end{equation}
where $a_\mathrm{p}$ is the semi-major axis of the particle. The same relation
was quoted by \citet{2017MNRAS.466.1170M} based on results by \citet{2015MNRAS.447..747L}.
A similar relation was given by \citet{2013ApJ...763..107L} without the $e_\mathrm{bin}$ term.

\begin{figure}
    \centering
    \resizebox{0.8\hsize}{!}{\includegraphics{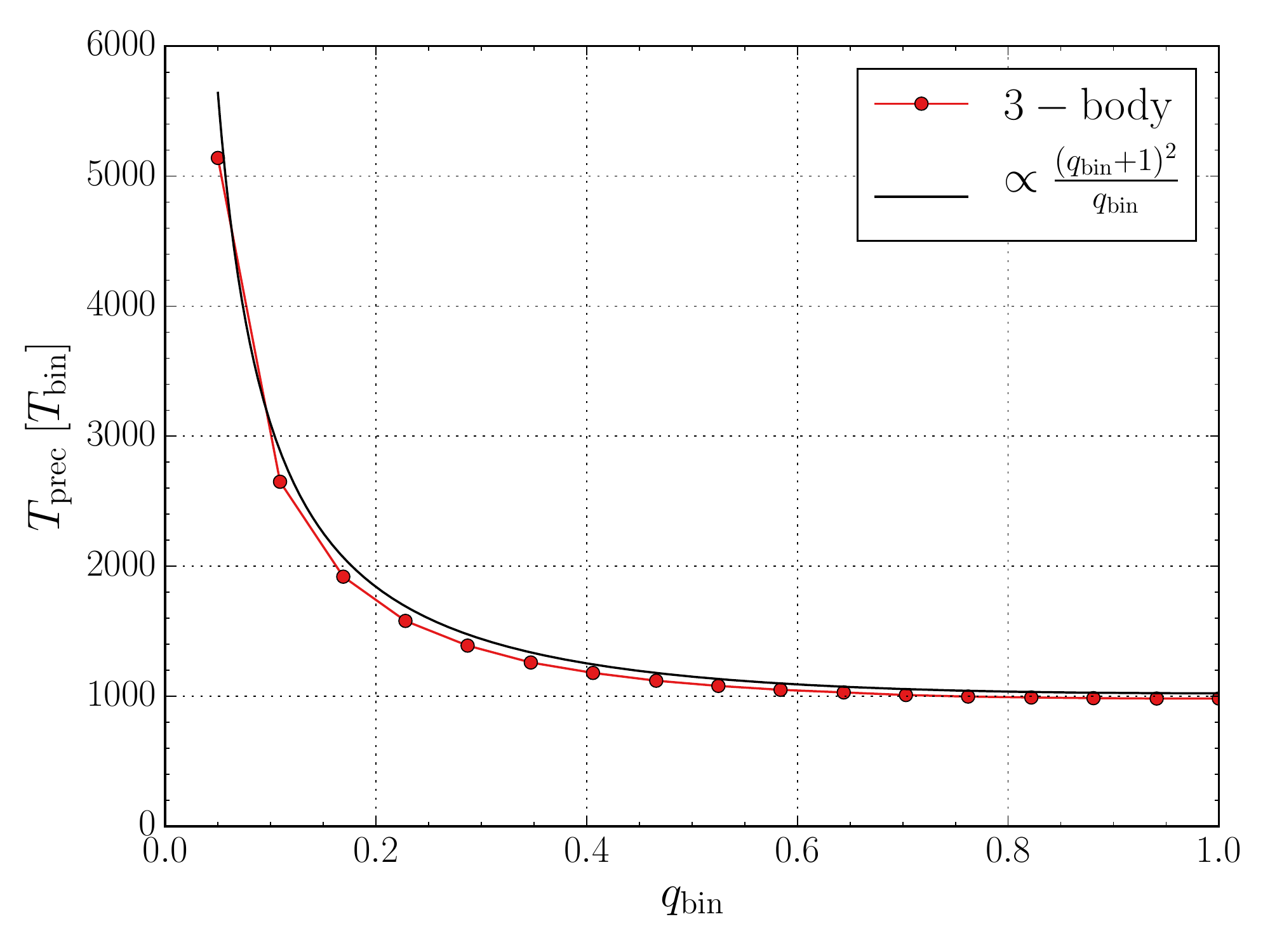}}\\
    \resizebox{0.8\hsize}{!}{\includegraphics{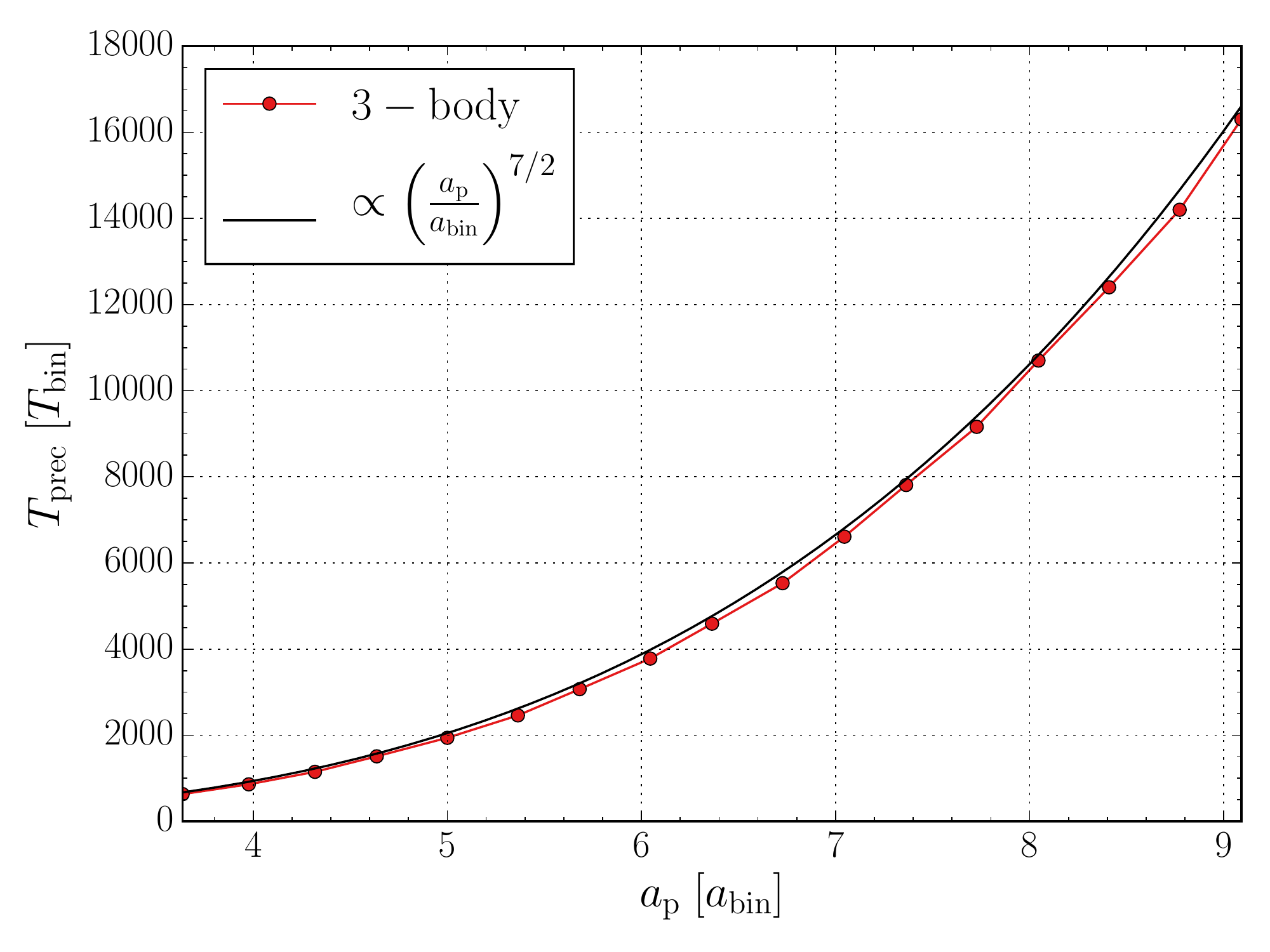}}\\
    \resizebox{0.8\hsize}{!}{\includegraphics{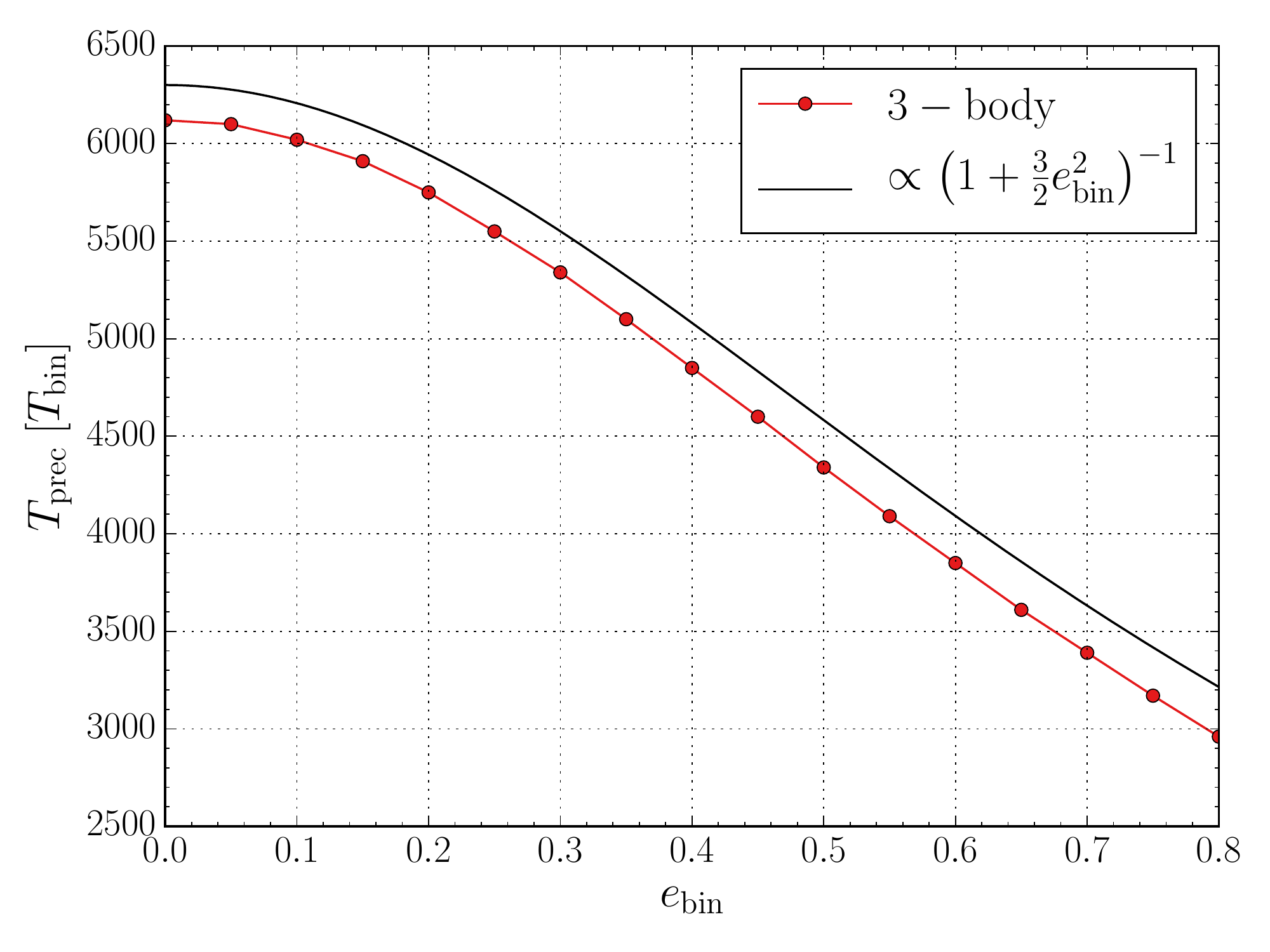}}\\
    \resizebox{0.8\hsize}{!}{\includegraphics{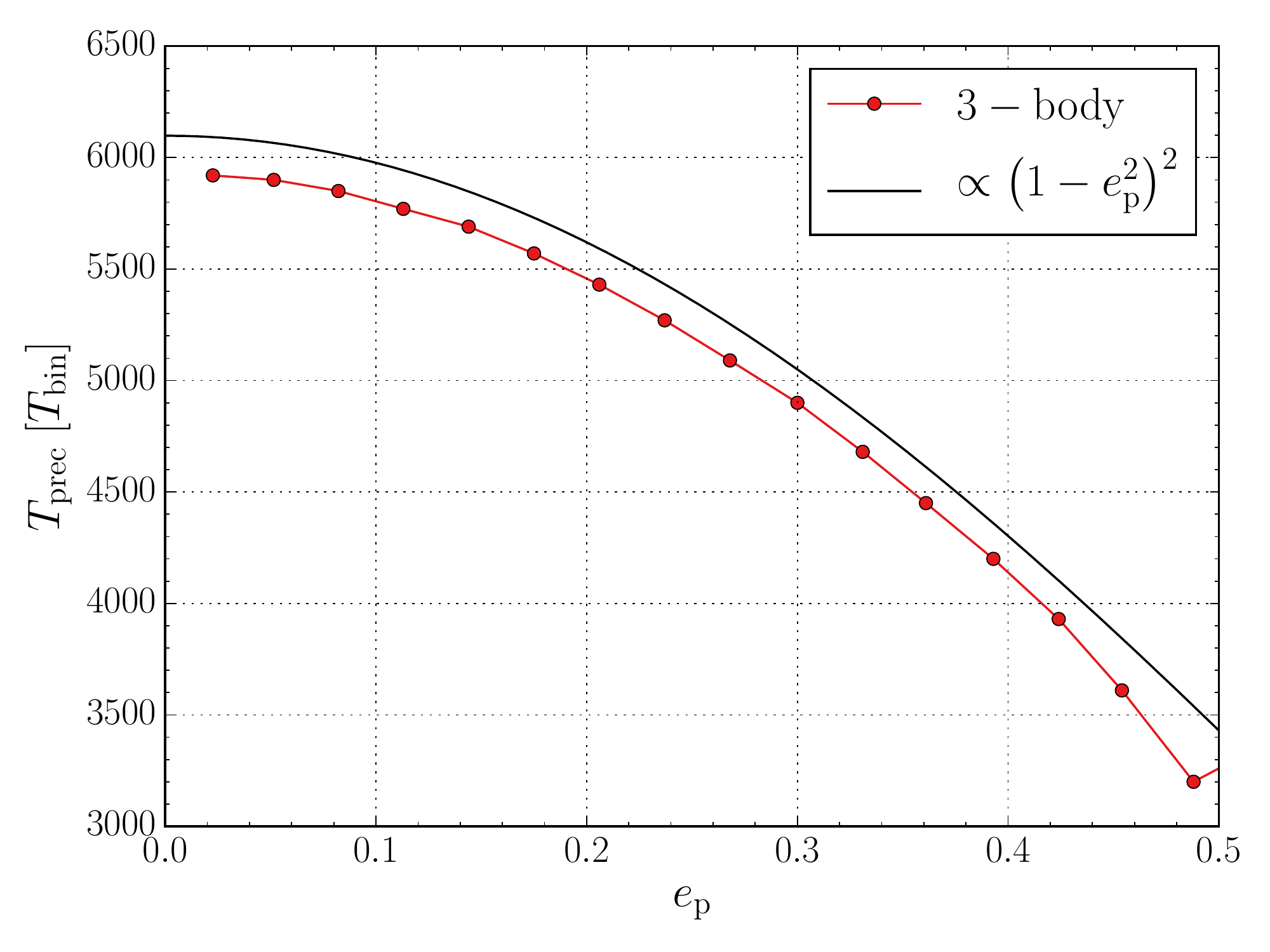}}%
    \caption{Scaling of the precession period of a particle orbiting a binary
    stars. As base binary parameter we chose the Kepler-16 parameters
    and for the planet $a_\mathrm{p} = \SI{1.50}{au}$ and $e_\mathrm{p}=0.05$.
    As an exception we chose for the 1st panel $a_\mathrm{p} = \SI{1.0}{au}$.
The black curves refer directly to eq.~(\ref{eq:particle_precx}) and the corresponding scaling is indicated.}
    \label{img:particle_prec}
\end{figure}

In order to verify the applicability of expression (\ref{eq:particle_prec}) when
varying of binary
and planet parameters, we performed series of three-body simulations with very
low planet mass ($10^{-6} M_\odot$),
where we varied individually $q_\mathrm{bin}$, $e_\mathrm{bin}$, and $a_\mathrm{p}$.
In addition we varied the planet eccentricity $e_\mathrm{p}$.
For the simulations we used the parameter of the Kepler-16 systems as a reference
(see Table~\ref{tab:kepler_16}) and integrated the system for several
\num{10000} binary orbits.
Our results of these three-body simulations are shown in Fig.~\ref{img:particle_prec}.
We find that the precession rate scales with $q_\mathrm{bin}$, $a_\mathrm{p}$, and $e_\mathrm{bin}$ exactly as expected
from relation (\ref{eq:particle_prec}). 
The fourth panel of Fig.~\ref{img:particle_prec} indicates that the  
precession period also depends on the particle eccentricity as
$T_\mathrm{prec} \sim (1 - e_\mathrm{p}^2)^{2}$, where we plotted the average eccentricity of the
orbit which is equivalent to the particle's free eccentricity.
The agreement holds up to about $e_\mathrm{p}^2 \sim 0.5$ for the used $a_\mathrm{p}$.
Clearly, for higher values of $e_\mathrm{p}$ the particle's
orbit will be unstable as this leads to close encounters with the binary. The value of $e_\mathrm{p}$ where
this happens depends on the distance from the binary star. For lower $a_\mathrm{p}$ the range will be more limited.
In summary, we find for the precession period of a particle around a binary star
\begin{equation}\label{eq:particle_precx}
    T_\mathrm{prec} = \frac{4}{3}
                            \frac{(q_\mathrm{bin} + 1)^2}{q_\mathrm{bin}}
                            \left(\frac{a_\mathrm{p}}{a_\mathrm{bin}}\right)^{7/2}
                            \frac{\left(1 - e_\mathrm{p}^2\right)^{2}}
                                 {\left(1 + \frac{3}{2} e_\mathrm{bin}^2\right)} \,
                            T_\mathrm{bin} \,.
\end{equation}
This relation is plotted in Fig.~\ref{img:particle_prec} as the solid black line. 
In addition to the scaling behaviour we find that the numerical results are about $3-5\%$ 
lower than the theoretical estimate. The agreement of the numerical results with the theoretical
prediction becomes better for larger $a_\mathrm{p}$ because eq.~(\ref{eq:particle_precx}) is
derived from an approximation for large $a_\mathrm{p}/a_\mathrm{bin}$.
The full relation (\ref{eq:particle_precx}) can be inferred directly from eq.~(11) in
\citet{2015ApJ...802...94G}.

\begin{figure}
    \resizebox{0.8\hsize}{!}{\includegraphics{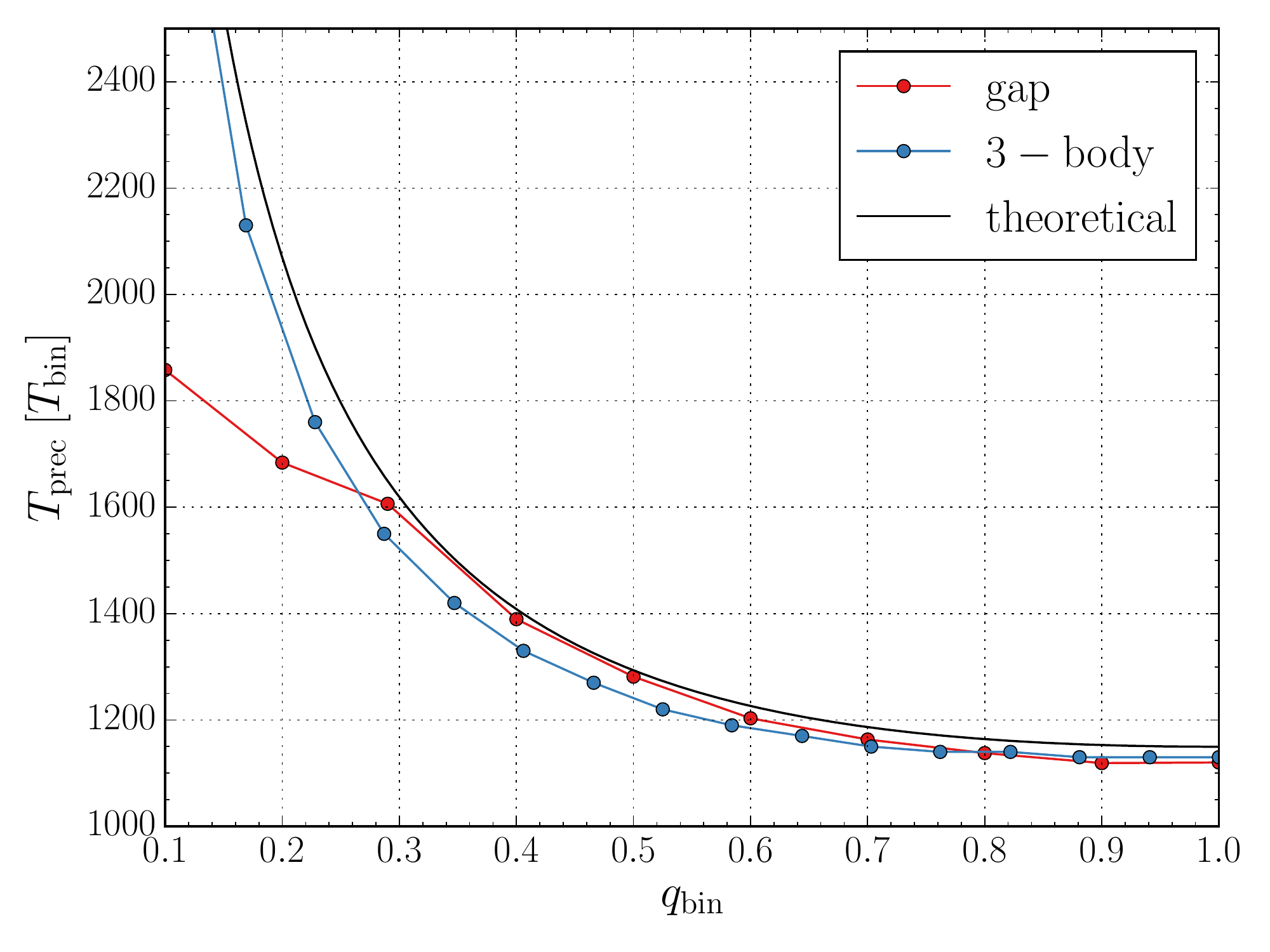}}
    \caption{Comparison of the precession period of the inner disc with the
    precession period of free particle. For the particle we used $a_\mathrm{p} = 4.9\,a_\mathrm{bin}, e_\mathrm{p} = 0.25$. 
    To obtain the shown agreement the
    particle's semi-major axis $a_\mathrm{p}$ has to be roughly 20 percent
    longer than
    $a_\mathrm{gap}$. The theoretical line was calculated with equation~\eqref{eq:particle_precx}.
    }
    \label{img:mass_ratio_particle}
\end{figure}

Now we compare the particle precession rate to our disc simulations. 
The best option for this comparison is to check the scaling of the precession rate for
models where $q_\mathrm{bin}$ has been varied because 
for binary mass ratio greater than $q_\mathrm{bin} = 0.3$
the gap size and eccentricity do not change much (purple points in  Fig.~\ref{img:e_T_gap_vs_a_gap}).
This is supported by the small variation of the gap radii in Fig.~\ref{img:gap_peri_mass_ratio}.
For higher mass ratios the gap size is approximately $a_\mathrm{gap} =
4.0\,a_\mathrm{bin}$ and $e_\mathrm{gap} = 0.27$. 
As seen from eq.~(\ref{eq:particle_precx}), a test particle with these orbit
elements would have a precession period that is far too short.
However, if we assume that the test particle has a semi-major axis of $a_p = 4.9\,a_\mathrm{bin}$
(roughly 20 percent more than $a_\mathrm{gap}$), the
precession period of the gap and the particle match very well for different $q_\mathrm{bin}$
(Fig.~\ref{img:mass_ratio_particle}), at least for the higher mass ratios. 
In general, however, due to the strong sensitivity of the precession period with
$a_\mathrm{p}$, 
an exact agreement will be difficult to obtain. 

\end{appendix}

\end{document}